\documentclass[12pt]{iopart}
\usepackage{graphicx}
\usepackage{amssymb}
\newcommand{\cut}[2]{} %first bracket provide reasoning!
\newcommand{\sign}[0]{\mathrm{sign}}
\newcommand{\Freeenergy}[0]{F}

\newcommand{\Factorgraph}[0]{\mathcal{F}}

\newcommand{\vectorin}[1]{\vec{#1}}

\newcommand{\Numberoffields}{{\mathcal{M}}}
\newcommand{\populationindex}{a}
\newcommand{\Indicator}{\mathbb{I}}
\newcommand{\atanh}{\mathrm{atanh}}
\newcommand{\sech}{\mathrm{sech}}
\begin{document}

\title[NNNI models on random graphs]{Next nearest neighbour Ising models on random graphs}

\author{Jack Raymond$^1$ and K.Y. Michael Wong$^2$}
\address{$ˆ1$ Dipartimento di Fisica, Universita di Roma ''La Sapienza'', Piazzale Aldo Moro 2, Roma I-00185, Italy}
\address{$ˆ2$ Department of Physics, Hong Kong University of Science and Technology, Hong Kong, China}
\ead{jack.raymond@physics.org}
%\thanks{}
%\subjclass{Condensed Matter, Disordered Systems}%
%\keywords{Cavity Method, Spin Glass, ANNNI, Vannimenus Model, b-Matching, Extremal Optimization}%

\date{18 June 2012}%
%\dedicatory{}%
%\commby{}%
% ----------------------------------------------------------------
\begin{abstract}
This paper develops results for the next nearest neighbour Ising model on random graphs.
Besides being an essential ingredient in classic models for frustrated systems, second neighbour interactions interactions arise naturally in several applications such as the colour diversity problem and graphical games. 
We demonstrate ensembles of random graphs, including regular connectivity graphs, that have a periodic variation of free energy, with either the ratio of nearest to next nearest couplings, or the mean number of nearest neighbours.
When the coupling ratio is integer paramagnetic phases can be found at zero temperature.
This is shown to be related to the locked or unlocked nature of the interactions.
For anti-ferromagnetic couplings, spin glass phases are demonstrated at low temperature.
The interaction structure is formulated as a factor graph, the solution on a tree is developed.
The replica symmetric and energetic one-step replica symmetry breaking solution is developed using the cavity method.
We calculate within these frameworks the phase diagram and demonstrate the existence of dynamical transitions at zero temperature for cases of anti-ferromagnetic coupling on regular and inhomogeneous random graphs.
\end{abstract}
%\submitto{\JPG}
\maketitle
% ----------------------------------------------------------------

\section{Introduction} \label{sec:introduction}

Many systems exhibit complicated low temperature phases that are described at a
phenomenological level by classical spin state lattice models. A necessary
ingredient in the description is often frustration in interactions, and one of the
simplest origins of frustration is a competition between nearest and next nearest
neighbour couplings. The simplest models can already demonstrate a range of phenomena
with interesting long range correlation and modulation patterns. At the same time
many problems in optimization have a similar structure, an objective function with
solutions equivalent to the ground states of some next nearest neighbour Ising model.
In these optimization problems the ground state space structure can be related to
issues of algorithmic complexity in resolving decision problems and sampling.

Theoretical developments based on the cavity and replica method have allowed greater
insight into these problems in both traditional solid state physics, theoretical
computer science and other cross disciplinary fields~\cite{Mezard:SGT,Nishimori:SP}.
This includes dilution effects that mimic the finite range of interaction of spins,
with much recent focus on random graphical models. Random graph models allow
dilution effects to be studied within a mean-field analysis. More recently, much
development of the theory of the ground state space of Ising models on random graphs
has occurred in the guise of constraint satisfaction
problems~\cite{Krzakala:GS,Montanari:NLT}. Random graph models, including
inhomogeneous connectivity, provide benchmarks for optimization problems such as
kSAT and colouring, and a basis for powerful coding methods.

Two famous models in physics involving frustrated couplings are the Axial Next
Nearest Neighbour Ising (ANNNI) Model~\cite{Selke:ANNNI}, and the Next Nearest
Neighbour Ising model (NNNI)~\cite{Domb:2D}. The difference between the two is that
couplings in the former are frustrated only along a particular axis, whereas in the
latter interactions are symmetric along all axes. The thermodynamics in the bulk
depend on the ratio of the nearest to next nearest couplings, and the lattice
structure. In these models the ground state is normally dominated by ferromagnetic
and modulated $\langle2\rangle$ ordering. The competition of the nearest and next
nearest interactions leads to a weakening of long range correlations, and hence a
decreased Curie Temperature, or even the absence of a transition at some special
coupling ratios. A simple argument based on the local balancing of ferromagnetic and
modulated tendencies provides an accurate prediction for some of these ratios
~\cite{Domb:2D,Stephenson:IM}. At non-zero temperature long range correlations are
changed, and other modulation patterns may be thermodynamically dominant, or
relevant as meta-stable attractors for dynamics. At some special ratios one can
raise the temperature and observe an order from disorder effect~\cite{Reimers:OBD}.
A Devil's staircase phenomenon can also be found varying the coupling ratio at fixed
temperature~\cite{Bak:MFT}.

In understanding the phase diagram, locally tree-like models have become popular.
The Vannimenus model was proposed as the Bethe-Peierls approximation to the ANNNI
model ~\cite{Vannimenus:PDIM,Moreira:MS,Yokoi:SA}, with analogous studies in the
case of the NNNI model~\cite{Inawashiro:IM,Katsura:BL,Ganikhodjaev:ES}. These
reproduce many thermodynamic features of the finite dimensional models including
special ratios. Other frustrated next nearest neighbour models have been studied on
Bethe Lattices and Cayley trees coinciding quantitatively with ANNNI or NNNI models
in some parameterizations. Amongst these are Husimi Cactii Ising spin models, Potts
and continuous spin NNN models and third nearest neighbour models
~\cite{daSilva:IMBL,Ganikhodjaev:PD,Melin:SGB,Monroe:PD,Chandra:SL,Ostilli:PDIM,Moreira:IMTNN}.

The purpose of this paper is to develop an approach to the NNNI model utilizing
insights from hard optimization problems and the cavity method. Our original
motivation was the diversity colouring problem, an algorithmically hard problem 
%of resource allocation 
that finds its modern applications in distributed storage
on networks that are typically
disordered~\cite{Wong:MU,Bounkong:CRG,Pelizzola:SMS}. A statistical physics approach
for the case to the anti-ferromagnetic Potts model (Q-colour states) has been
previously studied, demonstrating spin glass and paramagnetic phases~\cite{Wong:MU}.
In this paper the Ising spin case is studied for general coupling ratios, and a
one-step replica symmetry breaking analysis is undertaken to demonstrate a more
precise description of the phase space that is expected to generalize in many ways
to Potts spin models. 

A closely related application of NNNI models is in the analysis of pure Nash Equilibria in graphical games~\cite{Kearns:GMGT,Ramezanpour:SPGG}, the distributed storage problem may also be posed in such a framework. In this context each node (player) on a graph may choose a strategy (such as $\pm 1$) so as to maximize his objective that depends on the action of his neighbours, this objective is often a simple pairwise function. Since the optimal strategy depends on the joint distribution of the neighbours' strategies there is a second neighbour interaction through the intermediate player. Certain types of Nash equilibria can be shown to correspond to the ground states of a NNN model, and other thermodynamic properties can have game theoretic interpretations. 

In section \ref{sec:the_model} we begin our analysis with a formulation of a
graphical model that enables the solution on a tree to be presented. In section
\ref{sec:free_energy_on_a_tree} we develop the free energy for a tree as a simple
exact solution of the graphical model. In section \ref{sec:special_ratios} we
identify a simple relationship between coupling ratios and connectivity, identifying
special ratios based on local properties in the graphical model. In section
\ref{sec:small_system_studies} we then present ground state statistics for some
loopy graphs of small size sampled from the linear ensemble, which demonstrates
clearly the significance of the special ratios.

In section \ref{sec:cavity_method} replica symmetric (RS) and energetic one-step
replica symmetry breaking (1RSB) forms of the cavity method are developed.
Equivalence between our model and classes of constraint satisfaction problems (CSPs)
are demonstrated. In section \ref{sec:results_of_the_cavity_method} the phase
diagram is presented as a function of temperature, connectivity and coupling
ratio~\cite{Mezard:CMZT} at the level of replica symmetry for cases of
anti-ferromagnetic nearest and next nearest couplings. The phase diagram allows a
paramagnetic phase at high temperature, which at the special coupling ratios can
persist to zero temperature. At low temperature we find a spin glass solution. For
graphs sampled with regular or linear connectivity we consider special coupling
ratios and show that with variation of connectivity, there may exist both
paramagnetic and strongly correlated extensive entropy phases at zero temperature.
Paramagnetic behaviour is shown to follow a predictable periodic pattern with mean
connectivity. In regimes with a competition of nearest and next nearest couplings
both dynamical and continuous transitions from paramagnetic to spin glass phases are
shown. The replica symmetric and zero temperature one-step replica symmetry breaking
solutions for the spin glass are found to be unstable; whereas the paramagnetic
solution is the unique stable solution in some parts of the parameter space. We
conclude with a discussion in section \ref{sec:discussion}.

Appendices include additional numerical results on small graphs, an investigation of
the support for the energetic 1RSB method, and details of the numerical methods used
to solve the cavity equations.

\section{The Model} \label{sec:the_model}

We consider a spin state model with an interaction structure described by a graph
$G(V,E)$, consisting of a set $V$ of $N$ vertices and a set $E$ of $M$ edges. Each
vertex $i$ in $V$ is associated with a spin variable $S_i = \pm 1$. Two vertices, if
connected by an edge, are called nearest neighbours, and denoted $\langle i,j
\rangle$. If two distinct vertices are both nearest neighbours of a third vertex,
then they are next nearest neighbours, and denoted $\rangle i,j \langle$. We consider
simple graphs, so that there are no self-loops, and no double edges -- vertices are
never nearest neighbours of themselves and never twice nearest neighbours of each
other. However, vertices may be second neighbours by multiple paths, and
$\mathcal{N}_{\rangle i,j \langle}$ denotes the number of such paths. Every path
contributes additively to the coupling. In our asymptotic analysis we will neglect
the effects of neighbourhoods in which vertex pairs are both nearest and next nearest
neighbours (due to a loop of length three), and cases where $N_{\rangle i,j\langle}$
exceeds one (due to loops of length four).

Given the set of first and second neighbours, we write the Hamiltonian
\begin{equation} \mathcal{H} = 2\lambda \sum_{\langle i,j\rangle} S_i S_j  +
\sum_{\rangle i,j\langle} N_{\rangle i,j\langle} S_i S_j \label{eq:Hamiltonian} \;,
\end{equation} which is an Ising spin model with the ratio of first and second
neighbour couplings defined by $\lambda$. The second neighbour couplings are always
anti-ferromagnetic, positive $\lambda$ indicates anti-ferromagnetic nearest neighbour
couplings and negative $\lambda$ indicates ferromagnetic nearest neighbour couplings.
Frustration, the key feature of the model, may be present even where $\lambda=0$,
though we concentrate in this paper on the case of positive $\lambda$.

\begin{figure}[h!tbp] \begin{center}
\includegraphics[width=\linewidth]{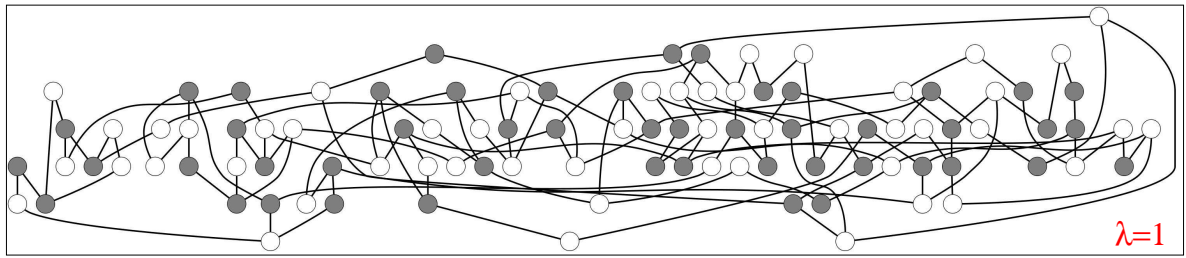}
\caption{\label{fig:groundstateexample} In our model spins interact according to a
random graph structure. We illustrate a ground state on a graph of $N=100$ vertices,
sampled from a linear connectivity ensemble. All vertices have connectivity two or
three, mean connectivity is $2.56$, and $\lambda=1$. Black indicate $+1$ spin
assignments, and white $-1$.} \end{center} \end{figure}

When considering macroscopic properties we consider a graph sampled from an
ensemble, an example is shown in figure \ref{fig:groundstateexample}, for which we
also illustrate a ground state assignment of the spins. We consider mainly an
ensemble of random graphs subject to a linear connectivity distribution constraint,
parameterized by mean connectivity ${\bar C}$. A graph from this ensemble is
equivalent to a random regular connectivity graph when the mean connectivity is an
integer, or else contains vertices with connectivity $\lfloor{\bar C}\rfloor$ and
$\lceil{\bar C}\rceil$ such that the connectivity distribution is described by
\begin{equation} P(C) = \delta_{C,\lfloor {\bar C}\rfloor}(\lceil {\bar C}\rceil -
{\bar C}) + \delta_{C,\lceil {\bar C}\rceil} ({\bar C} - \lfloor {\bar C}\rfloor)
\;. \label{eq:LinConnDist} \end{equation} Here $\lfloor x \rfloor$ ($\lceil
x\rceil$) notation is used to denote the nearest integer smaller (respectively
larger) than or equal to $x$.

Some topological definitions are useful: a loop is a closed sequence of vertices
connected by edges; a leaf is a vertex of connectivity one; a tree is a connected graph
without loops. This terminology extends naturally to the factor graph defined
later.

Assuming the linear ensemble, the model can be parameterized by mean connectivity
${\bar C}$, the coupling ratio $\lambda$, and we also consider its dependence on the
inverse temperature $\beta$.

\subsection{The factor graph representation}
\label{ssec:the_factor_graph_representation}

The factor graph representation is a bipartite graph representation of the model and
can be derived following the cluster variational method at distance one, or
equivalently a simple application of the region graph method~\cite{yedidia,Kschischang:FG}. Alternatively, for those familiar with the junction tree method, one can realize that ignoring long loops the width of the graph is locally two -- as such one can define pseudo-states to create a locally tree-like graph in the Markov sense -- this is precisely the graph we formulate.
The factor graph encodes all nearest and next nearest neighbour couplings amongst
spins in terms of multi-body interactions amongst some generalized variables. Under
this framework any neighbourhood in the graph $G(V,E)$ without loops is mapped to a
region in the factor graph without short loops between generalized variables; so
allowing locally tree-like methods to be applied.

We first transform the Hamiltonian (\ref{eq:Hamiltonian}) to a sum of local
multi-body interactions. For each vertex in the graph, $i$, we define a multi-body
interaction dependent on the spin $S_i$ and the set of nearest neighbour spins
$S_{nn(i)}$. The multi-body interaction at factor $i$ is \begin{equation} H_i(S_i,
S_{nn(i)}) = \frac{1}{2}\left(\lambda S_i + \sum_{j\in nn(i)} S_j \right)^2 -
H^0_i(C_i,\lambda) \label{eq:FactorHami} \;, \end{equation} with $nn(i)$ the set of
all vertices that are the nearest neighbours of $i$, and $C_i=|S_{nn(i)}|$ is the
vertex connectivity. The second term normalizes the local ground state energy to
zero, \begin{equation} H^0_i(C_i,\lambda) = \min_{S_i, S_{nn(i)}} H_i(S_i,
S_{nn(i)}) \label{eq:FactorHamNorm} \;. \end{equation} The Hamiltonian is
\begin{equation}
 H(\vec{S}) = \sum_i H_i(S_i,S_{nn(i)})
 \label{eq:FactorHam}
 \;,
\end{equation} which is equivalent to (\ref{eq:Hamiltonian}) up to a quenched
additive term, and is conveniently bounded below by zero.

Generalized variables, which we can henceforth call g-spins, are introduced in
one-to-one correspondence with the edges of $G(V,E)$. A g-spin, located at edge
$(i,j)$ is an ordered spin pair: created from copies of the spins $S_i$ and $S_j$.
Factors are introduced in one-to-one correspondence with the vertices in the
original graph. A factor $i$ is connected by edges to the $C_i$ generalized
variables $\{(S_i,S_j)| j\in nn(i)\}$. Factor $i$ incorporates intersection
constraints and the multi-body energetic term $H_i$ (\ref{eq:FactorHami}). The
intersection constraint at $i$ requires that, for any Hamiltonian or temperature,
the elements $S_i$ in each neighbouring g-spin are identical. Note that although the
state space expands to $(2^2)^M$ in the new model, there are only $2^N$ assignments
consistent with the intersection constraints as expected. Frequently we will be
interested in whether the state has two spins aligned, which we call a dimer, or
misaligned, which is called a non-dimer. Note the intersection constraint depends on
the parity of the constituent spins, but the energy does not.

\begin{figure}[h!tbp] \begin{center}
\includegraphics[width=\linewidth]{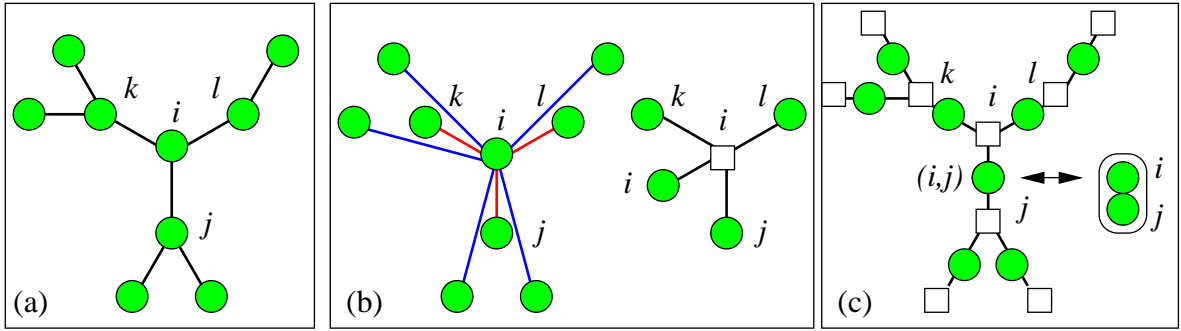}
\caption{\label{fig:factorgraph}(a) A locally tree-like subgraph of $G(V,E)$ is
demonstrated, we label the central vertex and its nearest neighbours; vertices are
shown by circles and have connectivity 1, 2 or 3 in the example. (b) With circles
representing spin variables the interactions can be represented either by two types
of binary couplings (left), or by multi-body interactions (right). The multi-body
interaction labeled $i$ encodes $H_i(S_i,S_{nn(i)})$. Spin $i$ interacts in three
other multi-body interactions (not shown) since it is also in the sets $S_{nn(j)}$,
$S_{nn(k)}$ and $S_{nn(l)}$. (c) g-spins variables can be associated with the edges
of the graph $G(V,E)$ and are represented as circles. Each factor encodes an
energetic multi-body interaction, and an intersection constraint. The graph is
locally tree-like, unlike the interaction structures implied by figure (b).}
\end{center} \end{figure}

To summarize, the factor graph consists of a set of $N$ factor vertices, a set of
$M$ variable vertices, and a set of edges representing their dependencies. Each
variable is called a g-spin and has four states. The relationships with the original
graph are visualized in figure \ref{fig:factorgraph}, where factors are represented
by squares and generalized variables by circles. The factor graph is found
deterministically from $G(V,E)$: \begin{itemize} \item[1)] Each factor vertex
corresponds to an element in $V$, vertex labels $i$ define factors uniquely;
\item[2)] Each variable vertex corresponds to an element in $E$, directed edge
labels define g-spin(s) uniquely $(i,j)$. \item[3)] An edge exists from generalized
variable $(j,k)$ to factor vertex $i$ only if $i=j$ or $i=k$. \end{itemize} By
construction every variable vertex in the factor graph is connected to exactly two
factors, and each factor vertex is attached to $C_i$ variable vertices.

\subsubsection{Cavity graphs} \label{sssec:cavity_graphs} The cavity graph
$\Factorgraph_{i}$ is a subgraph of the full factor graph formed by removing one
factor $i$ and its adjacent edges. In this subgraph g-spins previously attached to
factor $i$ have become exceptional: they are connected to only one factor. The
factor graph $\Factorgraph_{j \rightarrow i}$ is that part of the cavity graph
$\Factorgraph_i$ connected to g-spin $(S_i,S_j)$. The $C_i$ cavity graphs derived
from $\Factorgraph_i$ ($\Factorgraph_{*\rightarrow i}$) are disjoint if $G(V,E)$ is
a tree.

\begin{figure}[h!tbp] \begin{center}
\includegraphics[width=\linewidth]{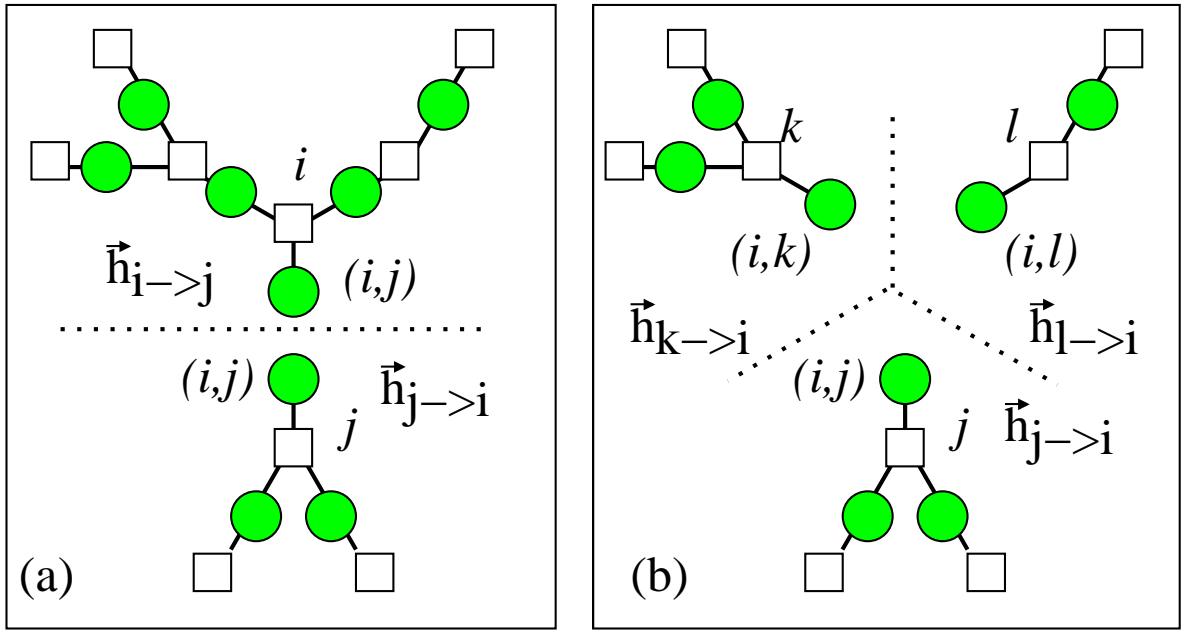}
\caption{\label{fig:cavitygraph} Beginning with a subgraph of the full factor graph
figure \ref{fig:factorgraph}(c), this figure demonstrates the construction of the
free energy from cavity graph fields. (a) Knowing the fields on two factor graphs
convergent on a single edge $(i,j)$ these can be converged to give the edge free
energy (\ref{eq:freeenergyij}). (b) Calculation of (\ref{eq:freeenergyi}) is
possible given the $C_i$ factor graphs convergent on a factor $i$. The probability
of the root state marginalised over all other g-spins is encoded in the vector
$\vec{h}_{* \rightarrow *}$ in a tree.} \end{center} \end{figure}

To describe relationships on the factor tree we use the terminology of trees. In a
factor graph $\Factorgraph_{j \rightarrow i}$ we say that generalized variable
vertex $(i,j)$ is the root of the graph. The root (ancestor) has $C_j-1$
descendants, which are the vertices attached to factor $j$ excluding the root
itself. Factor graphs of the full model are shown in figure \ref{fig:factorgraph},
with some additional notation relating to the cavity method shortly to be
discussed.

\subsection{Auxiliary Hamiltonian} In developing our analysis the variety of energy
levels applicable in the multi-body interaction can be problematic. An auxiliary
Hamiltonian is proposed and utilized in the 1RSB method. The local ground state
energy continues to be zero energy, but all excitations are fixed to energy one;
(\ref{eq:FactorHami}) is mapped to the new form by \begin{equation}
H_i(S_i,S_{nn(i)}) = \mathbb{I}\left(H_i(S_i,S_{nn(i)})>0\right)
\;.\label{eq:FactorHami2} \end{equation} using the indicator function $\mathbb{I}$
that evaluates to $1$ if the argument is true and zero otherwise. Note that (\ref{eq:FactorHami}) and (\ref{eq:FactorHami2}) are interchangeable only if local excitations are absent. Thermodynamically this requires both that the temperature is zero, and that the ground state is of zero energy (unfrustrated), which is not true for many of the ensembles studied herein. This Hamiltonian is exploited only in the zero temperature analysis within this paper, use of
(\ref{eq:FactorHami}) is to be assumed unless otherwise stated.

\subsection{Cut-Poissonian connectivity distribution}
\label{app:cut-poisson_poissonian_and_other_ensembles}

The Cut-Poisson ensemble is also considered in experiments. The marginal probability
distribution is given by \begin{equation} P(C) = \mathbb{I}(C \geq 2)\frac{\exp
\{2-{\bar C}\}({\bar C}-2)^C}{(C-2)!} \;, \label{eq:cutpoisson} \end{equation} The
connectivity at any site is two plus a Poisson distributed random number.

Both ensembles presented in (\ref{eq:cutpoisson}) and (\ref{eq:LinConnDist}) have minimum node connectivity two, and have modest variance in node connectivity. The ensembles show a degree of variability in results, whilst avoiding some more simple sources of variation such as the size of graph cores. Henceforth use of the linear ensemble (\ref{eq:LinConnDist}) is to be assumed unless otherwise stated.

\section{Free energy on a tree} \label{sec:free_energy_on_a_tree}

The partition function where $G(V,E)$ is a tree can be determined recursively and
efficiently by the Bethe-Peierls recursion. The cavity method allows an extension of
this method to locally tree-like graphs in section \ref{sec:cavity_method}. This
formalism is equivalent to that employed in \cite{Wong:MU}, in which it was applied
to Potts spins with $\lambda=1$.

The partition function can be factorized about an arbitrary g-spin $(S_i,S_j)$,
using g-spin probabilities and partition functions on cavity trees. The two cavity
graphs $\Factorgraph_{i \rightarrow j}$ and $\Factorgraph_{j \rightarrow i}$ are
trees, and non-interacting given $(S_i,S_j)$. Thus the partition function of the
full graph is \begin{equation} Z(\Factorgraph) = Z_{(i,j)}(\Factorgraph) =
\sum_{S_i,S_j} \left[ Z_{i \rightarrow j} P_{i \rightarrow j}(S_i,S_j)  \right]
\left[ Z_{j \rightarrow i} P_{j \rightarrow i}(S_j,S_i)  \right]
\label{eq:treepartitionij} \;, \end{equation} $Z_{i \rightarrow j}$ is the partition
function on the corresponding cavity tree, summing over all g-spins including
$(S_i,S_j)$. We define $P_{i \rightarrow j}(S_i,S_j)$ to be the probability of a
g-spin $(S_i,S_j)$ on a cavity graph $\Factorgraph_{i \rightarrow j}$, marginalising
over all other g-spins in the cavity graph.

To determine the partition function on the cavity tree we decompose the problem
iteratively; the partition function on the factor graph $\Factorgraph_{j \rightarrow
i}$ can be written in terms of partition functions on descendant factor graphs
$\{\Factorgraph_{k\rightarrow j} | k \in nn(j) \setminus i\}$; the notation means
$k$ are the factors neighbouring $j$, excluding $i$. Referring to the schematic of
figure \ref{fig:cavitygraph} and references~\cite{yedidia,Kschischang:FG}, the
recursion is \begin{equation} P_{j \rightarrow i}(S_j,S_i) \propto \prod_{k\in
nn(j)\setminus i} \left[\sum_{S_k} P_{k\rightarrow j}(S_k,S_j) \right]
\exp\left\lbrace - \beta H_j(S_j,S_{nn(j)}) \right\rbrace \;,
\label{eq:PSiSjrecursion} \end{equation} and \begin{equation} Z_{j \rightarrow i} =
\sum_{S_i,S_j} \prod_{k\in nn(j)\setminus i} \left[\sum_{S_k} Z_{k\rightarrow j}
P_{k\rightarrow j}(S_k,S_j) \right] \exp\left\lbrace -\beta H_j(S_j,S_{nn(j)})
\right\rbrace \label{eq:Zrecursion} \;. \end{equation} The multi-body interaction is
kept in a general form, whereas the intersection constraints are implicit in the
appearance of the same spin $S_j$ in each g-spin probability.

The three degrees of freedom describing the probability distribution
(\ref{eq:PSiSjrecursion}) can be expressed generally in terms of a coupling and two
fields \begin{equation}
 P_{j \rightarrow i}(S_j,S_i) = P_{J_{j \rightarrow i}, h^{f}_{j \rightarrow i},
 h^{b}_{j \rightarrow i}}(S_j,S_i)
\label{eq:PSiSjparameterization}
 \;,
\end{equation} with \begin{equation}
 P_{J, h^{f}, h^{b}}(S_j,S_i) \propto \exp\left\lbrace \beta \left( J S_i S_j + h^f
 S_i + h^b S_j \right) \right\rbrace
\label{eq:Pjtoi_siteindependent} \;, \end{equation} and $\beta$ introduced as a
known scaling. The superscripts are used to distinguish the ancestor (forward) spin
field $h^f$ and descendant (backward) spin field $h^b$.

The probability distribution recursion (\ref{eq:PSiSjrecursion}) parameterized by (\ref{eq:PSiSjparameterization}) may be summarized as a non-linear mapping between the three-component cavity fields denoted by \begin{equation} \vec{h}_{j \rightarrow i} = \left(J_{j \rightarrow i}, h^{f}_{j \rightarrow i}, h^{b}_{j \rightarrow i} \right) \label{eq:fieldvector} \;, \end{equation} allowing (\ref{eq:PSiSjrecursion}) to be written \begin{equation} \vec{h}_{j \rightarrow i} = {\hat T}(\{ \vec{h}_{k \rightarrow j} | k \in nn(j) \setminus i\},1/\beta) \;. \label{eq:T} \end{equation} Note that the graph dependence is only in the arguments of ${\hat T}$, and not in the mapping ${\hat T}$ itself. We have written here explicitly the dependence on the temperature $1/\beta$, for the convenience of taking the zero temperature limit in the cavity method.

\subsubsection{Free energy} \label{sssec:free_energy} Using
(\ref{eq:treepartitionij}), the free energy can be calculated from the partition
function at vertex $(i,j)$ as shown figure \ref{fig:factorgraph}(a)
\begin{equation}
 - \beta \Freeenergy_{ij} = \log Z_{i \rightarrow j} + \log Z_{j \rightarrow i} -
 \beta \Freeenergy_{E}(\vec{h}_{i \rightarrow j},\vec{h}_{j \rightarrow i})
\label{eq:freeenergyedge} \;, \end{equation} where we call the edge free energy
\begin{equation} \begin{array}{lll} - \beta \Freeenergy_{E}(\vec{h}_1,\vec{h}_2) &=&
\log \left(\sum_{S_i,S_j}  P_{\vec{h}_1}(S_j,S_i) P_{\vec{h}_2}(S_i,S_j)\right) \;.
\label{eq:freeenergyij} \end{array} \end{equation} An alternative calculation of the
partition function shown figure \ref{fig:factorgraph}(b) is achieved by an expansion
about factor $i$, first creating the factor graph $\Factorgraph_i$
\begin{equation}
\fl Z(\Factorgraph) = Z_i(\Factorgraph) = \sum_{S_i} \prod_{j\in nn(i)} \left[ Z_{j
\rightarrow i} \sum_{S_j} P_{j \rightarrow i}(S_j, S_i) \right] \exp \left\lbrace -
\beta H_i(S_i,S_{nn(i)})\right\rbrace \label{eq:treepartitioni} \;. \end{equation}
From this we have an alternative definition of the free energy \begin{equation}
-\beta \Freeenergy_i = \sum_{j\in nn(i)} \log Z_{j \rightarrow i} - \beta
\Freeenergy_{V}(\{\vec{h}_{j \rightarrow i} | j \in nn(i)\},1/\beta,H_i)
\label{eq:freeenergyvertex} \;, \end{equation} defining the vertex free energy
\begin{equation} \fl    -\beta \Freeenergy_{V}(\vec{h}_1 \ldots \vec{h}_{C},1/\beta,H) =
\log\!\!\left(\!\mathrm{Tr}_{S} \!\prod_{j=1}^{C}\left[ P_{\vec{h}_c}(S_c,S_0)\right] \exp
\left\lbrace - \beta H(S_0,\{S_{1},\ldots,S_{C}\}) \!\right\rbrace \!\right)
\label{eq:freeenergyi} \;\!. \end{equation} The trace $\mathrm{Tr}_{S}$ is a sum over
all spin variables, and $H$ is the Hamiltonian function.

In the tree the number of factors is $N$ and the number of variables is $M=N-1$. By
summing the $\Freeenergy_{i}$ and subtracting $\Freeenergy_{i j}$, we have an
expression dependent only on $\vec{h}$ for every edge
\begin{equation} \fl    (N-M)
\Freeenergy = \sum_i \Freeenergy_{V} \left(\{\vec{h}_{j \rightarrow i}| j \in
nn(i)\},1/\beta,H_i\right) - \sum_{(i,j)} \Freeenergy_{E}\left(\vec{h}_{i
\rightarrow j},\vec{h}_{j \rightarrow i} \right) \label{eq:freeenergyNoZij} \;.
\end{equation}

Another informative quantity is the free energy change when merging $C_j-1$ descendant trees to generate a new tree (\ref{eq:treepartitionij}). This can be written as a function of the $C_j-1$ descendant fields \begin{equation} \Delta F_{j \rightarrow i} = {\hat T}_F (\{ \vec{h}_{k \rightarrow j} | k \in nn(j) \setminus i\},1/\beta) \label{eq:DeltaF} \;, \end{equation} where we define the site independent function
\begin{equation}
\fl {\hat T}_F (\{ \vec{h}_{1} \ldots \vec{h}_{C-1}, 1/\beta) = - \frac{1}{\beta}\log \mathrm{Tr}_S \prod_{c=1}^{C-1} \left[P_{\vec{h}_{c}}( S_c,S_0) \right] \exp\left\lbrace -\beta H(S_0,\{S_1,\ldots,S_C\}) \right\rbrace \;. \label{eq:Tf} \end{equation}

\subsubsection{Marginals and order parameters}
\label{sssec:marginals_and_order_parameters} From the cavity fields $\{\vec{h}\}$ we
can derive relevant marginals. Taking the fields convergent on a particular neighbourhood we can define the joint probability distribution over the variables about a vertex
\begin{equation}
\fl P(S_i,S_{nn(i)}) \propto \exp \left\lbrace \beta \left( \sum_{j\in nn(i)} ( J_{j \rightarrow i} S_i S_j +
h^b_{j \rightarrow i} S_j + h^f_{j \rightarrow i} S_i) - H_i(S_i,S_{nn(i)}) \right)
\right\rbrace \label{eq:PSi} \;, \end{equation}
and a similar disribution may be defined for edges
\begin{equation}
\fl P(S_i,S_j) \propto \exp \left\lbrace \beta \left(
(J_{i \rightarrow j} + J_{j \rightarrow i}) S_i S_j + ( h^f_{i \rightarrow j} +
h^b_{j \rightarrow i}) S_j + ( h^b_{i \rightarrow j} + h^f_{j \rightarrow i}) S_i
\right) \right\rbrace \label{eq:PSiSj} \;. \end{equation}
A marginal probability for variable $i$, $P(S_i)$, can be found marginalising over either (\ref{eq:PSi}) or (\ref{eq:PSiSj}), which are constrained to give the same result. Summing the marginals, we
can find the order parameters and standard extensive observables. The
magnetization is \begin{equation} m_F = \frac{1}{N} \sum_{i=1}^N \sum_{S_i} S_i
P(S_i) \;, \label{eq:m_F} \end{equation} and we take the spin glass order parameter
to be \begin{equation} q_F = \frac{1}{N} \sum_{i} \left(\sum_{S_i} S_i P(S_i) -
m_F\right)^2 \;. \label{eq:q_F} \end{equation} Finally it is useful to define the
probability that a particular edge is in a dimer state, represented by a dimer
magnetization \begin{equation} m_D = \frac{1}{M} \sum_{\langle i,j \rangle}
\sum_{S_i,S_j} S_i S_j P(S_i,S_j) \;, \label{eq:m_D} \end{equation}

Let the free energy density be $f=\Freeenergy/N$, then the energy density is
\begin{equation} e = \frac{1}{N} \sum_{i=1}^N \sum_{S_i,S_{nn(i)}} H_i(S_i,S_{nn(i)}) P(S_i,S_{nn(i)}) \label{eq:energydensity} \;, \end{equation}
and entropy density \begin{equation} s = \beta (e - f) \label{eq:entropydensity} \;. \end{equation}

\subsection{Spin-symmetric solutions and the two-state model}
\label{ssec:spin-symmetric_solutions_and_the_two_state_model} The recursive
decomposition on a tree (\ref{eq:T}) terminates in a set of single g-spin cavity
graphs on the leaves. With no external field, a cavity graph consisting of a single
g-spin is described by a symmetric field 
\begin{equation} 
\vec{h}_{k\rightarrow j} =
(-\lambda,0,0)^T \;;\; Z_{k\rightarrow j} = 4\cosh(\lambda) \label{eq:leaffield} \;.
\end{equation} 
The zero values, $h^f=h^b=0$, reflect the spin symmetry of the
Hamiltonian, and the mapping (\ref{eq:T}) does not break this symmetry. The
recursion (\ref{eq:T}) is then non-zero in only one component $J_{k\rightarrow j}$.
In a symmetric solution, which we later also call a paramagnetic solution, whether a
g-spin $(S_i,S_j)$ is dimer ($S_i=S_j$) or non-dimer ($S_i\neq S_j$) determines the
recursion properties; the sign of constituent spins is irrelevant to the
thermodynamics. 

For a general graph we find every symmetric solution of the "four-state" model (the model) in
one to one correspondence with a solution of a "two-state" model defined with two state g-spin variables $\{{\tilde S}_{ij}=\pm 1: ij \in E\}$, and by a Hamiltonian
\begin{equation} 
  H(\{{\tilde S}\}) = \frac{1}{2} \sum_i \left(\sum_{j \in nn(i)}
  {\tilde S}_{ij}\right)^2 + \lambda \sum_{ij} {\tilde S}_{ij} 
  \label{eq:Ham2state}
  \;. 
\end{equation}
The topology of the two factor graphs is unchanged other than in the presence in the two state model of factors corresponding to local fields on the variables ${\tilde S}_{ij}$. The ``hard'' intersection constraints are notably absent in the two state model. The quantity
\begin{equation} 
  P_{j \rightarrow i}({\tilde S}_{ij}) \propto \exp \left\lbrace
  \beta J_{j \rightarrow i} {\tilde S}_{ij} \right\rbrace
  \label{eq:2stateparameterization} 
  \;
\end{equation}
becomes the relevant recursively defined object in the Bethe-Peierls or cavity method.
The connection between solutions of the two-state model and the
four-state model is 
\begin{equation} 
  P_{j \rightarrow i}(S_j,S_i| \Factorgraph, 4\;
  states) = \frac{1}{2}P_{j \rightarrow i}({\tilde S}_{ij} = S_i S_j|\Factorgraph, 2\;
  states) \;. 
  \label{eq:4to2state} 
\end{equation} 
Since the marginal properties, and consequently many extensive properties such as energy, can be written in terms of these cavity probabilities. We can observe that if we identify ${\tilde S}_{ij}$ with $S_i S_j$ then local and thermodynamic properties derived from the marginals are equivalent. Furthermore we find the difference between the free energies is accounted for entirely by a constant entropic term $({\bar C}-2)/(2\beta)$. In this sense the properties of the symmetric solution, other than its susceptibility to symmetry breaking, can be understood in the context of the simpler two-state model.

This equivalence is rather surprising given that the models are not identical. One can make a two-to-one transformation of variables from the spins $S_i$ to a set of dimers $\{S_iS_j\}$, but the new state space is subject to constraints such that $1=\prod_{ij: ij \in L}{\tilde S}_{ij}$, for every loop $L$. This set of linear constraints on the variables are not independent, but are reducible to a set of $M-N+1 \sim ({\bar C}/2 - 1)N$ independent linear constraints, that depend on the details of the graph structure. It seems however, that these constraints are irrelevant to the moments of the symmetric solution, and affect the entropy only as a simple function of $\beta$ (independent of graph structure). For now we say only that since the loops are long (by the locally tree like assumption), and correlations decay rapidly in pure states (the fixed point purports to describe a single pure state), the effect of these constraints may be pushed to the boundary with respect to any neighbourhood. Since the nature of the constraints is not to favour either dimers or non-dimers per-se, we may anticipate a self-averaging effect at the boundary, and as such no bias would be present at, or persist from, the boundary. As such it is not surprising they have a weak effect on marginals even at low temperature. This argument relies on the assumption of weak correlation between a neighbourhood and its boundary at distance $\log(N)$, which is expected to be reasonable in a pure state (at a stable fixed point for the probabilities, stable also against symmetry breaking). The effect on the entropy of the change of representation is discussed further in section \ref{ssec:occupation_problem_equivalence}. 

\subsubsection{Boundary conditions and symmetry breaking}
\label{ssec:boundary_conditions_and_symmetry_breaking} For a given boundary
condition, a well studied property on trees, related to stability analyses in random
graphs, is the stability in the core of a large tree to perturbations on the leaves.
Typically one can define a variable at uniform distance from all leaves, and judge
the stability of its marginal towards perturbations of the leaves -- this can
indirectly establish the presence of long range correlations, susceptibility and
other useful thermodynamic properties. Infinitesimal perturbations on a field
$\delta \vec{h}^T=(\delta J, \delta h^f, \delta h^b)$, allow a linearized
description \begin{equation} \delta \vec{h}_{j \rightarrow i} = \sum_{k \in nn(j)
\setminus i}  \left[\frac{\partial {\hat T}}{\partial \vec{h}_{k \rightarrow j}}
\right]  \delta \vec{h}_{k \rightarrow j} \label{eq:linearperturbation} \;.
\end{equation} To establish asymptotic properties on trees we are interested in
whether these perturbations introduced on the leaves decay under recursion about a
given solution, such as the symmetric one. The two-state model stability analysis
can differ from the stability analysis of the symmetric solution of the full model
in that symmetry breaking perturbations of $\delta h^{f},\delta h^b$ can be
considered in the latter.

\subsection{The zero temperature recursion}
\label{ssec:the_zero_temperature_recursion}

We wish to study the limit of zero temperature, where some simplifications are
involved. We assume a decomposition of the field into energetic and entropic parts
that are assumed to evolve on separable scales in the limit of large $\beta$,
\begin{equation} \vec{h} = \vec{h}^E - \frac{1}{\beta} \vec{h}^S + O(1/\beta^2)
\label{eq:htoeandspart} \;. \end{equation} Separating the leading order term in
temperature from (\ref{eq:T}) gives \begin{equation} \vec{h}^E -
\frac{1}{\beta}\vec{h}^S = {\hat T}^{E}(\{\vec{h}_c^E\}) - \frac{1}{\beta}{\hat
T}^{S}(\{\vec{h}_c^E\},\{\vec{h}_c^S\},1/\beta) \;, \label{eq:Tzerorecursion}
\end{equation} and allows a simplified representation of the energetic recursion
\begin{eqnarray} J^{E}_{j \rightarrow i} &= J^E(\{\vec{h}_{k \rightarrow j}\}) &=
-\lambda - \frac{1}{4}\sum_{S_i,S_j} S_i S_j D(S_j,S_i,\{\vec{h}_{k \rightarrow j}
\}) \label{eq:TzerorecursionJ} \\ h^{f,E}_{j \rightarrow i} &= h^{f,E}(\{\vec{h}_{k
\rightarrow j}\}) &= -\frac{1}{4}\sum_{S_i,S_j} S_i D(S_j, S_i, \{
\vec{h}_{k\rightarrow j} \}) \label{eq:Tzerorecursionhf} \\ h^{b,E}_{j \rightarrow
i} &= h^{b,E}(\{\vec{h}_{k\rightarrow j} \}) &= \sum_k h^{f,E}_{k\rightarrow j} -
\frac{1}{4} \sum_{S_i,S_j} S_j D(S_j,S_i,\{\vec{h}_{k\rightarrow j} \})
\label{eq:Tzerorecursionhb} \;, \end{eqnarray} where \begin{equation}
\fl D(S_0,S_C,\{\vec{h}_1,\ldots,\vec{h}_{C-1}\}) = \min_{S_1 \ldots S_{C-1}}
\left\lbrace \frac{1}{2}\left(\sum_{k=1}^{C} S_k \right)^2 + \sum_{k=1}^{C-1}
\left((\lambda - J_k) S_0 - h^b_k\right) S_k\right\rbrace \;. \label{eq:D}
\end{equation}

\subsubsection{Restrictions on the space of energetic fields} In \ref{app:zero_temperature_field_support} we show that for any graph the distribution of fields generated self-consistently by (\ref{eq:TzerorecursionJ})-(\ref{eq:Tzerorecursionhb}) is bounded at leading order in $1/\beta$. The two-state model solution is described by $J_{j \rightarrow i}$ (\ref{eq:2stateparameterization}), and a simple bound on these elements is found given connectivity $C_j$ and an unconstrained set of descendant fields, namely,
\begin{equation}
\fl J_{j \rightarrow i} \in [\min_{\{\vec{h}_{k \rightarrow j}\}} J^E(\{\vec{h}_{k \rightarrow j}\}), \max_{\{\vec{h}_{k \rightarrow j}\}} J^E(\{\vec{h}_{k \rightarrow j}\})] = [-(C_j-1)-\lambda,(C_j-1)-\lambda] \;. \label{eq:2statebound} \end{equation}
In most cases this bound can be improved recursively by constraining the descendant fields self-consistently.

For rational values of $\lambda$ additional degeneracy of the energy levels can greatly modify the thermodynamic properties of the system. In the simplest case of integer $\lambda$ the energy levels are even integers on any cavity graph: self-consistency requires the energetic field components to be integers. Since only a small bounded range of energies is available locally, energetic fields on every cavity graph must be from within a finite set. As argued in \ref{app:zero_temperature_field_support} we require a set closed under recursion (\ref{eq:TzerorecursionJ})-(\ref{eq:Tzerorecursionhb}) and inclusive of any boundary conditions.

\section{Special Ratios} \label{sec:special_ratios}

Consider the properties of a factor of connectivity $C$ within a factor graph, and
assume a ground state of zero energy. A global ground state is certainly achieved if
at every factor a local ground state is achieved. To achieve zero energy about a
factor (\ref{eq:FactorHami}) requires a specific number of g-spins to be dimers in
the neighborhood, but does not depend on the sign of the spin enforced by the
intersection constraint. In the zero energy solution this number reflects a
balancing of the next nearest neighbor anti-ferromagnetic interaction with the
dimerizing ($\lambda<0$, favoring $S_i=S_j$) or anti-dimerizing ($\lambda>0$)
tendency of nearest neighbor interactions. The number of dimers in the neighborhood
of a factor required to achieve a local ground state may be written as
\begin{equation} N_{C,\lambda} = \mathrm{argmin}_{X \in \{0,\ldots,C\}} \left\lbrace
\left(X - \frac{C-\lambda}{2}\right)^2 \right\rbrace \label{eq:dimeroccupation} \;.
\end{equation} The minimizing argument must be integer, and will be unique
everywhere unless $C-\lambda$ is a positive odd integer. For a given $\lambda$, the
set of vertex connectivities that are unlocked, producing degenerate solutions to
(\ref{eq:dimeroccupation}) are \begin{equation} \mathbb{C}(\lambda) = \{C | C -
\lambda = 2 x + 1, x \in \mathbb{N}_0 \} \label{eq:specialconnectivities} \;.
\end{equation} Some integer values of $\lambda$, we call the special ratios, may
produce this {\em dimer-degeneracy} of local ground states, labeled by two
consecutive integer values of $x$. We will call a factor that meets the special
ratio criteria an unlocked interaction, whereas any other interaction will be called
locked~\cite{Zdeborova:CSPIS}.

We say that a ratio is special with respect to a graph $G(V,E)$, if amongst the vertices $V$ of the graph, a finite fraction $(0,1]$ are unlocked. The set of ratios that allow unlocked clauses in a graph is \begin{equation} \Lambda(G) = \{\lambda | C_i = \mathbb{C}(\lambda) \;\hbox{for some}\; i \in V\} \label{eq:specialratios} \;. \end{equation} Therefore we can say that a special ratio graph, is one in which the coupling ratio corresponds to an element in the set $\Lambda(G)$. By contrast a special ratio ensemble for large $N$ would be one in which replacing $G$ by a typical graph $\lambda$ is in the corresponding set.
Amongst the graph ensembles we study, it is only the case of regular ensembles for which every vertex is unlocked, in every other case the fraction of locked and unlocked will vary with mean connectivity.
For clarity we can discuss the implication for a regular graph with all vertices of connectivity $C$.

We present a mean field argument to indicate the significance of special ratios.
Consider the effects of adding a spin $i$ to the system. Assume that in the absence
of spin $i$ the neighbors have freedom to take either $+1$ or $-1$ independently to
achieve a ground state: the total number of ground states before adding the spin is
$2^C$. After adding a spin the number of ground states is reduced to
$N_{gs}=\sum_{x=N_{C,\lambda}} C! / [(C-x)!x!]$ for each state of $S_i=\pm 1$. The
number of assignments consistent with a ground state changes on addition of the spin
by a factor $r_{gs}=2 N_{gs}/2^C$.

\begin{figure}[h!tbp] \begin{center}
\includegraphics[width=0.45\linewidth]{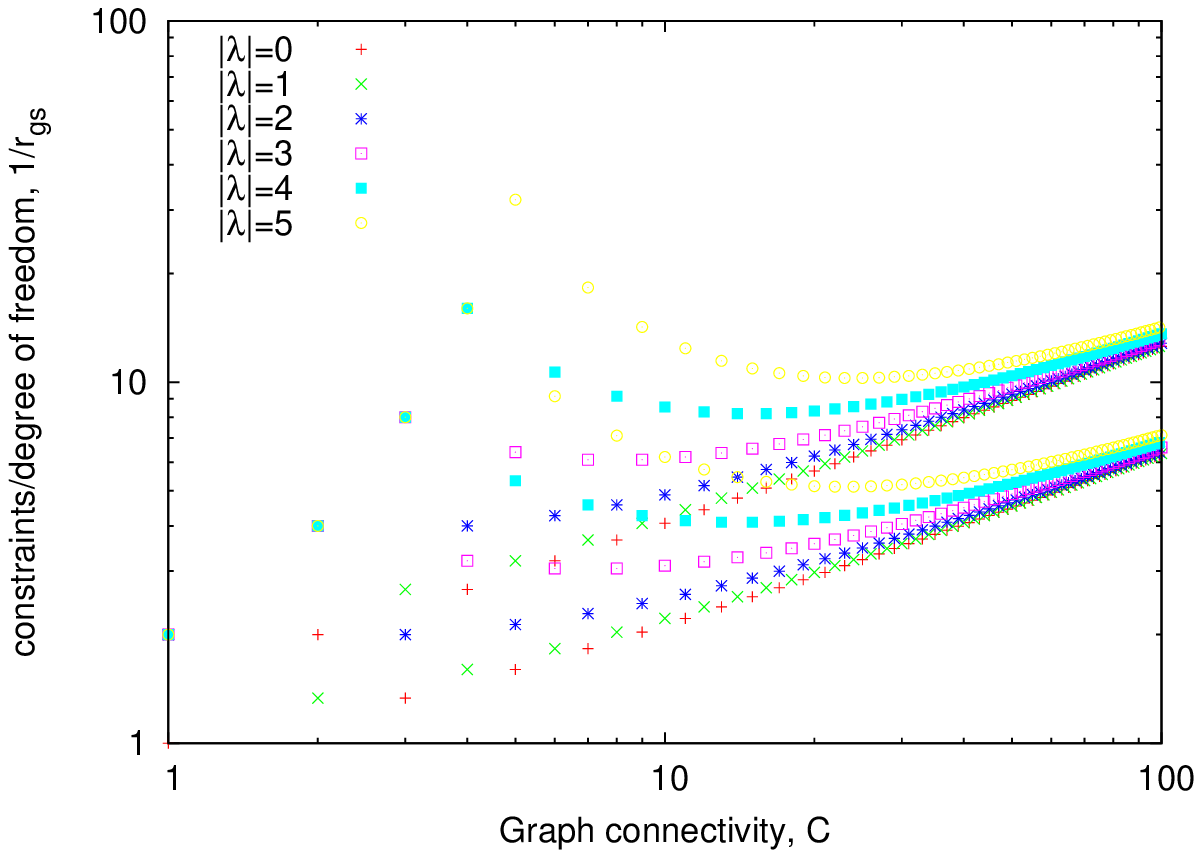}
\includegraphics[width=0.45\linewidth]{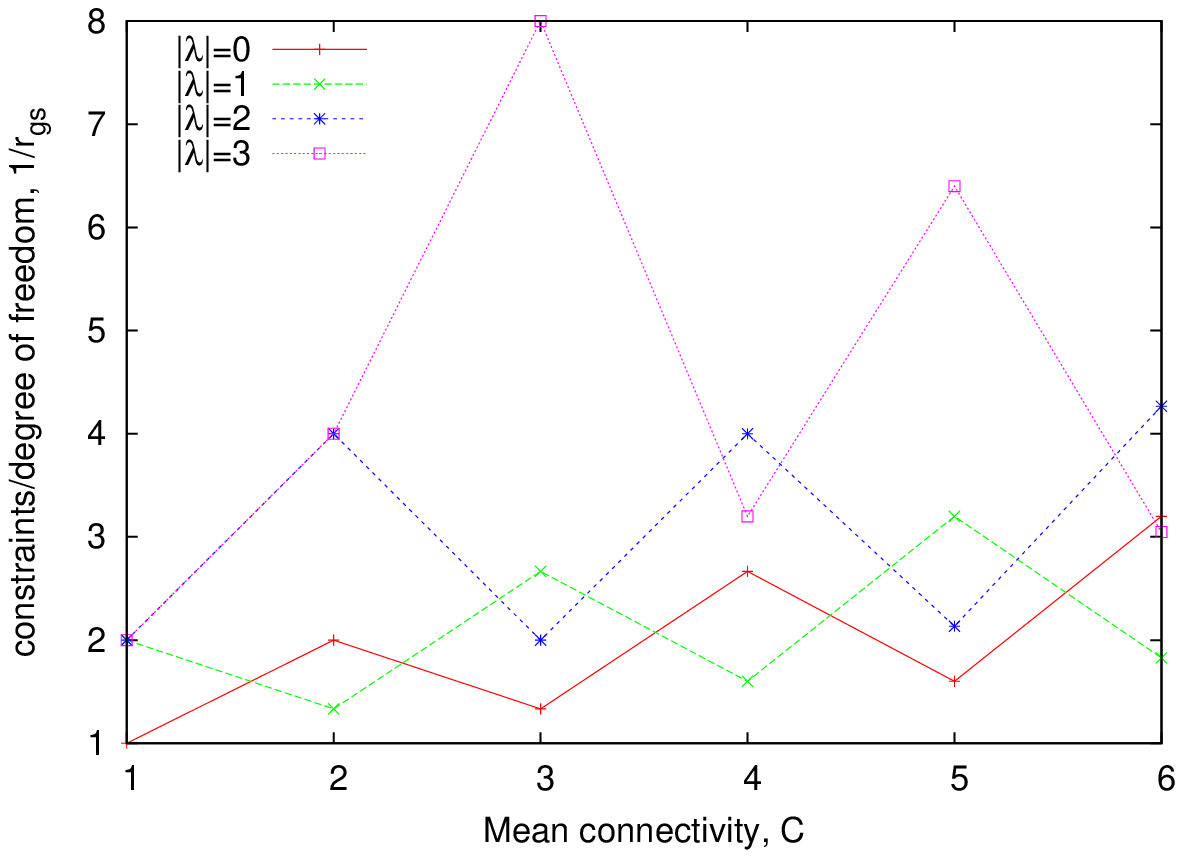}
\caption{\label{fig:SpecialRatios} Number of constraints per degree of freedom in
regular graphs of various connectivity. (left) Asymptotically the ratio oscillates
between asymptotes for the cases of $N_{C,\lambda}$ determined by locked and
unlocked interactions. (right) For small $C$ and $\lambda$ the oscillating pattern
is apparent in linear graph ensembles, ensembles with sufficiently small constraint
to variable ratios are good candidates for extensive ground state entropy.}
\end{center} \end{figure}

The factor $1/r_{gs}$ measures the number of constraints (implied by zero energy)
per degree of freedom, which is intuitively linked to the probability a zero energy
solution exists. The factor is high in the following two cases as shown in figure
\ref{fig:SpecialRatios}. First, when $C-1 < |\lambda|$ for most vertices, in which
case $r_{gs}$ equals $2^{1-C}$ for most vertices. Second, when $C$ is large $r_{gs}$
becomes $O(1/\sqrt{C})$. Especially relevant to the previous discussion is the
difference between odd and even connectivity nodes for given (integer) $\lambda$. In
special ratio ensembles two terms contribute to the sum $N_{gs}$, and the entropy of
the system is relatively high. As will be later shown for linear ensembles,
$r_{gs}\gtrsim 1/2$ appears to be required for a zero energy ground state of
extensive entropy.

This argument would indicate that, for integer $\lambda$, the greatest flexibility
of the zero energy state space follows a pattern determined by the connectivity
composition of the nodes. For example, if $|\lambda| = 0,2$ one can anticipate
significantly higher entropy for a zero energy ground state in a graph with odd
rather than even connectivity. The opposite is true for $|\lambda| = 1$. This opens
the unusual possibility that as $\lambda$ is varied the entropy may oscillate, as
will the free energy at finite temperature. Similarly, in the linear ensemble the
proportion of odd and even connectivity nodes varies cyclically with ${\bar C}$. One
can speculate that as constraints are relaxed and enforced when $\lambda$ or ${\bar
C}$ changes, a periodic pattern of phases may occur. Indeed we will show this to be
the case.

\section{Small graph studies} \label{sec:small_system_studies}

In this section we study the energy and entropy density of ground states for $N\leq
200$ numerically. The linear connectivity ensembles chosen for presentation have
small integer valued ratios ($\lambda$) and small mean connectivity (${\bar C}$).
These features allow a confirmation of the trends identified in section
\ref{sec:special_ratios}. Further results are presented in
\ref{app:ground_state_experiments}.

\subsection{Sampling methodology and algorithms} We generate simple graphs of $N$ vertices and $M$ edges by a variation on the configuration model~\cite{Janson:RG}. The ratio of the number of edges to the number of vertices in every sample is restricted to exactly ${\bar C}/2$, and graphs of size $N<40$ are restricted to be connected (every vertex is reachable from every other vertex along some sequence of edges).
An exhaustive enumeration method was applied to determine all ground states for small graph samples ($N<40$). Deterministic greedy search and stochastic sampling, using the extremal optimization (EO) method ~\cite{Boettcher:NWO,Boettcher:NR}, was applied to larger samples ($N=O(100)$). EO is an example of a stochastic local search method, performing a biased random walk in the space of spin assignments to discover ground states. This method was chosen because of the similarity between our model and algorithmically challenging constraint satisfaction problems where it has been successfully applied.

Exhaustive enumeration of ground states was implemented by a branch-and-bound
algorithm, exactly determining all ground states. Our EO parameters are
calibrated against these exact results for small systems, and by further
self-consistent analysis. For final data collection we adapted an algorithm provided
by Stefan Boettcher that had been applied to a sparse spin glass
problem~\cite{Boettcher:NR}. Three parameters $\tau,T_\tau,R_\tau$ control the
quality of the ground state estimate obtained and are given alongside the data in
the corresponding figures. Parameter $\tau$ controls the probability to take locally
suboptimal search directions, in a manner intuitively similar to the inverse
temperature in Monte Carlo sampling. Parameter $R_\tau$ is the minimum number of
independent searches per graph sample, and is adaptively increased where variation
between search result outcomes is large (hard regimes). Each search is initialized
at a point in state space sampled uniformly at random. Parameter $T_\tau$ controls
the maximum number of spin reassignments $T_\tau(1000+((N+1)/5)^3)$ per search. The
cubic scaling with $N$ is chosen self-consistently with the amount of time required
to find the ground state, for near optimal choices of $\tau$, in hard instances. To
adapt the algorithm to determine the entropy was found to require a sampling time
scaling as $N^{3.5}$ for relatively small systems $N \lesssim  50$, we attempted
only ground state search for the larger systems.

Throughout the parameter space for the system sizes presented we achieved good results by EO for $\tau\lesssim 2$, a parameter range in which locally suboptimal search trajectories are explored. The parameter $\tau=1.6$ provided consistently good results; $T_\tau$ or $R_\tau$ were chosen sufficiently large that only a very small fraction of ground states appeared to be incorrectly identified.
As an extreme contrast we studied greedy search methods, including the case $\tau\rightarrow\infty$ (numerically approximated as $\tau=10$) of EO. For $\tau\rightarrow\infty$ we descend the energy landscape until no downward steps are available. This works well if the landscape is smooth and has a unique minimum: more generally to compensate for the trapping of the dynamics we can consider small $T_\tau$ and large $R_\tau$.

In the regimes where the number of constraints per degree of freedom is smallest, and we achieve zero energy, search time is fast for all algorithms -- even linear in system size.
Where the ground state energy is non-zero the number of samples required to reach a ground state increases significantly for all algorithms tested, with greedy search failing dramatically. For EO, with a well chosen $\tau$, the search time is slowest precisely in marginal cases, at the threshold between zero and non-zero energy.

\subsection{Results} \begin{figure}[h!tbp] \begin{center}
\includegraphics[width=0.75\linewidth]{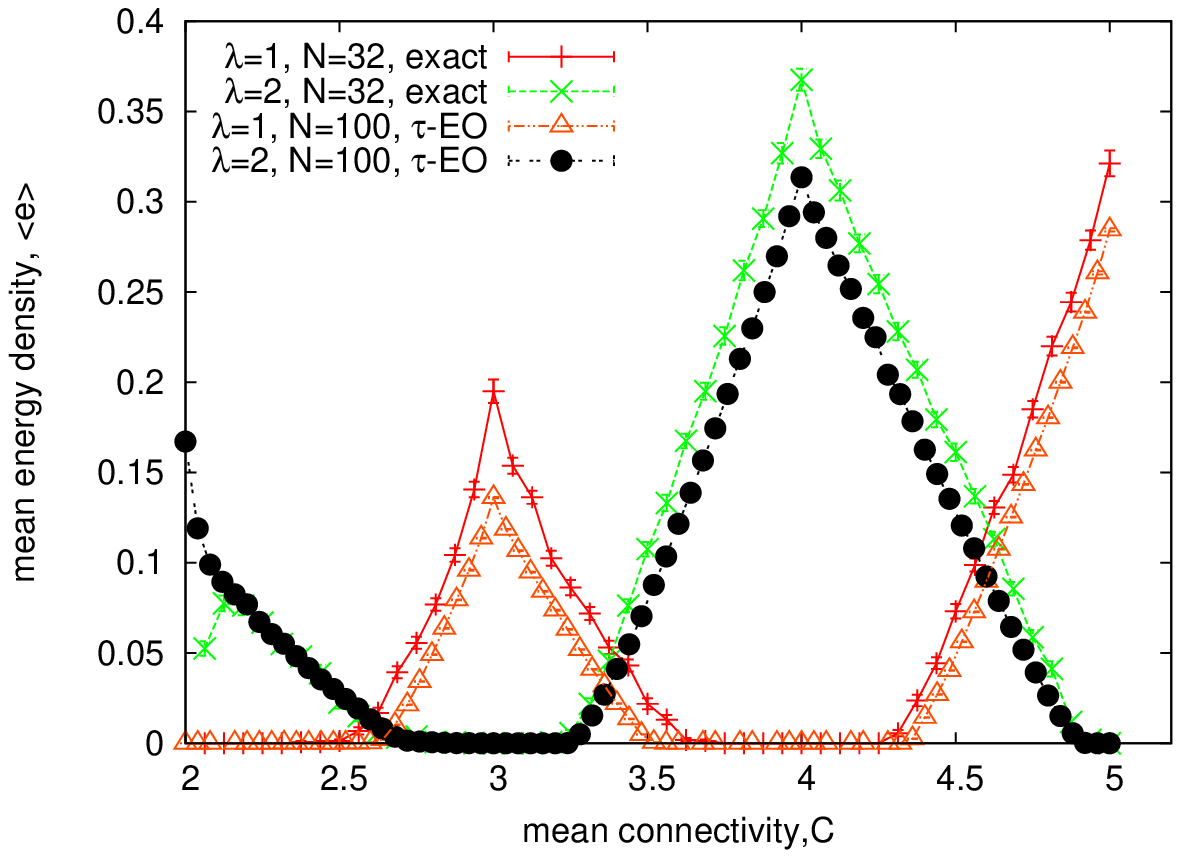}
\includegraphics[width=0.75\linewidth]{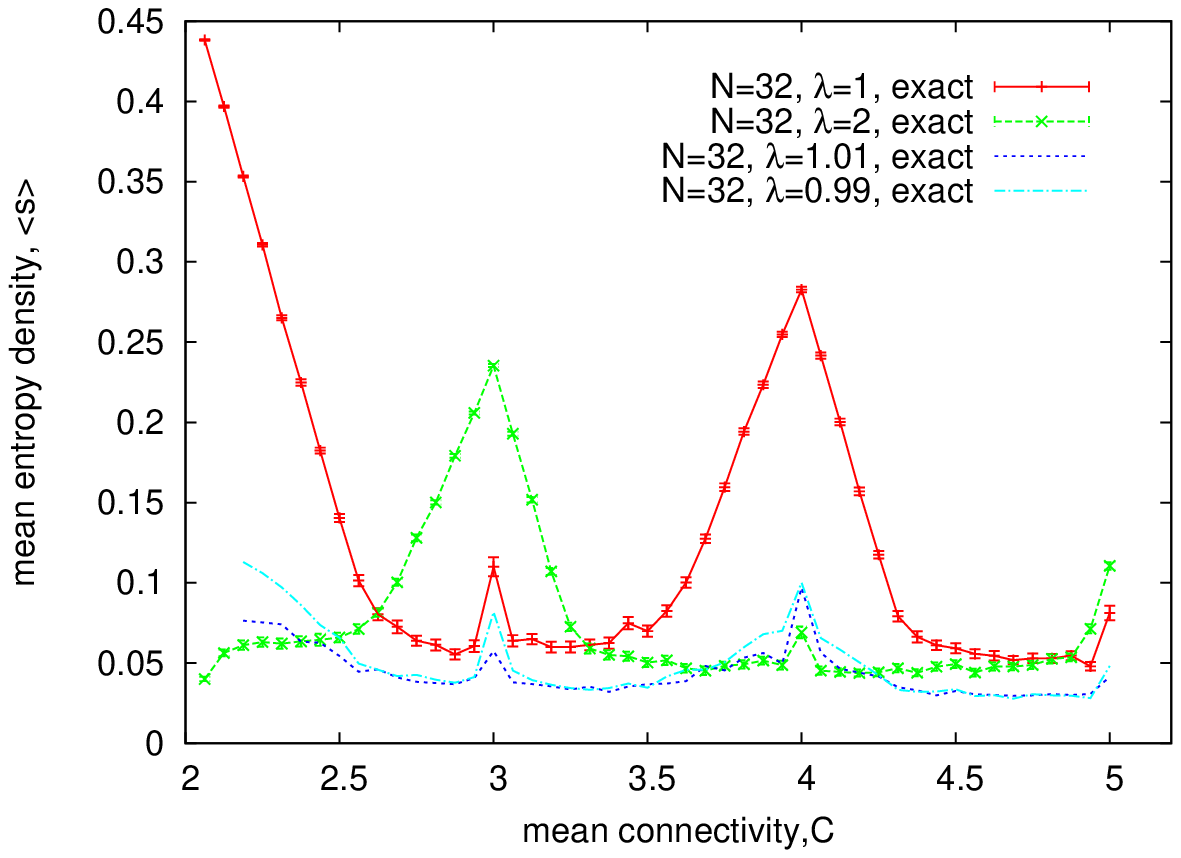}
\caption{\label{fig:EnergyEntropyN32} Statistics over $100$ graph samples of size
$N=32$, and $1000$ graph samples of size $N=100$, for ground state energy (upper)
and entropy (lower). Error bars are omitted for the entropy where $\lambda=1\pm
0.01$ for clarity.} \end{center} \end{figure}

Energy and entropy results for the sampling methods on graphs of size
$N=32$ and $100$ are presented in figure \ref{fig:EnergyEntropyN32}. As connectivity
and $\lambda$ are varied a non-monotonic trend exists in energy and entropy, in
agreement with the special-ratio argument of section \ref{sec:special_ratios}. The
pattern becomes more pronounced as system size increases. For $\lambda=1$ we see the
energy minimized, and entropy maximized, about even connectivity. For $\lambda=2$
the trend is similar but for odd connectivity, increasing to $\lambda=3$ the pattern
again reverses. A well-known feature of frustrated Ising models is the possibility
of extensive ground state entropy, and for finite systems we see parameterizations
that allow large entropy, in correspondence with regimes for which the energy equals
the known lower bound (zero).

For $N=32$ we demonstrate the entropy for $\lambda=1 \pm \delta \lambda$, with
$\delta \lambda$ chosen as $0.01$, the dimer degeneracy in the ground states is
broken, with a large reduction in entropy relative to the results for the special
ratio $\lambda=1$. However, the energy level splitting at each factor is
proportional to $\delta \lambda$, the energy curve follows closely the unperturbed
result.

For ${\bar C}\rightarrow 2$ and $|\lambda|=1$ we have the energy approaching zero
and the entropy density in excellent agreement with the asymptotic result for the
$1$-dimensional next nearest neighbor Ising model, where $s = \log(1+\sqrt{5})/2$
~\cite{Baxter:ESM}. By contrast for $\lambda=2$ there is a difference in behaviour
of the energy for $N=32$ and $N=100$ approaching ${\bar C}=2$ in figure
\ref{fig:EnergyEntropyN32}. This is explained by a difference in the graph sampling
method: for $N=32$ we have rejected all graphs except those that are connected (we
approach a result dominated by a single unfrustrated loop); for $N=100$ we allow
fragmented graphs (we approach a result dominated by disconnected frustrated loops).
Asymptotically for ${\bar C} = 2$ and either graph ensembles the energy density must
approach zero, since the number of loops is at most $O(\log N)$, and each will raise
the ground state energy by at most $O(1)$. In the limit ${\bar C}\rightarrow 2$
additional excitations might be anticipated as a fraction of the number of sites of
connectivity greater than $2$.

\begin{figure}[h!tbp] \begin{center}
\includegraphics[width=0.75\linewidth]{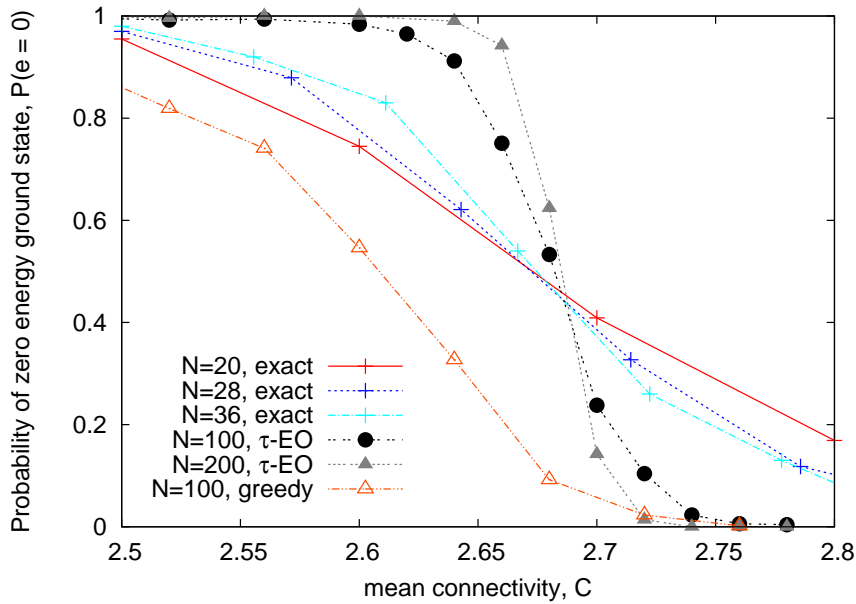}
\caption{\label{fig:PEnergy0} The probability that a sampled graph ($\lambda=1$)
allows a zero energy solution. EO has parameters
$(\tau,T_\tau,R_\tau)=(1.6,100,7)$, the greedy result is achieved by EO at
parameters $(\tau,T_\tau,R_\tau)=(10,10,300)$. With $100$ graph samples for $N<40$,
$1000$ samples for $N=100$, $500$ samples for $N=200$.} \end{center} \end{figure}
Also apparent in figure \ref{fig:EnergyEntropyN32} are points where the energy
becomes non-zero continuously, implying continuous phase transitions. In figure
\ref{fig:PEnergy0} we give the probability that a graph will have a zero energy
solution. Indeed, the drop in the probability becomes increasingly abrupt with
increasing system size about ${\bar C}\sim 2.69$. Results are shown from the
exhaustive method that visit all states, and the heuristic sampling method,
EO and a greedy search method. To compensate
for the fact that the greedy method will be frequently trapped we consider many
random minimizations (small $T_\tau$ and large $R_\tau$), but results are poor.
Close to the transitions, the failure of the greedy method, and increasing search
times can be taken as evidence for the appearance of many local minima that trap the
dynamics in the vicinity of suboptimal configurations. The observation that the
phase transition phenomena is correctly captured by EO, whilst the number of
states visited (including repeats) is far short of the full state space $2^{N}$,
demonstrates its efficiency.

For $2<{\bar C}<5$, $\lambda=1$ and $2$, we have found numerical evidence
for transitions from satisfied (energy 0) to unsatisfied (positive energy) phases in
every integer interval of ${\bar C}$ as shown in figures. Numerically we find the
transitions to be sharply defined for system size $N=O(100)$, as in figure
\ref{fig:PEnergy0}, with one exception which is the transition in interval ${\bar C}
\in (2,3]$ and $\lambda=2$. In this case there is strong variability between sample
energies and no clear cut-off for the system sizes studied. An important difference
between this transition and all others will become clearer in light of the
forthcoming cavity method.

\section{Cavity method} \label{sec:cavity_method}

The Replica Symmetric (RS) solution is presented in the context of the cavity
method~\cite{Krzakala:GS,Mezard:BLSG}. The replica symmetry broken framework is
provided at the level of energetic considerations, in the zero temperature
limit~\cite{Montanari:NLT,Mezard:CMZT}.

\subsection{The replica symmetric (RS) solution}
\label{ssec:replica_symmetric_solution}

The cavity method exploits the fact that the next shortest path connecting any two
neighboring g-spins, after breaking the immediate path, is infinite in the large
system limit, and that instances where the cavity graph rooted in a g-spin are not
locally tree-like are statistically negligible. Assuming a pure state describes the
thermodynamics, the point to multi-point correlations decay rapidly with distance,
and so the estimate of probability $P_{i \rightarrow j}(S_i,S_j)$ is independent of
$P_{j \rightarrow i}(S_j,S_i)$ and other locally disjoint cavity graphs. The
recursion (\ref{eq:T}) remains valid with the factorization of probabilities implied
by this asymptotic independence.\cut{}{, so does the free energy decomposition
(\ref{eq:freeenergyNoZij}) on a tree.}

\subsubsection{RS recursion condition}
\label{sssec:recursion_conditions_for_replica_symmetry}

We can make use of the self-averaging property for large $N$ to determine the fields
by only a local recursion~\cite{Mezard:BLSG}. In the recursion the fields which are
recombined at a generic factor can be represented by samples independently from a
common distribution $P(\vec{h})$. Self-consistency of this distribution requires
\begin{equation} 
\fl P(\vec{h}) = \frac{1}{{\bar C}}\left\langle C \int
\prod_{k=1}^{C-1} \left[ d \vec{h}_{k} P(\vec{h}_k)\right] \delta\left(\vec{h}_{k} -
{\hat T}(\{\vec{h}_{k} | k =1,\ldots C-1\}) \right) \right\rangle_C \;.
\label{eq:RS_population_recursion} 
\end{equation}

In the case of a regular graph we can identify one solution as
$P(\vec{h})=\delta(\vec{h}-\vec{h}*)$. By symmetry of the Hamiltonian we expect the
fixed point $\vec{h}*$ to be a symmetric field. Finding an analytic solution for
modest values of $C$ and zero temperature is possible, more generally it is
straightforward to determine a numerical solution for any $C$, temperature, and
$\lambda$.

A general method to establish a RS solution is population
dynamics~\cite{Mezard:BLSG}. This involves an iterative procedure over a set of
$\Numberoffields$ fields defined by
$\vec{h}_{POP}^t=\{\vec{h}_1,\ldots,\vec{h}_\Numberoffields\}$, after $t$ iterative
steps. The approximation to the probability distribution over fields is
\begin{equation} P^{t}(\vec{h}) = \frac{1}{\Numberoffields}\sum_{\vec{h}' \in
\vec{h}^t_{POP}} \delta(\vec{h}-\vec{h}') \label{eq:RShistogram} \;. \end{equation}
An estimate at step $t+1$ is created by iteratively updating once every element in
the set $\vec{h}^t_{POP}$ according to (\ref{eq:T}) by a stochastic sampling process
from within the set. After a short transient a stable solution is achieved -- one in
which the moments converge up to small statistical fluctuations, $\Numberoffields$
being a suitably large number. Detailed issues of the initial conditions and
convergence are dealt with in
\ref{app:population_dynamics_implementation}.

\subsubsection{RS free energy} \label{ssec:free_energy_for_replica_symmetry} To
construct the free energy we can calculate the additional free energy when creating
a typical graph of size $N+1$ from a graph of size $N$. The free energy shift
averaged over the growth processes gives the free energy density estimate. This is
done by averaging the additive vertex free energy term, subtracted by the average
edge free energy term to account for double counting of the free energy associated
with the edges. That is, as in the case of a tree, \begin{equation} f =  \sum_C P(C)
f_V(1/\beta,C) - \frac{\bar C}{2} f_E \label{eq:RS_freeenergy} \;, \end{equation}
where \begin{eqnarray} f_V &=& \int \prod_{k=1}^{C}\left[d\vec{h}_k
P(\vec{h}_k)\right] \Freeenergy_{V}(\{{\vec{h}}_k\}, 1/\beta, H_C)
\label{eq:RS_freeenergy1}\;,\\ f_E &=& \int d\vec{h}_1 P(\vec{h}_1) d\vec{h}_2
P(\vec{h}_2) \Freeenergy_{E}({\vec{h}}_1,\vec{h}_2) \label{eq:RS_freeenergy2} \;.
\end{eqnarray} This is analogous to (\ref{eq:freeenergyNoZij}), composed of vertex
(\ref{eq:freeenergyi}) and edge free energies (\ref{eq:freeenergyij}); $H_C$ is the
Hamiltonian applicable to a factor of connectivity $C$. The population dynamics
solution again requires a replacement of the integration processes in
(\ref{eq:RS_freeenergy1}) and (\ref{eq:RS_freeenergy2}) by random sampling, which is
subject to fluctuation effects related to the population size $\Numberoffields$.

\subsection{RS solution stability} \label{ssec:stability_and_symmetry_breaking}

A first stability test of a candidate solution $P^\infty(\vec{h})$ at the steady
state $t\rightarrow \infty$, considers the evolution of a perturbation
$P^\infty(\vec{h}) + \epsilon (P^t(\vec{h})-P^\infty(\vec{h}))$. The second
distribution $P^t(\vec{h})$ can be chosen to maintain the normalization but with the
moments of $P^\infty(\vec{h})$ distorted by small amounts.

In our study we focus on inhomogeneous graphs. We require a method compatible with
population dynamics to test the stability of the fields towards linear
perturbations. We achieve this by adding small random perturbations to every field
and consider their evolution with time through a joint distribution
$P^{t}(\vec{h},\delta\vec{h})$, maintaining the marginal over the first argument as
$P^\infty(\vec{h})$. A special case of the stability analysis can be undertaken for
regular graphs; this is discussed in \ref{ssec:regular_graph_stability}.

In this scenario the joint distribution of perturbations and fields at time $t+1$,
$P^{t+1}(\vec{h},\delta\vec{h})$, is found by updating the leading order terms
(\ref{eq:RS_population_recursion}), and the subleading order through
(\ref{eq:linearperturbation}). Under iteration the joint distribution can be
expected to converge towards a distribution concentrated on $\delta\vec{h}=(0,0,0)$;
otherwise if there is not a single pure state we expect the perturbations to
diverge. This can be quantified by the ratio of the second moment of the
perturbations after many iterations to the corresponding moment \begin{equation}
\eta = \lim_{t\rightarrow \infty} \frac{1}{t}\log \frac{\int d\vec{h} d\delta\vec{h}
P^{t}((\vec{h},\delta\vec{h})) \left[(\delta h^f)^2 + (\delta h^b)^2 + (\delta
J)^2\right]}{ \int d\vec{h} d\delta\vec{h} P^{0}(\vec{h},\delta\vec{h})
\left[(\delta h^f)^2 + (\delta h^b)^2 + (\delta J)^2\right]} \;. \label{eq:eta}
\end{equation} We are interested in the large $t$ limit, but numerically we take $t$
to be $O(100)$. In population dynamics, we expand the set $\vec{h}_{POP}$ to be
composed of a field and perturbation pair, the second moment thereby being the
statistic of $N$ samples \begin{equation} \vec{h}_{POP} = \{\left(\vec{h}_1,\delta
\vec{h}_1\right),\ldots,\left(\vec{h}_N,\delta \vec{h}_N\right)\} \;. \label{eq:H0}
\end{equation}

Special care should also be taken in the zero temperature population dynamics method
where we have energetic fields $O(1)$, and entropic fields $O(1/\beta)$, and
stability should be tested in both parts. However, a frequent manifestation that
indicates the inviability of RS solutions is the divergence of entropic fields under
recursion from random (symmetry broken) initial conditions. Non-convergence of the
entropic part, given that we expect continuity of the solution in $1/\beta$, is an
indication that RSB is required.

\subsection{The one-step replica symmetry breaking (1RSB) solution}
\label{ssec:one-step_replica_symmetry_breaking_solution} The next level of
approximation in the framework is the one-step replica symmetry breaking solution.
Our goal is principally to understand the critical behaviour associated with the
competition of nearest and next nearest neighbor interactions at zero temperature.
For this reason, as well as for issues of computational complexity, we focus on
integer values of $\lambda$, and linear connectivity ensembles with ${\bar C}>2$.
This section gives a sketch of the methodology which has been presented
systematically many times for generic locally tree like problems \cite{Mezard:CMZT}.
A standard stability analysis is considered as part of \ref{app:zero_temperature_field_support}.

In the method we restrict attention to energetic considerations at zero temperature,
ignoring terms of $O(1/\beta)$ both in the consideration of the distribution of pure
states, and by extension in the fields. The energetic field can be assumed to have
integer restricted components for each pure state~\cite{Mezard:CMZT,Montanari:IOS},
as was the case for a tree.

Under the 1RSB assumption we must consider that at every site there exist many pure
states. For every cavity graph the field, sufficient to describe a single pure
state, must be replaced by a probability distribution over pure states
$\mathcal{P}(\vec{f})$. At the energetic level, as shown in
\ref{app:zero_temperature_field_support}, a finite set
$\Gamma=\{\vec{h}_\populationindex|\populationindex=1\ldots|\Gamma|\}$ is able to
describe the space of all realizable fields reproducible under recursion
(\ref{eq:Tzerorecursion}), and is sufficient to describe the support of any
solution. For our recursion we are required to describe in principle a field for
every pure state, and the significance of each pure state should be weighted by its
free energy.

The fundamental assumption in the energetic 1RSB method relates to the free energies
of the pure states~\cite{Mezard:CMZT}. The set of lowest energy pure states are
assumed to have energies independent and identically distributed according to a
Poisson process, with density \begin{equation} P(E^\alpha) = \exp \left\lbrace \mu
(E^\alpha - E_{ref}) \right\rbrace \;, \label{eq:exponentialdistpurestates}
\end{equation} $E_{ref}$ being a normalizing constant. Equivalently we can
anticipate the number $\mathcal{N}(e)$ of pure states at a particular energy level
$e$ to be exponential in the system size, and described by an exponent that must be
a convex and monotonic function of energy called configurational entropy or
complexity \begin{equation} \Sigma(e) = \frac{1}{N}\log \mathcal{N}(e)
\label{eq:complexity} \;. \end{equation} This gives a functional interpretation for
$\mu$ as a variational parameter within the free energy, introduced by a Legendre
transform enforcing the identity (\ref{eq:complexity}). The complexity can be
interpreted as the entropy of locally stable states at a particular energy density.

Under recursion we expect that the pure states are each reweighted by the free
energy shift they induce self-consistently with the Poisson distribution. It is
argued that since in a pure state the fields on descendants are at long distances
and uncorrelated, the energy shift will equal that of a tree (\ref{eq:DeltaF}).
Therefore we should correct every mapping by a term with $\exp \{- \mu \Delta F\}$,
where $\mu$ acts now as a temperature over the pure states. Although the insight
lies at the level of pure states, it is convenient to represent the properties by a
field distribution without loss of generality. Since we have a finite set of fields
the distribution can be represented by a vector $\vectorin{\mathcal{P}}$. Each
component $\mathcal{P}_a$ can be interpreted as the probability of a given field as
being a (weighted) average over pure states on the cavity graph, \begin{equation}
\mathcal{P}(\vec{h}) = \sum_{a=1}^{|\Gamma|} \mathcal{P}_a \delta(\vec{h}_a -
\vec{h}) \label{eq:mathcalP} \;. \end{equation}

A relatively simple 1RSB solution might be obtained for a regular graph. Since we
are considering the large $N$ limit, the distribution of pure states is dependent
only on the local properties of the cavity graph. In the regular graph every cavity
graph is locally identical (tree-like), and so one case of (\ref{eq:mathcalP})
describes every neighborhood. A self-consistent relation to describe this
probability distribution over pure states is then \begin{equation}
\vectorin{\mathcal{P}} = T_{RSB}\left(\{\vectorin{\mathcal{P}}^{(c)} | c=1, \ldots,
C-1 \}\right) \;, \end{equation} where using the free energy shift ${\hat T}_F$ in
the zero temperature limit $\beta\rightarrow \infty$ applicable to independent
fields (\ref{eq:Tf}), the components of the new vector are determined by
\begin{equation}
\fl\begin{array}{lll}
T_{RSB,a}(\{\vectorin{\mathcal{P}}^{(c)}\}) &\propto&
\prod_{c=1}^{C-1} \left[\sum_{a_c=1}^{|\Gamma|} \mathcal{P}^{(c)}_{a_c} \right]
\Indicator\left( \vec{h}_a = {\hat T}(\{\vec{h}_{a_1}, \ldots, \vec{h}_{a_{C-1}}\})
\right) \\
&\times&\exp  \left\lbrace - \mu {\hat T}_F(\{\vec{h}_{a_1},\vec{h}_{a_{C-1}}\},0)
\right\rbrace \;. \label{eq:RSBrecursionCreg}
\end{array}
\end{equation} The indicator function
$\Indicator$ evaluates to one if the condition is met and zero otherwise. In the
regular graph we must have all the distributions on left and right identical.

In inhomogeneous graphs we must take into account that different cavity graphs will
result in distinct distributions over pure states. Hence, the parameter relevant in
the cavity method becomes a normalized distribution over cavity field distributions.
Numerically this function can be represented by a population of $\Numberoffields$
vectors $\vectorin{\mathcal{P}}_{POP}$, by analogy with the population of fields
used within the replica symmetric approach \begin{equation}
P(\vectorin{\mathcal{P}}) = \frac{1}{\Numberoffields} \sum_{\vectorin{\mathcal{P}}'
\in \vectorin{\mathcal{P}}_{POP}} \delta(\vectorin{\mathcal{P}} -
\vectorin{\mathcal{P}}') \label{eq:RSBpopulation} \;. \end{equation} At any site we
can consider independent samples from this distribution to reflect pure state
distributions incident upon any vertex in the graph. The recursion
(\ref{eq:RSBrecursionCreg}) becomes a componentwise mapping, whereas we would aspire
to undertake an integral, in the population dynamics method we evaluate the integral
approximately by sampling \begin{equation} P(\vectorin{\mathcal{P}}) = \int
\prod_{c=1}^{C-1} \left[d\vectorin{\mathcal{P}}^{(c)}
P(\vectorin{\mathcal{P}}^{(c)}) \right] \prod_{a=1}^{|\Gamma|}
\delta\left(\mathcal{P}_a - T_{RSB,a}(\{\vectorin{\mathcal{P}}^{(c)}\})\right) \;.
\label{eq:RSBrecursionirreg} \end{equation} The total free energy is again given by
(\ref{eq:RS_freeenergy}) weighted according to graph ensemble parameters, but with
the 1RSB expressions
\begin{equation}
\fl f_{V}(\mu,1/\beta) =
\frac{-1}{\mu}\lim_{\beta\rightarrow\infty}  \int \prod_{c=1}^C \left[d
P(\vectorin{\mathcal{P}}^{c}) \right] \log \prod_{c=1}^C \left[ \sum_{a_c}
\mathcal{P}^c_{a_c} \right] \exp  \left\lbrace - \mu \beta \Freeenergy_{V}
(\{\vec{h}_{a_c}\},1/\beta) \!\right\rbrace \;\!, \label{eq:RSBfi} \end{equation}
replacing (\ref{eq:RS_freeenergy1}) and
\begin{equation}
\fl f_{E}(\mu) = \frac{-1}{\mu}
\int d P(\vectorin{\mathcal{P}^{1}}) d P(\vectorin{\mathcal{P}^{2}}) \log
\sum_{a_1,a_2} \mathcal{P}^1_{a_1} \mathcal{P}^2_{a_2} \exp  \left\lbrace -\mu \beta
\Freeenergy_{E}(\vec{h}_{a_1},\vec{h}_{a_2}) \right\rbrace \;, \label{eq:RSBfij}
\end{equation} replacing (\ref{eq:RS_freeenergy2}). Again we are only interested in
the limit $1/\beta=0$ which restricts the argument in the exponent to integer
multiples of $\mu$. The variational dependence on $\beta$ from which energy and
entropy are determined through the term $f_{V}(\mu,1/\beta)$, is effectively
replaced by a variational dependence on $\mu$ in both parts. From this expression we
can construct the energy and complexity~\cite{Mezard:CMZT} \begin{equation} e(\mu) =
\frac{\partial}{\partial \mu} \mu f \;;\;\qquad \Sigma(\mu) =
\frac{\partial}{\partial \mu} f \;. \label{eq:energyandcomplexity} \end{equation}
Thermodynamically meaningful solutions for $\mu$ exist where $\Sigma(\mu)$ is
non-negative, and a convex function of $e(\mu)$.

\subsection{Cavity method for locked constraint satisfaction problems and
b-Matching} \label{ssec:occupation_problem_equivalence}

Ground states of energy zero in our g-spin framework can be considered as solutions to a constraint satisfaction problem (CSP). In the CSP framework each additive term in the Hamiltonian (\ref{eq:FactorHam}) is considered to be a constraint. Local assignments producing minimum energy solutions of (\ref{eq:FactorHami}) satisfy the corresponding constraint, whereas realization of any other excited energy level is a violation. Globally a g-spin assignment is satisfied only if the energy is zero.

\cut{}{\subsubsection{Equivalence of the symmetric solution and b-Matching}}

By solving the 2-state model, we obtain a spin symmetric (or paramagnetic) solution
to the full model, and we can show the 2-state model is equivalent to a well studied
class of constraint satisfaction problems known as
b-Matching~\cite{Bayati:ECM,Zdeborova:NMRG}.

In the b-Matching problem, the task is to determine a set of edge states such that at each vertex $i$ exactly $b$ out of the $C_i$ edges incident on the vertex are in state $1$, the rest being in state $-1$. By analogy each g-spin in the two-state model describes the state of edges as dimer or non-dimer, with the zero energy condition requiring $b=N_{C,\lambda}$ dimers about each factor (vertex), with $N_{C,\lambda}$ defined in (\ref{eq:dimeroccupation}). b-Matching has a trivial solution for $b=0$ or $C$ ($|\lambda|>C-1$), since then an optimal matching is simply all dimers or all non-dimers. Other cases are non-trivial and have only recently been solved on random regular graphs. Standard b-Matching applies where $N_{C,\lambda}$ contains a single element, therefore the results we discuss do not apply to special ratios (where degeneracy is allowed).

For random regular graphs, it has been shown that the satisfied solution always
exists asymptotically in $N$, implying that the ground state energy of the two-state
model is always zero. This result is most easily derived as a simple caveat to the
'contiguity' phenomena in random regular graph theory~\cite{Janson:RG}. Furthermore,
the solution space in the b-Matching problem in random regular graphs is known to be
replica symmetric at zero and finite temperature ~\cite{Bayati:ECM,Zdeborova:NMRG}.
At zero temperature, the fixed point for the cavity recursion is described in our
notation by \begin{equation} P_{j \rightarrow i}({\tilde S}_{ij}) \propto
\sqrt{x}\delta_{{\tilde S}_{ij},1} + \sqrt{1-x}\delta_{{\tilde S}_{ij},-1}
\label{eq:fixedpoint2state} \end{equation} where $x=N_{C,\lambda}/C$. The entropy
density derived via the cavity method is then~\cite{Zdeborova:CSPIS}
\begin{equation} s = \frac{2}{C}\log{C \choose C x} + \left[ x\log(x) +
(1-x)\log(1-x) \right] \label{eq:entropyzeroenergy}\;. \end{equation} It is a simple
matter to map the fixed point (\ref{eq:fixedpoint2state}) through
(\ref{eq:4to2state}) to obtain the four-state symmetric fixed point, and again
calculate the RS entropy. The entropy density is found to be reduced by $(C/2-1)\log
2$ and hence negative for all cases.

The following argument provides intuition on the contraction of the solution space
when transferring from the b-Matching problem to the Ising Hamiltonian problem at any
temperature or in any graph ensemble, that leads always to negative entropy in
locked problems at zero temperature. This transfer involves attaching an Ising spin
to every vertex, such that dimers and non-dimers are each formed in one of two
states. In the spirit of the cavity method, we consider the effect of adding a node
to the graph. This increases the entropy per node by $\log 2$. Since the solution
must be compatible with the intersection constraints, we have to subtract the
entropy due to the double counting of the entropy $\log(2)$ per link, or
equivalently ${\bar C}/2\log 2$ per node. Overall, in the four-state model space,
the entropy per node decreases by $({\bar C}-2) /2\log 2$, relative to the entropy
of the two-state model space.

The negative entropy indicates that the spin-symmetric solution cannot be the physical solution for any parameterization for which the corresponding 2-state model is a b-Matching problem on regular graphs. Whilst this result is stated only for random regular graphs, we find it to be quite generally true that the paramagnetic solutions are of negative entropy at low temperature if the interactions are locked, so that we must seek a spin-symmetry broken solution.

\section{Results of the cavity method}
\label{sec:results_of_the_cavity_method}

We present briefly some replica symmetric solutions, which qualitatively reproduce the phenomena of the small system experiments.
We discuss only the positive $\lambda$ and ${\bar C}>2$ scenario, where only solutions of zero magnetization are found.

\subsection{The paramagnetic solution}
\label{sec:paramagneticsolution}

We will describe in detail the paramagnetic solution for the case of regular graph ensembles, the salient features being relevant also for the case of the linear graph ensemble.
This paramagnetic solution for the case of regular graphs, by local homogeneity, is described by a single cavity field $\vec{h}$, and is compatible with the solution of the two-state model (\ref{eq:2stateparameterization}). The solution is a fixed point of the cavity equation (\ref{eq:T}), considering a single field we must solve a polynomial equation, the order of the polynomial grows linearly with ${\bar C}-1$.

To determine solution stability we can use the more precise method outlined in \ref{ssec:regular_graph_stability}, again avoiding a population dynamics approach. The eigenvalues describing the evolution of the mapping perturbations provide information on the stability towards spontaneous symmetry breaking (a linear instability) and towards replica symmetry breaking (a non-linear instability). Linear instability may arise for sufficiently negative $\lambda$, where an RS ferromagnetic solution emerges continuously at low temperature. We present results only for positive $\lambda$, where we find the solution is always linearly stable. The second type of instability is towards replica symmetry breaking and is relevant to positive $\lambda$.

\begin{figure}[h!tbp]
\begin{center}
\includegraphics[width=0.65\linewidth]{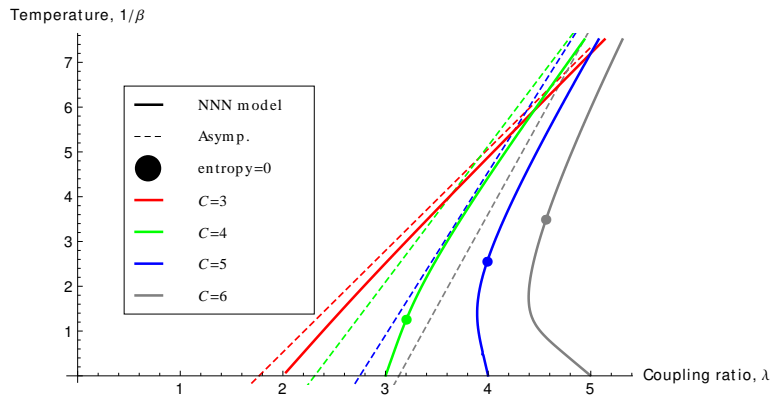}
\caption{\label{fig:TcPositive} Solid lines are the critical curves for the paramagnetic solution for regular connectivity graphs ${\bar C}=3,4,5,6$, curves emanate from $\lambda={\bar C}-1$ respectively. To the right of the curves the paramagnetic solution is unstable towards RSB, everywhere else the solution is stable. The paramagnetic solution is stable down to zero temperature for small $\lambda$. Markers for $C=4,5,6$ indicate the crtical temperature below which entropy is negative. For ${\bar C}\geq 5$ there is a reentrant behaviour for $\lambda \lesssim {\bar C}-1$, but limited to the negative entropy regime. Dashed lines indicate the asymptote (\ref{eq:nnTc2}).}
\end{center}
\end{figure}
In figure \ref{fig:TcPositive} we present the critical curves determined by the local stability analysis. For comparison, we plot the critical curves that would be anticipated from models of equal strength in nearest neighbor interactions, but without next nearest neighbor interactions.
For many standard random and sparse graph ensembles including the regular and linear cases, we anticipate that asymptotically in large $\lambda$ the critical temperature will approach from below the critical temperature of the Bethe spin glass, given by the solution of
\begin{equation}
1 = \sum_{C=1}^\infty P(C) \frac{C(C-1)}{{\bar C}} \tanh^2(2 \beta |\lambda|) \;,
\label{eq:nnTc}
\end{equation}
where $P(C)$ is the connectivity distribution describing the sparse model. This is because when $\lambda$ is large, interactions between spins are dominated by nearest neighbor anti-ferromagnetic couplings. In random graphs the presence of loops with odd number of edges causes frustration, and spin glasses are formed at low temperatures~\cite{Kanter:MFT}. There is however a perturbation on this result owing to next nearest interactions which is calculated in \ref{Appendix:expin1lambda}. The critical temperature found by an expansion in $1/\lambda$ is
\begin{equation}
  T_c = \frac{2}{\atanh(1/\sqrt{C-1})}\left(\lambda - \frac{3 C - 4}{2 \sqrt{C-1}} \right)\;,
\label{eq:nnTc2}
\end{equation}
in agreement at leading order $O(\beta\lambda)$ with (\ref{eq:nnTc}) and in good agreement with the curves of figure \ref{fig:TcPositive}.

\begin{figure}[h!tbp]
\begin{center}
\includegraphics[width=0.65\linewidth]{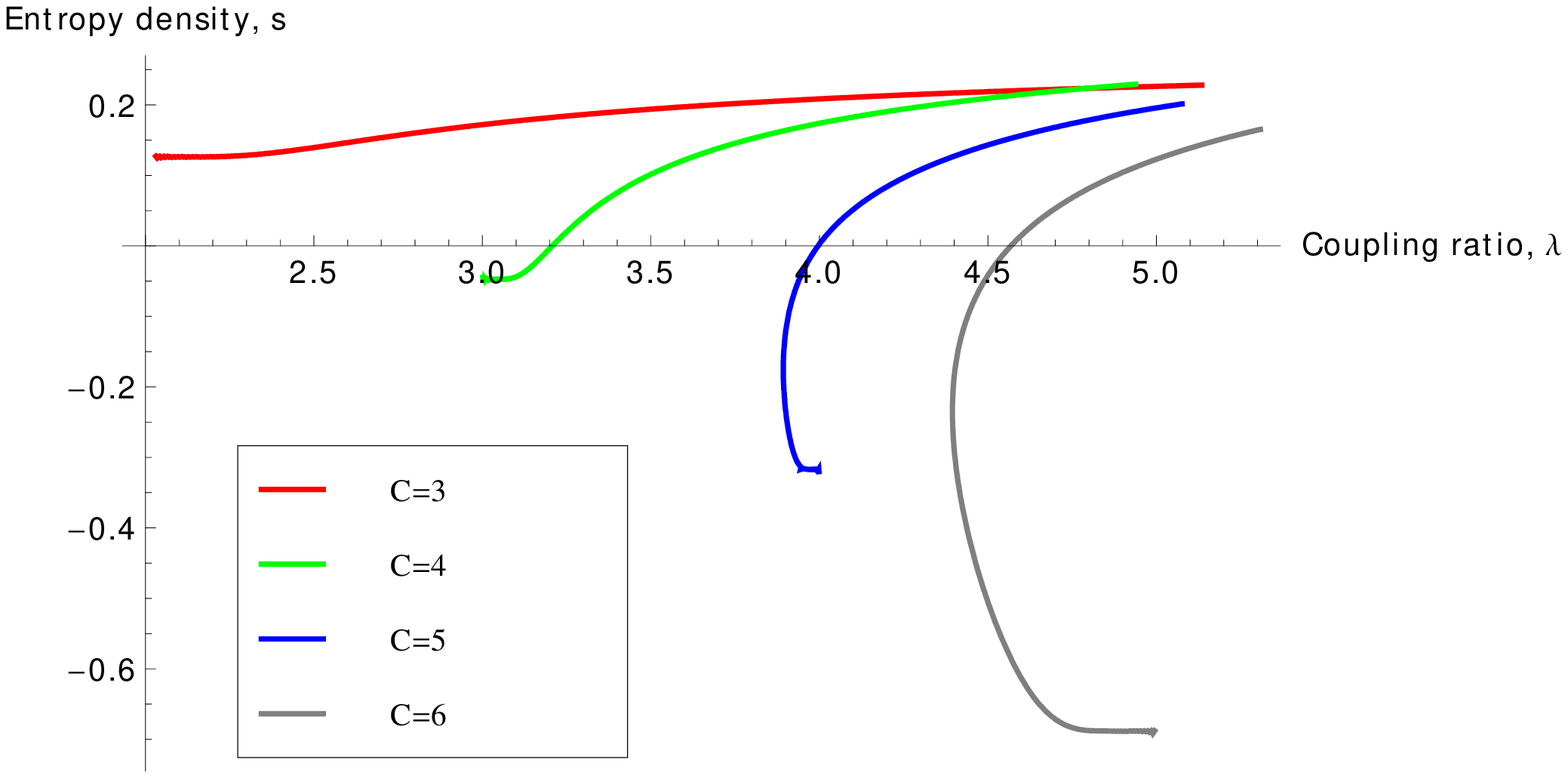}
\caption{\label{fig:EntropyPositive} Continuous lines indicate the entropy along the critical curves of figure \ref{fig:TcPositive}, curves corresponding to graph connectivity ${\bar C}=3,4,5,6$ emanate from $\lambda={\bar C}-1$.
\cut{}{Symbols indicate the positive entropy paramagnetic solutions at zero temperature, symbols are labeled by the corresponding graph connectivity. These solutions are restricted to a subset of the special ratio ensembles (integer $\lambda\leq 2$, and small $C$).}
}
\end{center}
\end{figure}

The curves approach $\lambda_c={\bar C}-1$ at zero temperature. It is noteworthy that for ${\bar C} \geq 5$ the curve becomes multi-valued for some $\lambda$, indicating a reentrant paramagnetic phase. However, as will be seen in figure \ref{fig:EntropyPositive}, the entropy demonstrates that the paramagnetic solution at this part of the curve is not viable.
It is also interesting to consider the stability as $1/\beta \rightarrow 0$. We find that where $\lambda<\lambda_c$ either the solution is marginally stable in locked ensembles (the eigenvalue approaches $1$ from below in the zero temperature limit), or stable in special ratio ensembles. For $\lambda>\lambda_c$ we find the solution is unstable, whereas for $\lambda=\lambda_c$ (which are also special-ratio ensembles) the solution is stable for ${\bar C} \leq 4$ and unstable for higher connectivity.

In figure \ref{fig:EntropyPositive} we present the entropy of the paramagnetic solution at the critical temperature(s) indicated by figure \ref{fig:TcPositive}. Along the critical curves we can have either a negative or positive entropy solution. For high enough temperature the local instability applies to a positive entropy paramagnet. This is consistent with the behaviour of nearest neighbor models, for which a continuous transition to a full-RSB phase is found~\cite{Kanter:MFT}. The curve for ${\bar C}=3$ is exceptional in that everywhere the entropy is positive. So this curve may be describing correctly a continuous transition even at low temperatures. Otherwise at low temperatures, we have an unphysical behaviour of the entropy and must invoke a different phase to explain behaviour. \cut{\ref{Appendix:expin1lambda} provides some insight into the dependence of entropy on weak next nearest neighbor interactions, the entropy is always reduced relative to a corresponding model with only nearest neighbor } interactions.

Despite the fact that for small $\lambda$ the paramagnetic solution is locally stable in the limit of zero temperature, we find that only if $\lambda$ and ${\bar C}$ are small, and $\lambda$ is a special ratio, can we have a positive entropy solution. In table \ref{table:entropyT0} we give an exhaustive list of those regular graphs with $C\geq 3$ for which entropy of the paramagnetic solution is positive at zero temperature to $3$ significant figures\\
\begin{table}[h!]
\begin{center}
\begin{tabular}{|c|c|c|c|c|c|c|c|c|}
\hline
 $|\lambda|$ & 0 & 0 & 0 & 1 & 1 & 1 & 2 & 2\\
 ${\bar C}$ & 3 & 5 & 7 & 4 & 6 & 8 & 3 & 5\\
 entropy density & 0.406 & 0.223 & 0.0896 & 0.269 & 0.123 & 0.0106 & 0.235 & 0.0925 \\
\hline
 \end{tabular}
 \end{center}
 \caption{\label{table:entropyT0}Zero temperature entropy of the paramagnetic solution on regular graphs, all positive entropy cases.}
 \end{table}
Thus amongst the locally stable paramagnetic solutions for regular graphs, it is only a small subset of special ratios for which the solution is viable. The subset identified is consistent with the zero energy and large entropy results found in the numerical studies of section \ref{sec:small_system_studies}. For the majority of ensembles in which the paramagnetic solution is locally stable, but the entropy is negative, we must expect an alternative solution as the temperature is lowered.

For inhomogeneous graph ensembles, the paramagnetic solution and its local stability have also been studied by means of population dynamics. The features found for the regular case seem to hold more generally. We find for $\lambda \lesssim {\bar C}-1$ the paramagnetic solution is locally stable down to very low, or effectively zero temperature. %For numerical reasons irrelevant to the analysis of this section, it is at low temperatures difficult to establish this threshold.
By contrast for $\lambda \gtrsim {\bar C}-1$ we have local instabilities. As for the regular graph ensemble, only a subset of special ratio ensembles have positive entropy locally stable solutions at low temperature.

\subsection{RS spin glass solutions at finite temperature by population dynamics}
\label{ssec:replica_symmetric_solutions}

The paramagnetic solution is the stable and unique solution for all parameterizations at sufficiently high temperature.
Initializing population dynamics with symmetric fields we always converge to this solution.
Another solution at finite temperature has broken spin symmetry, and we call it the spin glass solution since the magnetization is zero and the spin-glass order parameter (\ref{eq:q_F}) is non-zero.
As the temperature is lowered the paramagnetic solution may be locally unstable towards a spin glass solution, or both solutions may coexist at low temperature.

\begin{figure}[h!tbp]
\begin{center}
\includegraphics[width=0.8\linewidth]{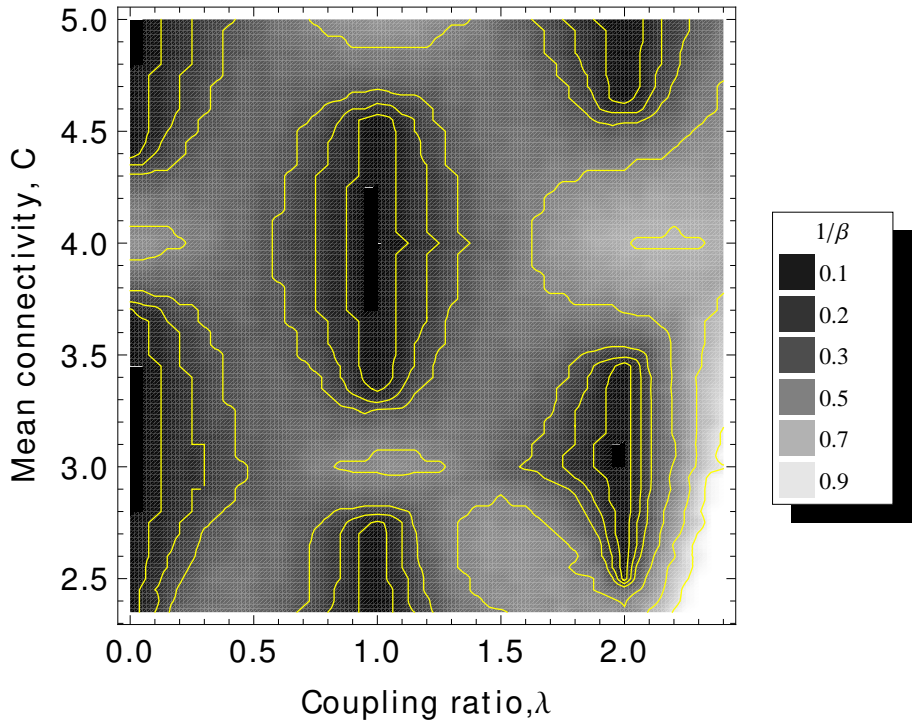}
\caption{\label{fig:Tc} The critical temperature $1/\beta$, below which a spin glass solution exists for the RS equations is shown as a contour map.
Parameters are sampled on a grid with increments $\Delta \lambda=0.05$, $\Delta {\bar C}=0.05$, the critical temperature found is represented by a grayscale: $1/\beta=0$ (black) to $1/\beta=1$ (white), contours are fitted by linear splines at the levels indicated by the legend.
For some combinations of integer $\lambda$ and ${\bar C}$ the spin glass solution is absent at all temperatures, we assign the critical temperature to zero in these cases.
A periodic pattern of critical temperatures is apparent, as $\lambda+{\bar C}$ approaches odd integer values we find local minima in the critical temperature.
}
\end{center}
\end{figure}

The extent of the spin glass phase is shown in figure \ref{fig:Tc} for a range of parameters.
Above the critical temperature, populations of $N=10000$ fields converge to the symmetric solution within $O(100)$ iterations from random symmetry broken initial conditions.
Very close to the transition longer timescales are relevant, so that the critical temperatures presented in the figure should be considered upper-bounds, but still correct quantitatively up to $O(10^{-2})$.

\begin{figure}[h!tbp]
\begin{center}
\includegraphics[width=0.57\linewidth]{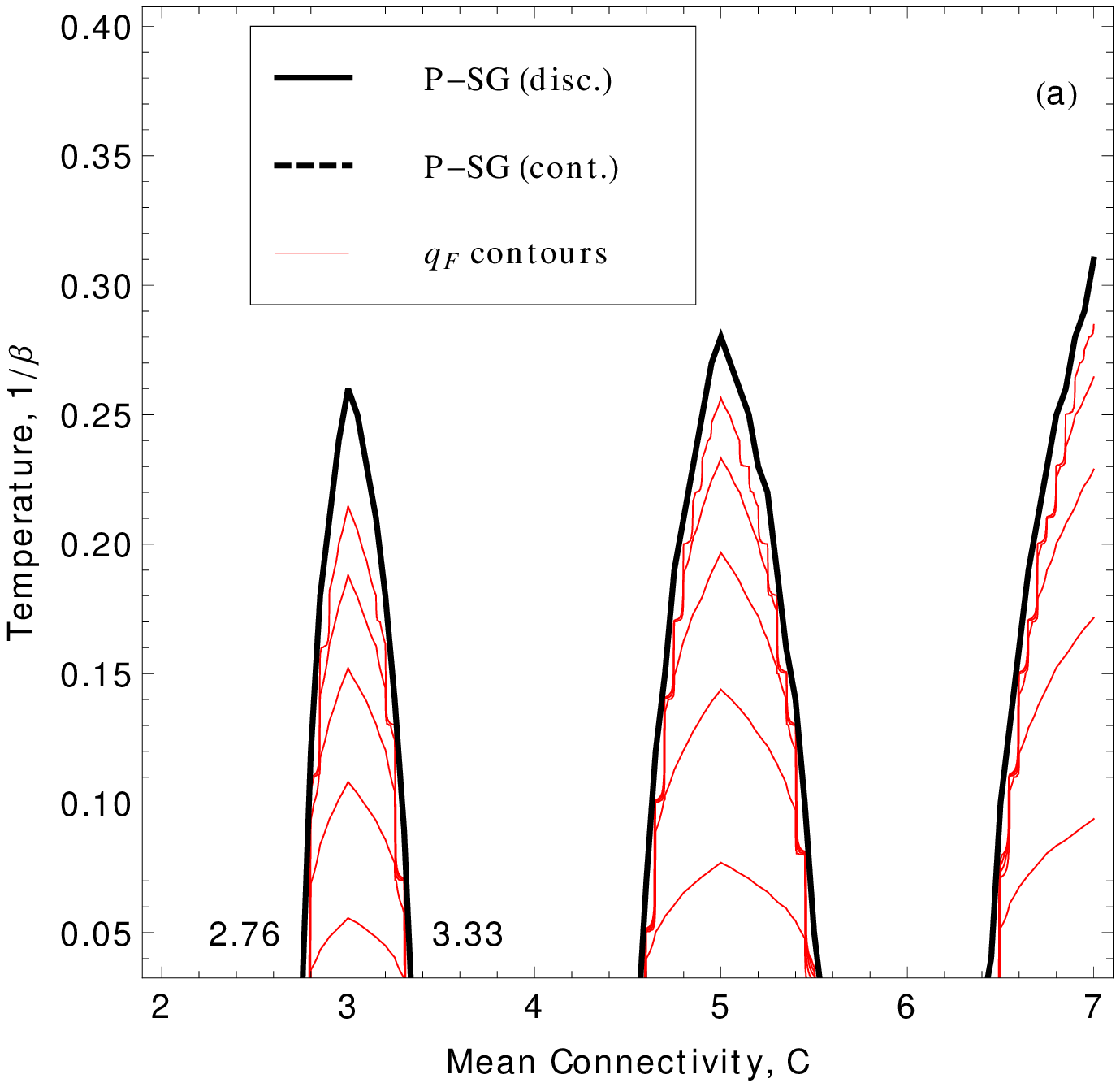}
\includegraphics[width=0.57\linewidth]{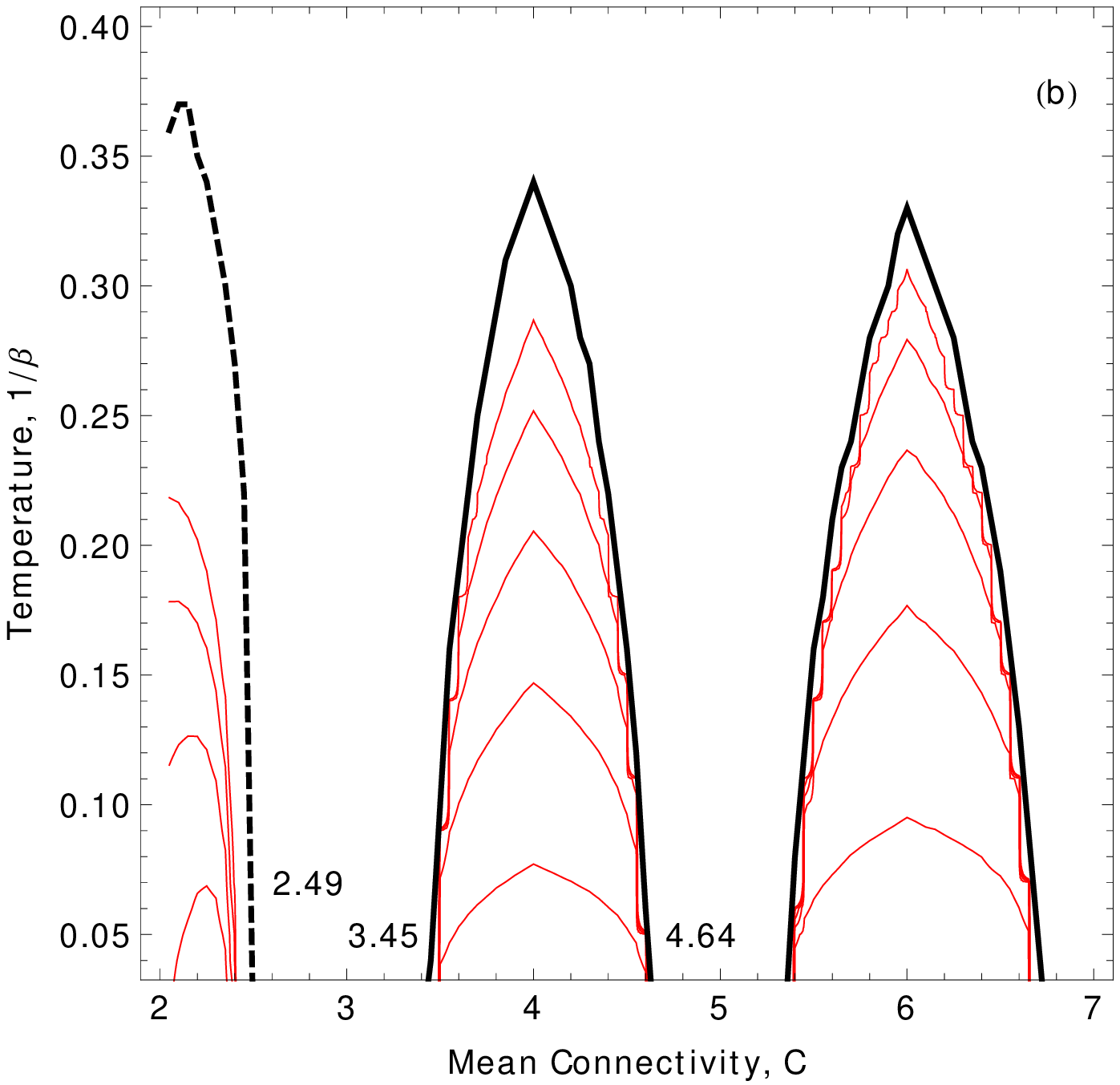}
\caption{\label{fig:Tc2} Spin glass phase properties for (a) $\lambda=1$ and (b) $\lambda=2$. Data is sampled with increments $\Delta(1/\beta)=0.01$, $\Delta{\bar C}=0.05$ and contours are fitted by linear splines.
The uppermost lines indicate the maximum temperature allowing a spin glass solution to be found from random initial conditions of population dynamics (an RS spin glass existence curve).
Solid lines indicate a discontinuous transition in $q_F$, dashed lines indicate a continuous transitions from $0$ and a corresponding local instability of the paramagnetic solution.
The solid (red) lines indicate contours in $q_F$ (\ref{eq:q_F}) ordered monotonically decreasing in temperature, $q_F=0.96,0.92,0.88,0.84,0.8$.
Labeled beside the x-axis are thresholds obtained from a zero temperature analysis, discussed in the main text, in apparent agreement with the $1/\beta \rightarrow 0$ limit of the spin glass existence curve.}
\end{center}
\end{figure}

The most interesting phenomena appear in the range $|\lambda|\lesssim C$.
This is the regime where nearest and next nearest neighbor couplings have comparable strength and the behaviour is influenced by proximity to special ratios.
In the case of integer $\lambda$ we observe that a paramagnetic solution can be the unique solution even at zero temperature.
Two cases are shown in figure \ref{fig:Tc2}.
For $\lambda=1$, the zero-temperature paramagnetic phase exists when even connectivity dominates, agreeing with the prediction of (\ref{eq:specialratios}) and reproducing the periodic pattern that characterizes the ground state energy in figure \ref{fig:EnergyEntropyN32}.
The scenario for $\lambda=2$ is similar, except that zero-temperature paramagnetism exists when odd connectivity dominates.

\begin{figure}[h!tbp]
\begin{center}
\includegraphics[width=0.57\linewidth]{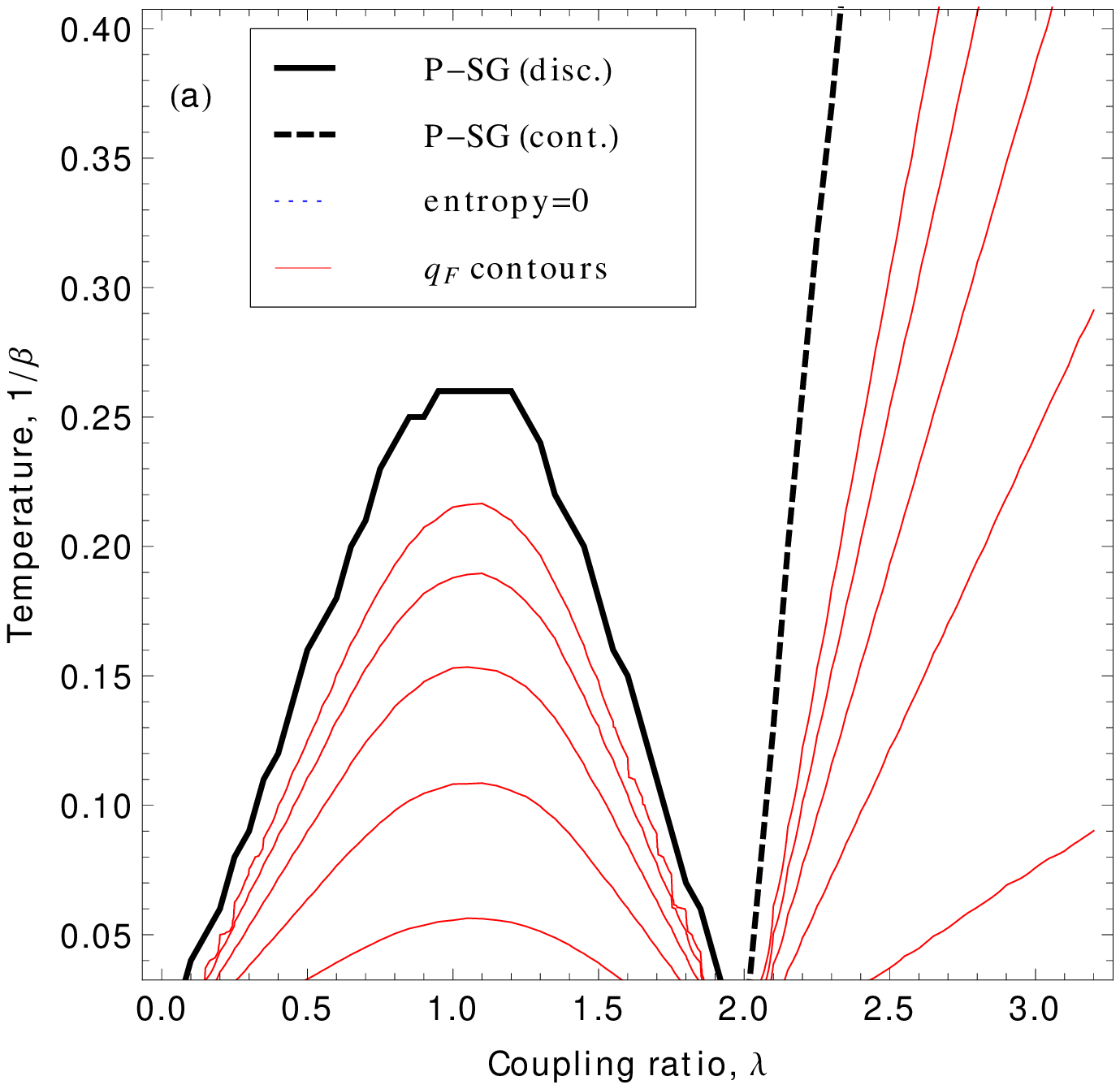}
\includegraphics[width=0.57\linewidth]{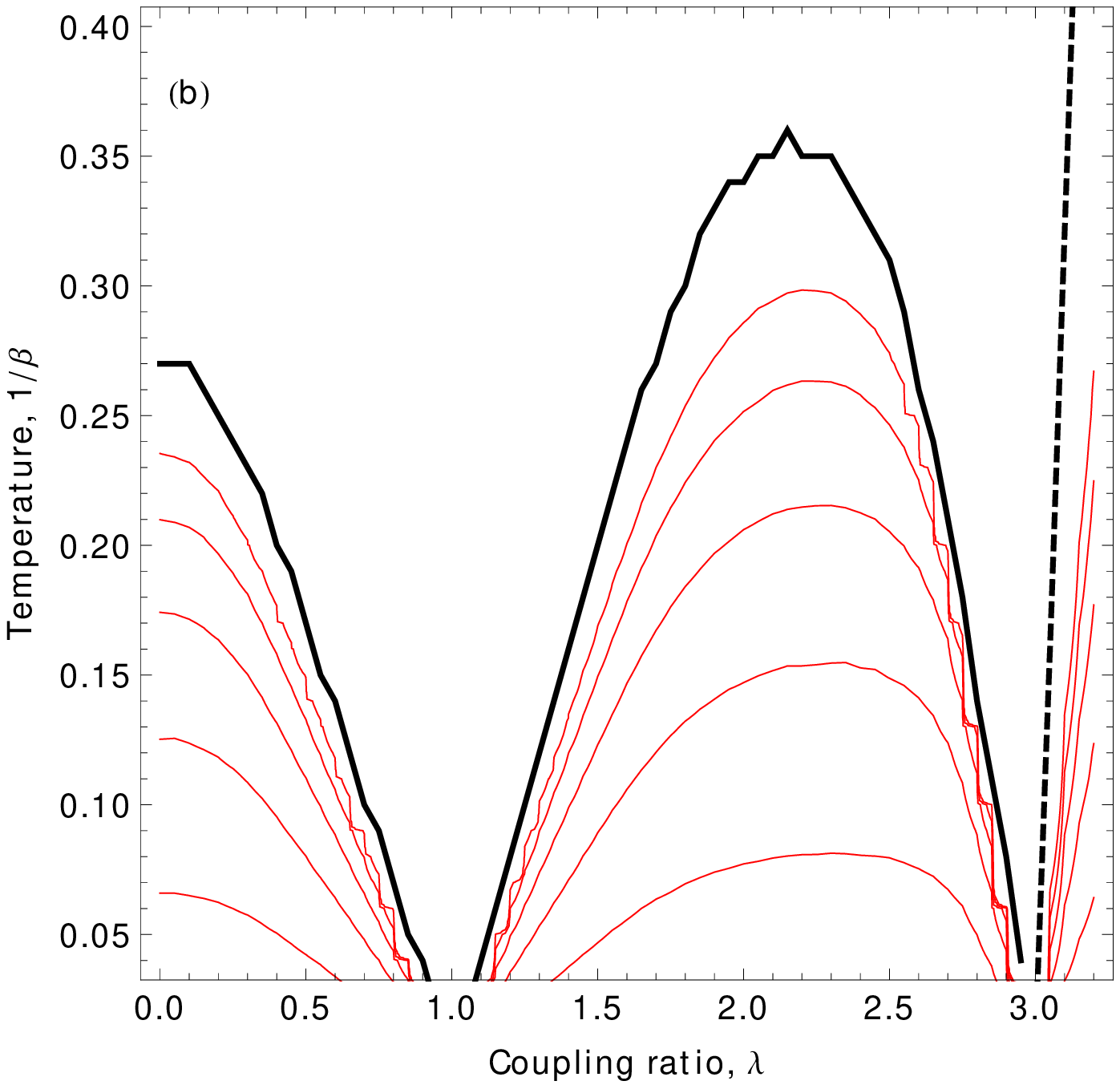}
\caption{\label{fig:Tc3} Spin glass phase properties for regular graphs  ${\bar C}=3$ (a) and ${\bar C}=4$ (b).
Data is sampled with increments $\Delta(1/\beta)=0.01$, $\Delta\lambda=0.05$.
Lines have same sense as figure \ref{fig:Tc2}.}
\end{center}
\end{figure}

Figure \ref{fig:Tc3} demonstrates the extent of the RS spin glass solution in regular graphs.
For non-integer $\lambda$, we have only locked interactions and find the spin glass phases exist below some finite critical temperature, this temperature being lowest around integer values.
The critical temperature is locally minimized at special ratio ensembles, and maximized at approximately their midpoints, following an odd-even pattern.
Due to the special ratios there is no spin glass solution even in the limit of zero temperature, for some combinations of integer $\lambda$ and small ${\bar C}$.

Not shown in figures \ref{fig:Tc}-\ref{fig:Tc3} are the cases of larger ${\bar C}$ and larger $\lambda$, where the critical temperature is typically much larger.
As for regular graphs, we find evidence that low or zero temperature paramagnetic solutions are inviable either due to local instability (${\bar C} \gtrsim \lambda - 1$), or unphysical values for quantities such as entropy. For $\lambda$ large relative to ${\bar C}$ we see again the emergence of a continuous transition following a critical curve consistent with a model dominated by nearest neighbor couplings as in (\ref{eq:nnTc}).

It is noteworthy to distinguish between continuous and discontinuous spin glass to paramagnetic transitions within the figures,
the spin glass order parameter either drops to zero discontinuously from a value $q_F \sim [0.7,1]$,
or goes continuously to zero as the temperature is increased.
In the vicinity of discontinuous transitions a coexistence region exists between the paramagnetic and spin glass solutions.
We observe that continuous transitions can occur for either: ${\bar C}\sim 2$, i.e. models dominated by chains; or $\lambda \gtrsim {\bar C}-1$, i.e. models dominated by nearest neighbor couplings.
Elsewhere the transitions were found to be discontinuous.

The RS population dynamics method for inhomogeneous graphs shows a similar local instability of the paramagnetic solution at large $\lambda$ to that uncovered in the analysis of regular graphs. In addition it demonstrates the discontinuous emergence of a replica-symmetric SG solution as the temperature is lowered or the ratio of locked to unlocked clauses is changed. The phase boundaries match qualitatively the mean field arguments of section \ref{sec:special_ratios} and the numerical results of section \ref{sec:small_system_studies}.
However, the RS spin glass solution is found to be locally unstable everywhere towards further levels of replica symmetry breaking, the specific heat and entropy behave unphysically in all spin glass solutions.
Furthermore, even where the paramagnetic solution is locally stable, it is frequently found to be a negative entropy solution below some critical temperature.
Higher level RSB is required for a correct description; but we develop the cavity method henceforth only for the limit of zero temperature.

\subsection{Replica symmetric solutions at zero temperature}
\label{ssec:replica_symmetric_solutions2}

In the zero temperature limit the RS population dynamics involve energetic $O(1)$ and entropic $O(1/\beta)$ fields. If these fields are initiated symmetrically, the population %COMMENT20120212 the population rather than the solution
can converge to a paramagnetic solution. The solution can be locally stable, and of positive entropy close to the special ratio values. However, as for finite temperature, the paramagnetic solution is often of negative entropy away from these points.

We can by contrast find a critical behaviour if the fields are initialized as $O(1)$ random numbers. The solutions are then paramagnetic ones only for odd (even) integer $\lambda$, and with sufficiently many even (or odd) connectivity vertices.
We mark in figure \ref{fig:Tc2} the critical values in ${\bar C}$ at which the paramagnetic solution becomes the unique solution achieved from all initial conditions. These points are in good agreement with the limits of the finite temperature results.

Outside this regime where the paramagnet appears to be the unique solution, we find the entropic fields diverge under iteration so that no solution can be found, with one exception. An RS spin glass solution is found for special ratio $\lambda=2$ and ${\bar C} \in (2.4,2.49]$.
This solution has spin-symmetric energetic fields, but spin-symmetry broken entropic fields; $q_F$ is non-zero, and goes continuously to zero at ${\bar C}=2.49$. Although population dynamics converges in this case, this RS spin glass solution is locally unstable to RSB, like the finite temperature spin glasses.

\subsection{Energetic 1RSB solutions}
\label{ssec:energetic_1RSB_solutions}

To gain insight beyond RS we look to 1RSB and focus on graphs with low connectivity and integer $\lambda$.
We will skip the discussions on the cases of non-integer $\lambda$, large integer $\lambda$, or large ${\bar C}$, since the state spaces in those cases are strongly constrained, as argued in section \ref{sec:special_ratios}, and so far we have not found evidence of zero temperature phase transitions in these regimes.

\begin{figure}[h!tbp]
\begin{center}
\includegraphics[width=0.75\linewidth]{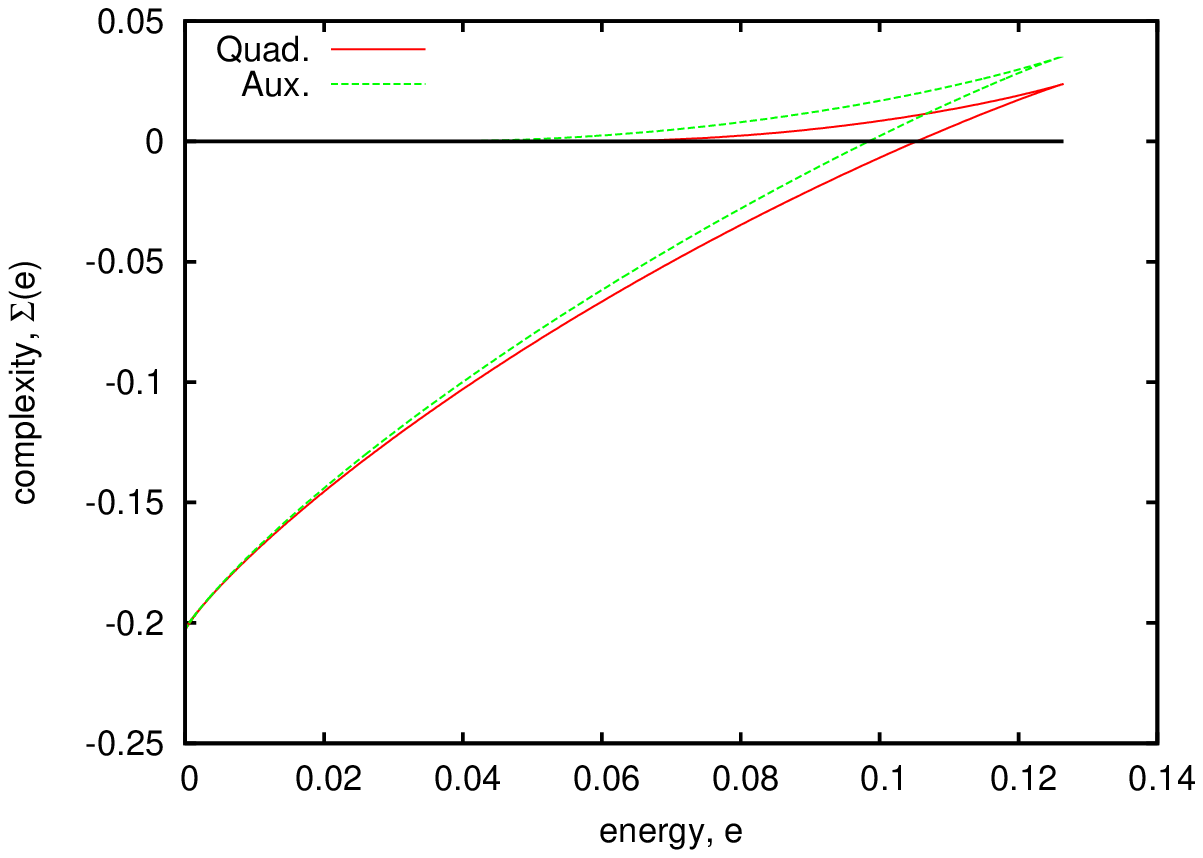}
\includegraphics[width=0.75\linewidth]{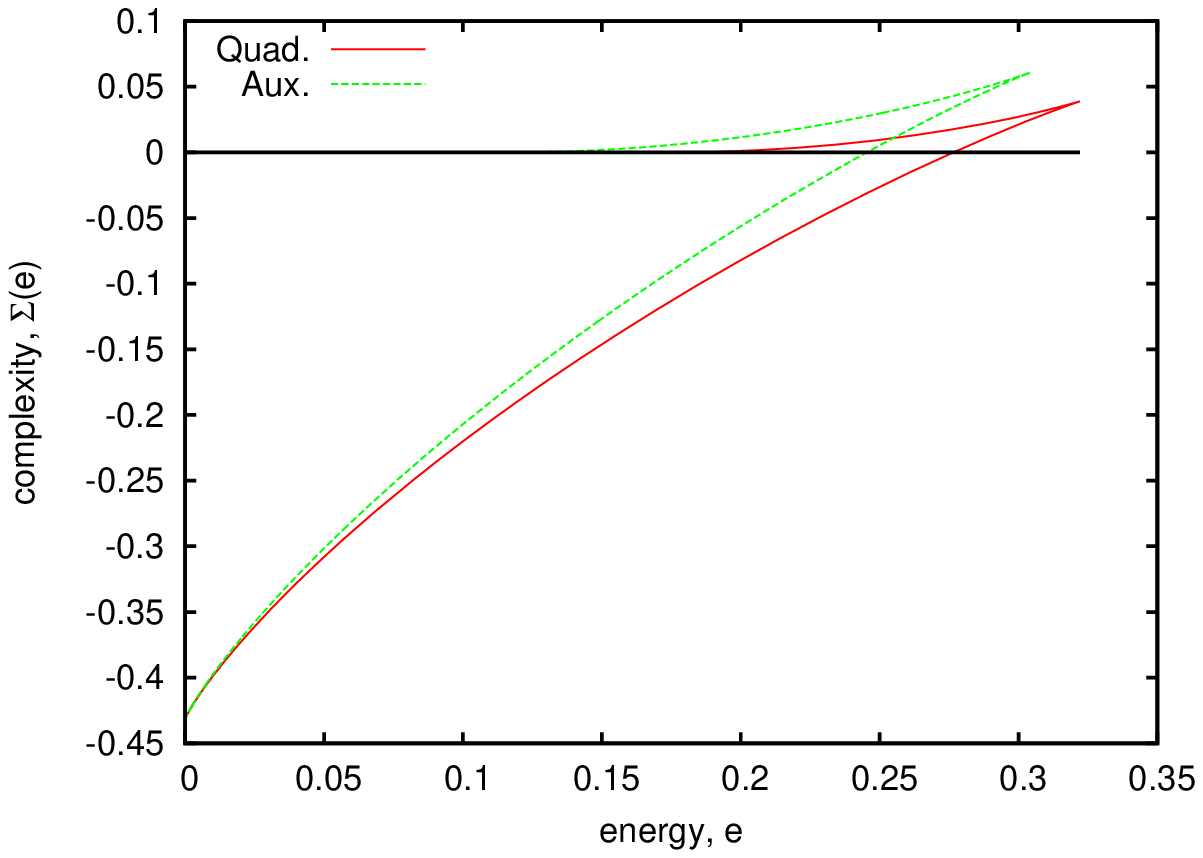}
\caption{\label{fig:complexity1} Complexity curves for the locked problems on regular graphs: $C=3,\lambda=1$ (top) and $C=4$,$\lambda=2$ (bottom).
Physically relevant solution for complexity is the convex upward part from the point of maximal complexity to the intersection at zero complexity.
Phenomenology of the full (\ref{eq:FactorHami}) and auxiliary (\ref{eq:FactorHami2}) Hamiltonian solutions are similar, but with significant errors in absolute terms.}
\end{center}
\end{figure}

To facilitate the calculation of the energy and complexity in (\ref{eq:energyandcomplexity}), we first compare the behaviour of the full Hamiltonian (\ref{eq:FactorHami}) and the Auxiliary Hamiltonian (\ref{eq:FactorHami2}).
Results for the complexity $\Sigma(e)$ are shown in figure \ref{fig:complexity1}. The new model provides a lower bound for the energies at given complexity, and an upper bound for the complexity at given energy, where it exists. There seems to be a good overlap in these two parameters so that it is useful. However, the absolute values predicted for the ground state energy, and spinodal point (solution of maximum energy), differ quantitatively and to consider a better auxiliary, for example 
%approximating the local energies by the set $\{0,1,\infty\}$ as opposed to $\{0,1\}$ 
 we may approximate the local energies by the set $\{0, 1, \infty\}$, where infty corresponds to forbidden energies when the local energies of the full Hamiltonian are 2 or above. As opposed to $\{0, 1\}$, this
might tighten the curves for such a purpose, at the cost of complexity.
%The high precision in figure \ref{fig:complexity1} can be attributed to the choice of integer ${\bar C}$; owing to the symmetry of the cavity trees within the pure state assumption for regular graphs, it is possible to greatly simplify the analysis.
We have also compared the results for non-integer ${\bar C}$, the agreement between the two Hamiltonians becomes more convincing. This is because figure \ref{fig:complexity1} presents the case of maximum frustration, in which local excitations (where the Hamiltonians differ) are necessarily realized. In regimes where the ground state energy is close to zero 
%the curves become 
the gap between the curves becomes
correspondingly tight.

The complexity curves in figure \ref{fig:complexity1} are typical of the regimes of locked constraints, say $(\lambda,{\bar C})=(1,3)$ or $(2,4)$.
In these cases $\Sigma(e)$ starts from a negative value at $e=0$, indicating that the zero-energy state is thermodynamically negligible.
At this point the configuration parameter $\mu$ approaches infinity.
When $e$ increases, $\Sigma(e)$ increases and the corresponding value of $\mu$ decreases from infinity.
The intersection with the energy axis yields the ground state energy at zero temperature.
When $e$ increases further the positive values of $\Sigma(e)$ ends in a cusp, indicating the absence of metastable states of higher energy.
This point indicates a dynamical transition, the spinodal point.
This means that for a dynamical process in which the temperature is lowered the state will be trapped in metastable states at this energy due to their configurational entropy.
Figure \ref{fig:complexity1} also shows the upper concave segment of the complexity curves, which are unphysical.

\begin{figure}[h!tbp]
\begin{center}
\includegraphics[width=0.75\linewidth]{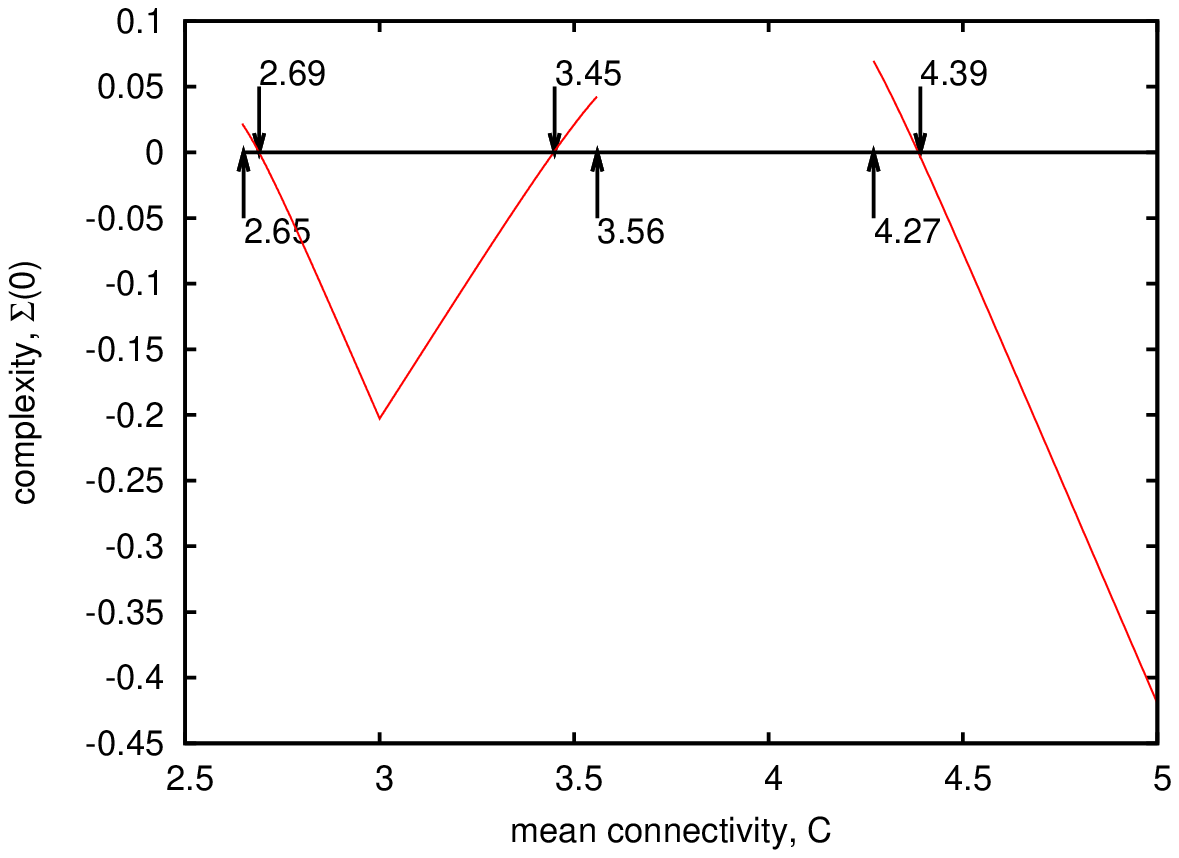}
\includegraphics[width=0.75\linewidth]{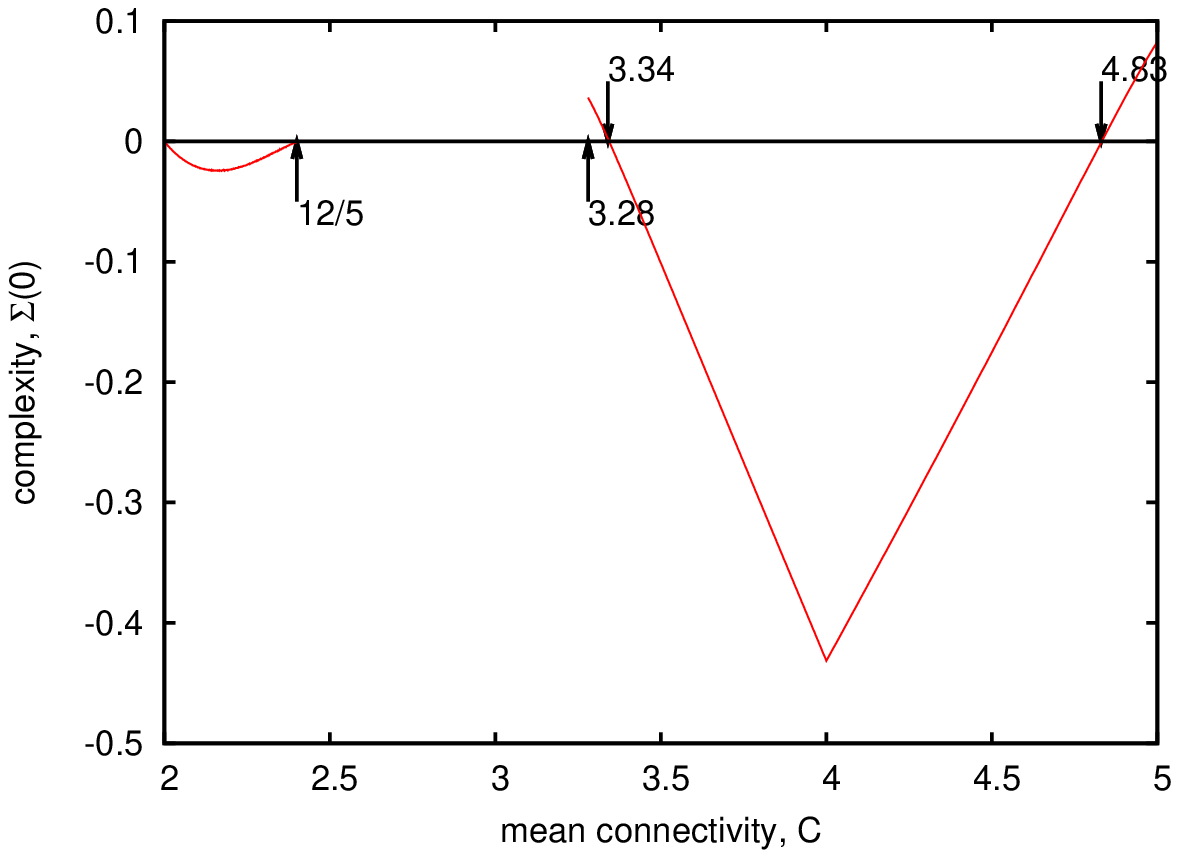}
\caption{\label{fig:complexityeq0} Zero temperature complexity curves under the assumption of zero energy, $\lambda=1$ (top) and $\lambda=2$ (bottom), and a linear connectivity ensemble.
Dynamical and thermodynamic transitions values are indicated by labeled arrows.}
\end{center}
\end{figure}

The stability analysis of 1RSB for regular graphs is discussed in ~\ref{app:population_dynamics_implementation}, and follows standard local tests called type I and type II~\cite{Montanari:NLT}.
The solutions described by the curves in figure \ref{fig:complexity1} are found to be type I stable throughout.
Testing type II instability it is found that they are unstable everywhere: both the metastable states, and the thermodynamic state, must be considered within a higher level of RSB to correctly describe all system properties.

To locate the transition points between spin glass and paramagnetic solutions, we plot $\Sigma(0)$ as a function of ${\bar C}$ in figure \ref{fig:complexityeq0}. Since energy is restricted to zero, the excitation structure that distinguishes the auxiliary and quadratic Hamiltonians are irrelevant, the results are the same.
Since $\Sigma(e)$ is an increasing function, negative values of $\Sigma(0)$ indicates a higher ground state energy $e$ is relevant,
and positive values of $\Sigma(0)$ indicate the presence not only of a zero energy ground state solution, but of exponentially many metastable spin glass states in the system.
The intersection of $\Sigma(e)$ with the ${\bar C}$ axis indicates the transition point between spin glass and paramagnetic phases,
and the points where the positive complexity disappears indicate dynamical transitions~\cite{Mezard:CMZT,Montanari:NLT}.
These transitions are labeled in figure \ref{fig:complexityeq0}.

A sequence of transitions can be observed when the value of ${\bar C}$ alternates between odd and even dominated regimes.
The curves of $\Sigma(0)$ have minima in the regular graphs with locked interactions, such as the point $(\lambda,{\bar C})=(1,3)$ indicating that the zero energy states are absent.
When one moves away from these points, the inclusion of unlocked interactions into the graph causes $\Sigma(0)$ to increase until it becomes positive at the transition points.
Comparing the RS and 1RSB predictions, we observe that the 1RSB results yield a broader spin glass regime.
For example, around $(\lambda,{\bar C})=(1,3)$, the RS and 1RSB spin glass regimes are given by $2.76\leq {\bar C} \leq 3.33$ and $2.69 \leq {\bar C} \leq 3.45$ in figures \ref{fig:Tc2} (a) and \ref{fig:complexityeq0} (left) respectively.
Besides the thermodynamically stable phases, the 1RSB analysis also predicts the existence of metastable spin glass phases at $2.65 \leq {\bar C} \leq 2.69$ and $3.45 \leq {\bar C} \leq 3.56$.
Comparing with the numerical results in figure \ref{fig:PEnergy0} we can see good agreement between the thermodynamic transition prediction and the point at which $P(e=0)$ goes to zero with increasing system size; metastable states would also explain the algorithmic slowdown of stochastic local search methods approaching this transition.

When ${\bar C}$ increases we observe that the paramagnetic phases narrow down until the entire paramagnetic phase coexists with a metastable spin-glass phase and there is no longer any dynamical transition.
An example can be found at $4.83\leq {\bar C} \leq 5$ for $\lambda=2$.
When ${\bar C}$ is sufficiently large, we expect that the paramagnetic phase at zero temperature disappears due to the increasing number of constraints, and $\Sigma(0)$ becomes negative everywhere.
Also for ${\bar C}\lesssim \lambda$ we find that $\Sigma(0)$ increases continuously from negative to zero without a dynamical transition.
This is apparent at $(\lambda,{\bar C})=(2,2.4)$ in figure \ref{fig:complexityeq0}.
We have only shown results for $\lambda=1$ and $\lambda=2$, similar properties may characterize other positive $\lambda$.

\begin{figure}[h!tbp]
\begin{center}
\includegraphics[width=0.57\linewidth]{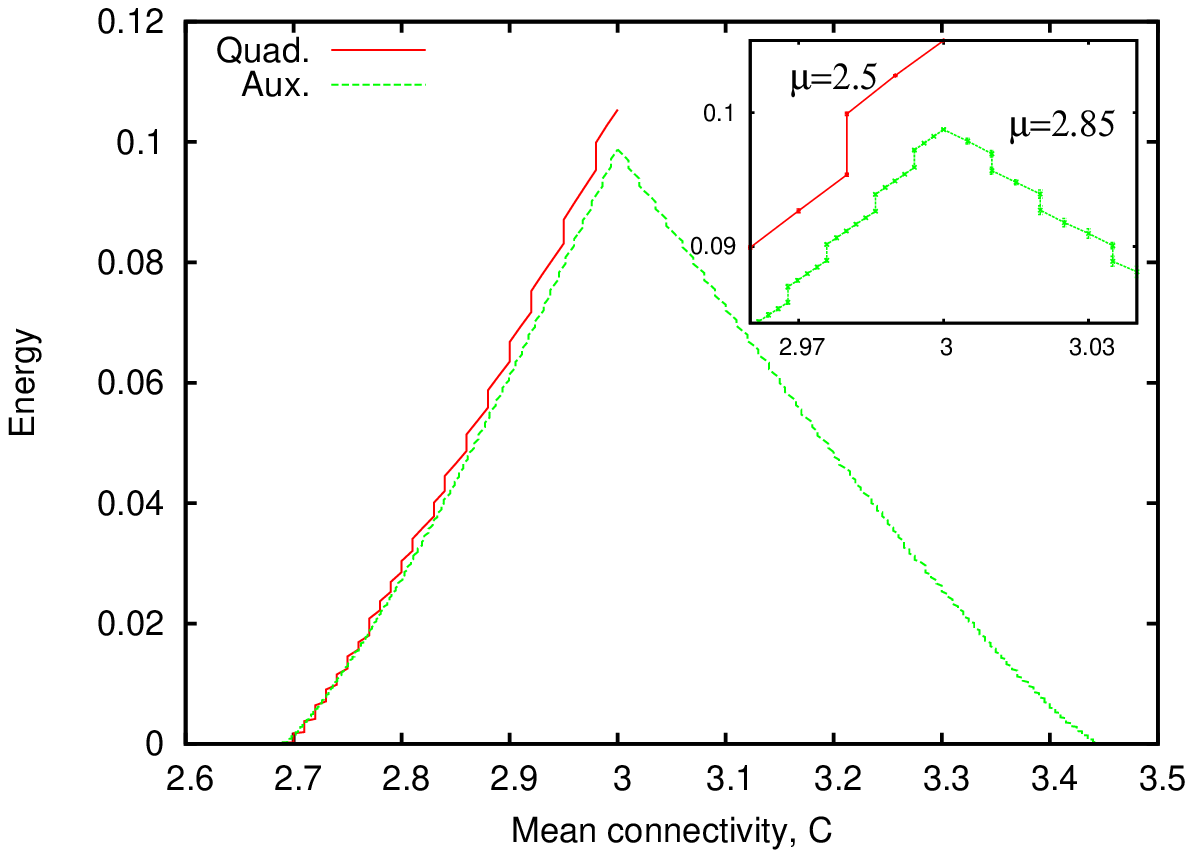}\\
\includegraphics[width=0.57\linewidth]{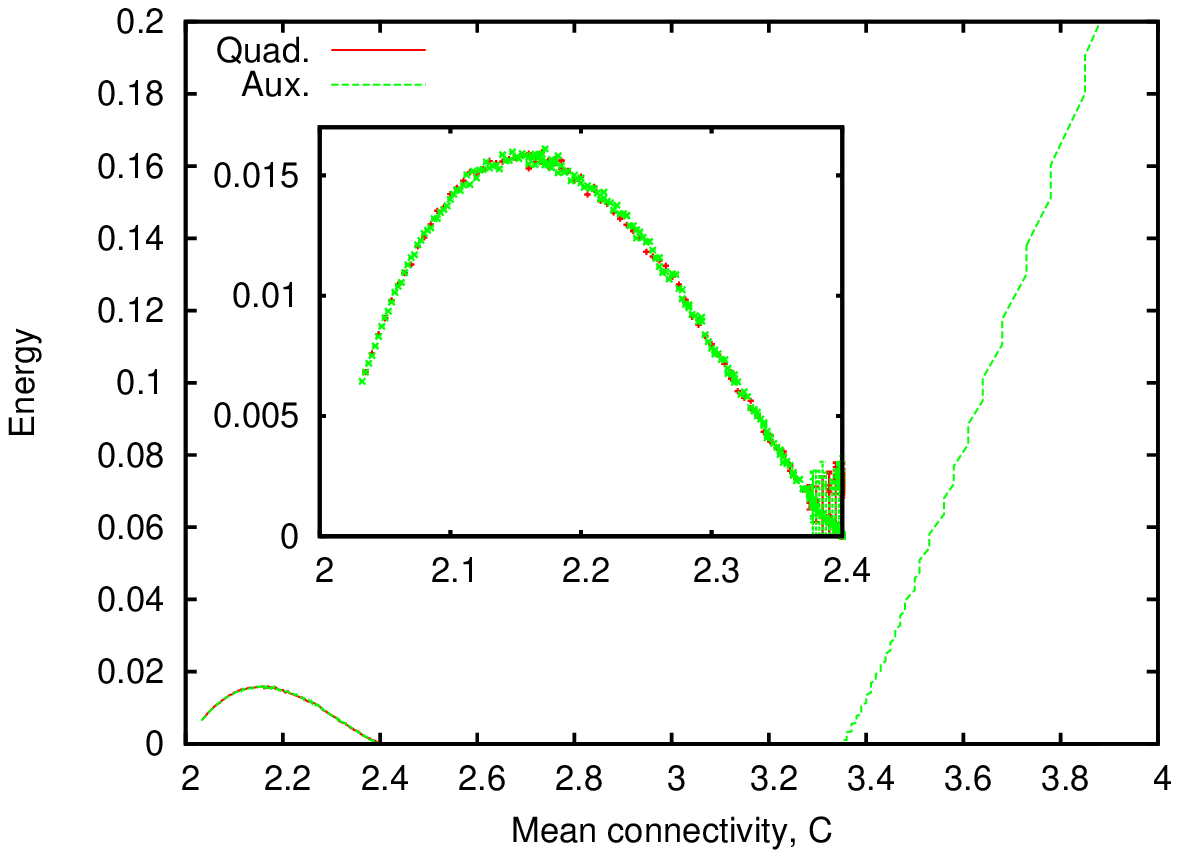}
\caption{\label{fig:energyRSB} Ground state energy for the linear ensemble with auxiliary and quadratic Hamiltonians (${\bar C}<3$).
(Top) $\lambda=1$. (Inset) Close up view of the region near ${\bar C}=3$. The configurational parameter relevant approaching ${\bar C}=3$ is marked for the two Hamiltonians.
(Bottom) $\lambda=2$, (Inset) Close up view of the region near ${\bar C}=2.4$. The transition at $2.4$ is due to a percolation transition and data is noisy, the difference between auxiliary and quadratic Hamiltonians is small in this regime.
Data is not collected in $2 \leq {\bar C}\leq 2.03$ for numerical reasons, the curve is anticipated to continue smoothly to $({\bar C},e)=(2,0)$.}
\end{center}
\end{figure}
In figure \ref{fig:energyRSB} with $\lambda=1$ we find the ground state energy for the 1RSB solution,
given by the energy at which the complexity first becomes positive on increasing $\mu$.
Some granularity in $\mu$ of size $0.01$ is visible in more computationally challenging regimes,
leading to the appearance of step-like artefacts.
Again the difference between the auxiliary Hamiltonian and full model is small.
The curve for small ${\bar C}$ indicates a very low value for the energy by contrast with other curves,
only $O(1/100)$ factors are excited.
A second spinodal line, tracing the structure highlighted by figure \ref{fig:complexity1},
is not drawn but follows a similar trend.
Comparing the energy to early experimental results, there is a systematic trend apparent in figure \ref{fig:EnergyEntropyN32} that the energy is decreasing with system size. As well as being qualitatively similar to the numerical results, the curves in figure \ref{fig:energyRSB} seems to be a lower bound to the finite system experiments as anticipated.

The analysis of the 1RSB and RS energetic fields, at zero temperature, indicate a spin glass transition at $\lambda=2$ and ${\bar C}=12/5$.
In the full RS analysis, considering also the entropic contributions, a spin glass transition at ${\bar C}=2.49$ is observed.
We argue that the value ${\bar C}=12/5$ has a topological significance and probably represents the correct transition point from a zero to non-zero energy solution.
This value corresponds to a percolation threshold. For ${\bar C}>12/5$ there begin to exist percolating clusters in the factor graph on which all factors have connectivity $3$.
Symmetric fields are generated and propagated by such a substructure.
Conversely, for ${\bar C}<12/5$ the graph is composed of connectivity $3$ factors interspersed by many factors of connectivity $2$.
The interactions along these chains restrict g-spins to non-dimer states, then for a given configuration of g-spins on the boundary, all internal spins become aligned anti-ferromagnetically.
In a random topology where this substructure percolates, many closed paths lead to frustration, a raised energy and spin-glass behaviour.
The details of this argument help to explain the exceptional regime of strong finite size effects seen in experiments (section \ref{sec:small_system_studies}).

\section{Discussion}
\label{sec:discussion}

We have studied the Ising model with nearest and next nearest neighbor antiferromagnetic interactions  on random graphs with varying dominance by odd or even connectivities.
The model is closely related to many problems in optimization, such as maximizing the diversity of colours on random graphs and matching.
Experimental observations are made by exhaustive branch-and-bound search for small systems and EO for larger systems.

Theoretical analyses were undertaken by the cavity method in both the RS and 1RSB approximations.
The results demonstrate the existence of both paramagnetic and spin glass phases at zero temperature.
Zero temperature paramagnetism can be explained by the high configurational freedom in special ratio ensembles due to unlocked interactions, and the transition to spin glasses, at small $\lambda$ and ${\bar C}$, can be attributed to the inclusion of locked interactions as the average connectivity moves away from the special ratios.

To deal with nearest and next nearest neighboring interactions on equal footing,
we have developed the formalism of generalized variables.
This enables us to conduct the thermodynamic analysis based on locally tree-like recursions,
resembling those used to study the equilibrium and dynamical properties in the Bethe lattice and Cayley tree.

It is also worthwhile to consider extending our algorithmic strategy to other computationally hard problems.
We first perform simulations on small systems using exhaustive enumeration,
which already reveals the effects of special ratios.
Since the system sizes they accessed are too small to accurately estimate finite size effects,
we use in larger graphs the heuristic EO.
The transition point found by EO is increasingly well defined with increasing system size.
Its power is evident in its ability to discover ground states for system sizes up to $O(100)$ in a short computation time.

As far as we know,
this study is the first in applying the 1RSB cavity method to Ising models on random graphs with next nearest neighboring interactions and inhomogeneous connectivity.
However,
both the RS and 1RSB solutions are not sufficient to exactly describe the spin glass phases at low temperature.
The description of the energetic 1RSB approximation,
which has been studied at high numerical precision in many graphical models,
becomes quite complicated owing to the multiple components of the cavity fields ($3$ for $\vec{h}$ in our case).
Nevertheless,
the approximations are in qualitative agreement with the simulation results for small graphs.

Looking forward to studying models with higher connectivity,
greater distance interactions,
or Potts spins,
it will be necessary to introduce simplified approaches to keep the problem tractable.
Our work also demonstrates some useful directions along this line.
Applying the population dynamics in the 1RSB framework we have demonstrated an efficient method to enumerate the support,
in the space of fields,
for the order parameter.
The use of the auxiliary Hamiltonian is also able to produce accurate results on the transition points,
and reasonable estimates at finite temperature and finite energy.
Other simplification strategies,
not yet explored in this paper,
include the Gaussian approximation for high connectivity graphs and message passing algorithms approximated with reduced number of variables~\cite{Bounkong:CRG}.
In exploring the possibility of Potts spins,
a proposal simplifying the complexity of calculating the marginals has been recently proposed,
using a belief propagation algorithm on generalized states~\cite{Pelizzola:SMS}.

We have studied regular graphs and inhomogeneous graphs with a combination of connectivity $\lfloor{\bar C}\rfloor$ and $\lceil{\bar C}\rceil$.
We expect that in other ensembles the prevalence of odd or even connectivity nodes will remain a dominating factor in determining the phase diagram,
provided that the connectivity is not too large and $\lambda$ is small by comparison with vertex connectivity.
For example,
if $\lambda=2$,
zero temperature paramagnetism will only be found in graphs with a significant fraction of odd connectivity nodes.
However,
in the cut-Poisson ensemble,
which has a nearly uniform distribution of odd and even connectivity nodes for any ${\bar C}$,
we do not see the periodic trends in free energy observed in the linear ensembles with variation of ${\bar C}$ and $\lambda$.
We anticipate a similar pattern for other standard connectivity distributions.

Special ratios should also affect the low temperature properties of graphs with $\lambda$ close to integer values.
Since $\lambda$ is not an integer,
genuine unlocked interactions no longer exist.
However,
since $\lambda$ is close to integer values the energy gap to the first excited state is small.
Thus the special ratio effects are manifested in a reduced critical temperature.

Besides the Ising models,
special ratio effects are also expected in $Q$-state  Potts models.
For example,
in the colour diversity problem,
which corresponds to $\lambda=1$,
the local configuration for $Q=4$ of a node is locked at $C=3$,
and unlocked for $C=4$,
giving rise to a spin glass to paramagnetic phase transition when ${\bar C}$ increases from $3$ to $4$ in a linear graph ensemble~\cite{Wong:MU}.
In general,
for $\lambda=1$,
the multi-body interaction is locked when $C+1$ is an integer multiple of $Q$,
and the corresponding ground state local configuration on g-spins translates to each of the $Q$ colours appearing $(C+1)/Q$ times,
with a total of $C+1$ colours decorating a node and its nearest neighbors.
Indeed,
we have calculated the ground state entropy in the RS ansatz for $\lambda=1$ and $2$,
and found that they are particularly high in special ratio ensembles.
One may wonder whether transitions between zero temperature paramagnetism and spin-glass phases can be observed with variation of $\lambda$ or ${\bar C}$ for some ensembles.
However for larger $Q$,
the special ratio ensembles are restricted to relatively high connectivity,
with many constraints per degree of freedom.
Even for those special ratio ensembles of lowest connectivity the paramagnetic solution becomes unfeasible,
the presence of next nearest neighbor interactions imposes stringent constraints on the colour configurations that renders the entropy negative.
Nevertheless,
the special ratio effects may again be exhibited in a modulation of some thermodynamic quantities with variation of $\lambda$ or ${\bar C}$,
such as the dependence of spin glass transition temperatures on the average connectivity.
This is a challenging topic for future studies.

\appendix

\section{Ground state experiments}
\label{app:ground_state_experiments}

\begin{figure}[h!tbp]
\begin{center}
\includegraphics[width=0.8\linewidth]{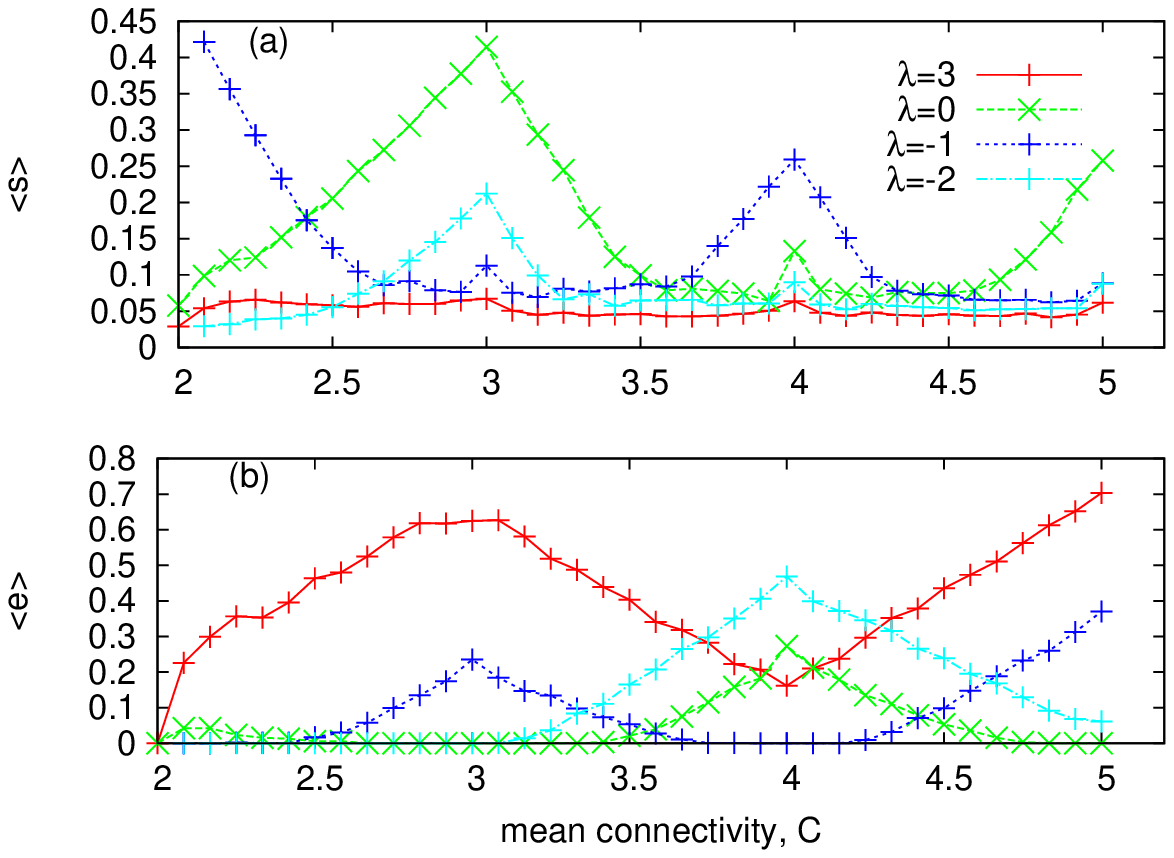}\\
\includegraphics[width=0.8\linewidth]{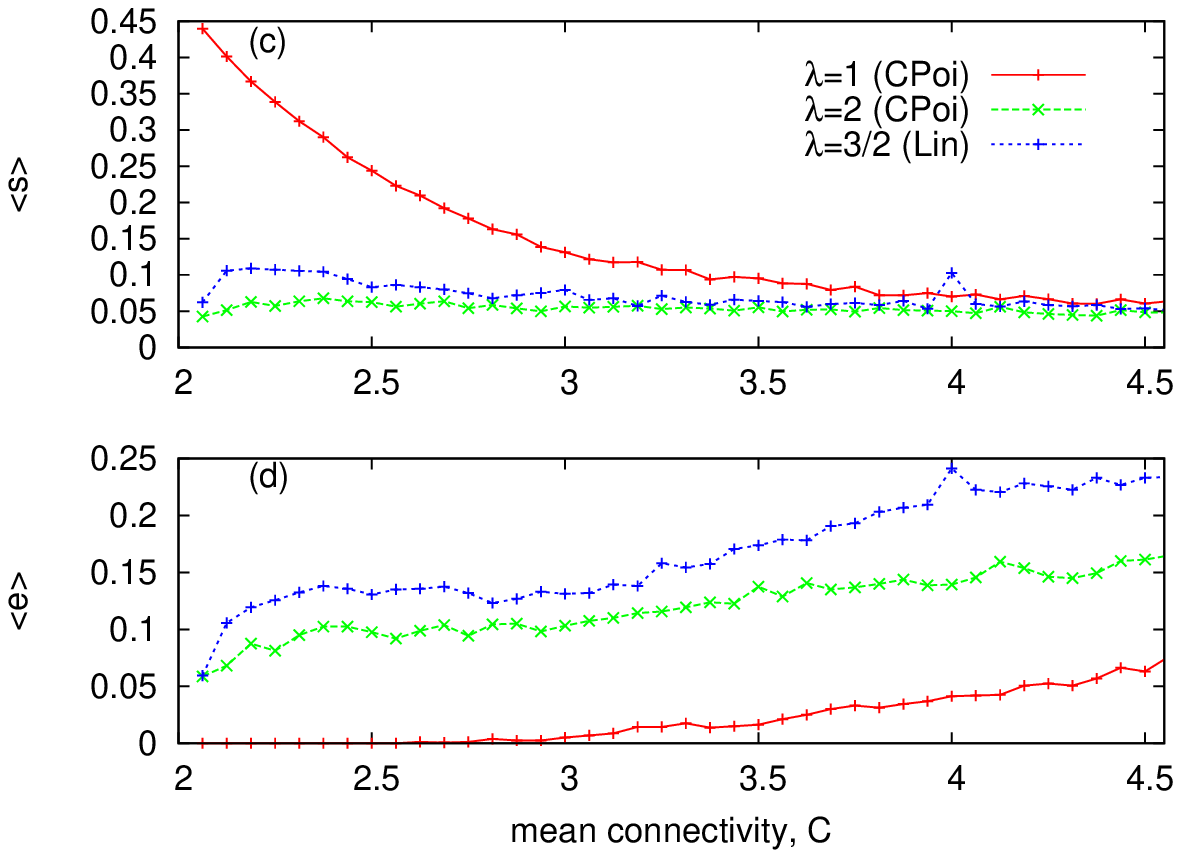}
\caption{\label{fig:moresimulations} (a) Mean entropy density $\langle s\rangle$ and (b) mean energy density $\langle e\rangle$ for linear ensembles.
Results of $\langle s\rangle$ and $\langle e\rangle$  for Cut-Poisson ensembles are shown in (c) and (d) respectively, $100$ graph samples in every case.
Error bars are typically smaller than the point size.}
\end{center}
\end{figure}

Results are presented to complement those discussed in section \ref{sec:small_system_studies}.
Figures \ref{fig:moresimulations}(a) and (b) demonstrate the entropy and energy statistics for linear ensembles of size $N=24$.
For $\lambda=0,3$ the energy is minimized at the special ratios,
while the entropy is maximized.
With increasing $\lambda$,
regimes of ground state energy zero become rarer,
and unattainable above some threshold in ${\bar C}$ in line with the mean field argument, RS and 1RSB results.

Figures \ref{fig:moresimulations}(c) and (d) show results for Cut-Poisson ensembles with $N=32$.
These figures demonstrate a nearly monotonic trend in entropy and energy statistics;
strong mixing of locked and unlocked factors for all ${\bar C}$ prevents the appearance of modulated behaviour due to dominance by even or odd connectivity.

\section{Zero temperature field support}
\label{app:zero_temperature_field_support}

Whereas the space of energetic fields is sampled statistically within the RS approach of the main text,
our implementation of the 1RSB cavity method requires an exact description of the space.
We establish properties of the space of energetic fields in this Appendix.
Attention is  focused on integer $\lambda$,
and we consider ensembles of minimum connectivity two so that boundary energetic fields can be ignored.
This allows us to restrict attention to spaces of fields defined only by the zero temperature mappings (\ref{eq:TzerorecursionJ})-(\ref{eq:Tzerorecursionhb}).

The exact knowledge of the support allows a higher fidelity representation of the probability distributions manipulated in the cavity method for either RS or 1RSB.
However,
the methods outlined prove to be infeasible in many graph ensembles due to the restriction of studying ensembles with some maximum connectivity $C_{max}$,
and the unfavorable scaling of the space of fields with the connectivity.
In principle a population dynamics approach can be applied to the 1RSB analysis for these cases~\cite{Mezard:BLSG}.

\subsection{Bounds on the space of energetic fields}
\label{appssec:bounds_on_the_space_of_energetic_fields}

The components of the energetic fields,
assuming that they are generated by recursion of the mappings (\ref{eq:TzerorecursionJ})-(\ref{eq:Tzerorecursionhb}) from some initial condition,
will be shown to be bounded.
This reflects the intuition that changing the state of the g-spin at the root of the cavity tree allows only a restricted number of energy shifts in the ground state energy regardless of boundary conditions.

For analyzing the zero temperature field support,
it is more convenient to introduce a new representation of $\vec{h}^E_{j \rightarrow i}$ to replace the representation ($J_{j \rightarrow i},h^f_{j \rightarrow i},h^b_{j \rightarrow i}$) used in the main text.
We can do this by decomposing the cavity probability according to
\begin{equation}
P_{j \rightarrow i}(S_j,S_i) = P_{j \rightarrow i}(S_i|S_j) P_{j \rightarrow i}(S_j)
\;.\label{eq:newdecomposition}
\end{equation}
where $P_{j \rightarrow i}(S_i|S_j)$ is the conditional probability of $S_i$ given $S_j$ on the cavity graph $\Factorgraph_{j \rightarrow i}$,
and $P_{j \rightarrow i}(S_j)$ the marginal on the same graph.
For Ising spins we can express $P_{j \rightarrow i}(S_j,S_i)$ in terms of three probabilities $P_{j \rightarrow i}(S_i|S_j=+1)$,
$P_{j \rightarrow i}(S_i|S_j=-1)$ and $P_{j \rightarrow i}(S_j)$.
In turn,
the fields $f_{j \rightarrow i}^{\pm}$ and $b_{j \rightarrow i}$ can be introduced to describe the three probabilities,
namely,
$P_{j \rightarrow i}(S_j) \propto \exp\{ \beta b_{j \rightarrow i} S_j\}$ and $P_{j \rightarrow i}(S_i|S_j=\pm 1) \propto \exp\{\beta f_{j \rightarrow i}^{\pm} S_i\}$.
Combining these cases,
we can write $P_{j \rightarrow i}(S_j,S_i)\propto \exp[\beta E_{j \rightarrow i}]$,
where in the limit of low temperature
\begin{equation}
\fl E_{j \rightarrow i}(S_j,S_i) \!=\! -\frac{(f_{j \rightarrow i}^+ \!+\! f_{j \rightarrow i}^-)S_i}{2} -\frac{(f_{j \rightarrow i}^+ + f_{j \rightarrow i}^-)S_j S_i}{2}
- \left(b_{j \rightarrow i} \!-\!\frac{[ |f^+_{j \rightarrow i}| - |f^-_{j \rightarrow i}|]}{2}\right)S_j
\label{eq:newdecomposition2}
\;.
\end{equation}
We henceforth work with fields only at the energetic level.
Comparing with (\ref{eq:PSiSjparameterization}) we obtain the transformation in the limit of low temperature given by
\begin{eqnarray}
b_{j \rightarrow i} &=& h^b_{j \rightarrow i} + \frac{\left( | h^f_{j \rightarrow i} + J_{j \rightarrow i} | - | h^f_{j \rightarrow i} - J_{j \rightarrow i} | \right)}{2}\\
f^{\pm}_{j \rightarrow i} &=& h^f_{j \rightarrow i} \pm J_{j \rightarrow i}
\;.
\label{eq:newfromold}
\end{eqnarray}

\subsubsection{Bounds on fields describing symmetric solutions}
\label{appssec:bounds_on_fields_describing_symmetric_solutions}

The symmetric solution is described with $f^{-}_{j \rightarrow i}=-f^{+}_{j \rightarrow i}$, and $b_{j \rightarrow i}=0$.
At the energetic level the recursions become identical to those of a two-state model (\ref{eq:4to2state}),
the parameter $J_{j \rightarrow i}\rightarrow \pm f^{\pm}_{j \rightarrow i}$ can be considered to describe either the two-state model or the symmetric solution of the NNN model.
$2 \beta J_{j \rightarrow i}$ becomes the logarithm of the ratio of likelihoods for dimer and non-dimer states of the ancestor g-spin given its descendants.

A simplified form of (\ref{eq:TzerorecursionJ}) applies for the two-state model,
taking ${\tilde S}=1$ or $-1$ to represent a dimer or non-dimer state
\begin{equation}
J_{j \rightarrow i} = J^E(\{J_{k \rightarrow j}\}) = - \lambda - \frac{1}{2}\sum_{{\tilde S}=\pm 1} {\tilde S} D^{(2)}({\tilde S},\{J_{k \rightarrow j}\})
\;,
\label{eq:Jbound}
\end{equation}
where (\ref{eq:D}) becomes
\begin{equation}
\fl D^{(2)}({\tilde S}_C,\{J_{k\rightarrow j}\}) = \min_{ \{{\tilde S}_k\} \setminus {\tilde S}_C } \left\lbrace \frac{1}{2}\left( \sum_{k=1}^{C-1} {\tilde S}_k \right)^2  + {\tilde S}_C \sum_{k=1}^{C-1}{\tilde S}_k + \sum_{k=1}^{C-1} (\lambda - J_{k\rightarrow j}) {\tilde S}_k\right\rbrace
\;.
\label{eq:2stateD}
\end{equation}
The first two quadratic terms arise from the next nearest neighbors interactions,
and the final two terms are nearest neighbor penalties.
A maximum and minimum value for an ancestor field $J_{j \rightarrow i}$ is found by freezing the descendant fields to uniformly large values of $\{|J_{k\rightarrow j}|\}$, either all positive or all negative in sign.
Doing this ${\tilde S}_k = \sign(J_{k\rightarrow j})$ and maxima and minima of the expression are derived giving the bounds (\ref{eq:2statebound}).
The bounds can be somewhat tightened by recursively reintroducing the bounds found on the ancestor,
into the descendants.

\subsubsection{Bounds on fields describing symmetry broken states}
\label{appssec:bounds_on_fields_describing_symmetry_broken_states}

Without restriction to a symmetric solution we can present bounds on the components  $f^{\pm}_{j \rightarrow i}$ and $b_{j \rightarrow i}$.
We can take first the case of $f_{j \rightarrow i}^\pm$,
determined as in (\ref{eq:newdecomposition2}) by
\begin{equation}
f^\pm_{j \rightarrow i} = - \frac{1}{2}\sum_{S_i} {S}_i E_{j \rightarrow i}(\pm 1, S_i)\;,\label{eq:fpmjtoi}
\end{equation}
where the energy is given by the low temperature limit of (\ref{eq:freeenergyi}),
\begin{equation}
\begin{array}{lll}
\fl E_{j \rightarrow i}(S_j,S_i) &=& \min_{S_1,\ldots,S_{C-1}}\left\lbrace (\lambda S_j + \sum_{k=1}^{C-1} S_k)S_i - \frac{1}{2} \sum_{k=1}^{C-1}(f_{k\rightarrow j}^+ + f_{k\rightarrow j}^-)S_j  \right.
\\
&+& \left.
\sum_{k=1}^{C-1} \left[ \frac{1}{2}\left(|f_{k\rightarrow j}^+| + f_{k\rightarrow j}^+ S_j -|f_{k\rightarrow j}^-|+f_{k\rightarrow j}^- S_j + \sum_{l=1}^{C-1} S_l \right)\right.\right.\\
&+& \left.\left.
\lambda S_j - b_{k\rightarrow j}\right] S_k \right\rbrace\;.
\end{array}\label{eq:EjtoiSSjSi}
\end{equation}
From (\ref{eq:fpmjtoi}) we note that $f_{j \rightarrow i}^\pm$ are determined by the energy change when the ancestor spin $S_i$ flips at a fixed value of $S_j$.
Since from (\ref{eq:EjtoiSSjSi}) the energy primarily depends on $S_i$ via the term $(\lambda S_j + \sum S_k)S_i$,
bounds for $f_{j \rightarrow i}^\pm$ are determined in the cases that the fields $f_{k\rightarrow j}^\pm$ are strong, so that the values of $S_k$ are fixed.
Taking the combinatino of $S_k$ that yields the maximum and minimum arguments,
we find that $f_{j \rightarrow i}^+$ and $-f_{j \rightarrow i}^-$ are subject to the same bounds as $J_{j \rightarrow i}$ in (\ref{eq:2statebound}).
For this representation of the cavity probabilities,
$f^\pm_{j \rightarrow i}$ are bounded the same way for symmetric (two-state) and symmetry broken (four-state) boundary conditions.

By contrast $b_{j \rightarrow i}$ is unbounded,
given unbounded descendant fields. For brevity we demonstrate the case of uniform connectivity $C$ on ancestor and descendant nodes.
However,
we can apply the  bounds established for $f_{j \rightarrow i}^\pm$ to the descendant fields $f_{k\rightarrow j}^{\pm}$.
$b_{j \rightarrow i}$ is given by
\begin{equation}
b_{j \rightarrow i} = -\frac{1}{2}\sum_{S_j} S_j \min_{S_i} \left\lbrace E_{j \rightarrow i}(S_j,S_i)\right\rbrace\;.
\end{equation}
Taking the limit that $b_{k\rightarrow j}$ are very strong so that $S_k$ are aligned with $b_{j \rightarrow i}$,
we obtain for non-negative $\lambda$
\begin{equation}
\fl b_{j \rightarrow i} = \sum_{k=1}^{C-1} f^{S_k}_{k\rightarrow j} - \lambda (2 X - C + 1) - \frac{1}{2}\left[|2X-C+1-\lambda| -|2X-C+1+\lambda| \right]\;,
\end{equation}
where $X$ is the number of descendant nodes with $S_k=1$.
The upper bound is reached when $X=0$ and $f^-_{k\rightarrow j}=\lambda+C-1$ for all $k$.
This yields
\begin{equation}
b_{j \rightarrow i} \leq (C-1)(C-1+\lambda) + (C-1)\lambda - \frac{1}{2}\left[|C-1+\lambda| - |C-1-\lambda| \right]\;.
\end{equation}
The lower bound,
as well as the case of negative $\lambda$,
can be derived analogously.
Summarizing,
\begin{equation}
\fl |b_{i \rightarrow j}| \leq H^b(|\lambda|,C_{max}) = \left\lbrace
\begin{array}{ll}
 (C-1)(C_{max} + 2 |\lambda|) - |\lambda|& \hbox{if}\; |\lambda| \leq C - 1 \\
(C-1)(C_{max} - 1 + 2 |\lambda|) & \hbox{otherwise}\;,
\end{array} \right.
\label{eq:4statebound}
\end{equation}
generalizing from the case of uniform connectivity we have $C_{max}$ which is the largest connectivity in the graph ensemble,
the extreme values of the bound are only achieved if all descendants are connectivity $C_{max}$ and the bounds (\ref{eq:2statebound}) saturated.

A combination of these two bounds places a bound on the components of $\vec{h}$,
the domain is restricted to a polyhedron within any ensemble of graphs of maximum connectivity $C_{max}$.
The polyhedron has a simple description,
but its volume can be compressed further by recursively reintroducing the bounds found on the ancestor into the descendants.
In so doing the bounds involve a complicated coupling of $\{f^{\pm},b\}$ in general,
and are evaluated only numerically.

\subsection{Integer $\lambda$ discrete set support}
\label{appssec:Integer_lambda_discrete_set_support}
Integer $\lambda$ cases are the focus of the main text.
At these values the energy level differences are restricted to be integer.
The field components,
which represent these relative differences on cavity trees,
are thereby restricted to integer values.
Since the components are also bounded,
only a short list of labeled fields can describe any possible pure state at any point in the graph,
at the energetic level.
Some small set of fields is sufficient to describe the support for the distributions describing either 1RSB or RS solutions.

Our 1RSB method utilizes this discreteness,
but involves summations over pure states.
A greater variety of ensembles,
and fidelity of results,
can be obtained if small sets can be used.
Our objective is to find the smallest set that describes correctly the solution support.

A sufficient set,
defined $\Gamma^{bound}_{C_{max},\lambda}$,
is the set of all fields with integer components,
consistent with the previously derived component-wise bounds.
An analogous set is defined when restricting to the symmetric fields,
$\Gamma^{bound,2}_{C_{max},\lambda}$.
Both these sets can support solutions,
and in some cases the same unique (paramagnetic) solution;
however,
it is much simpler to solve the recursions on the latter set,
owing to its small size.
The sets are completely defined by $C_{max}$ and $\lambda$, their sizes are
\begin{equation}
\fl |\Gamma^{bound}_{C_{max},\lambda}| = (2 H^b(C_{max},|\lambda|)+ 1) (2 C_{max} - 1)^2
\qquad|\Gamma^{(bound,2)}_{C_{max},\lambda}| = (2 C_{max} - 1)
\label{eq:firstboundSetSize}
\;.
\end{equation}

Without loss of generality in the application of the cavity method,
it is possible to restrict attention to self-consistent sets.
A self-consistent set contains only fields that can be generated by some combination of fields in the set under mapping (\ref{eq:Tzerorecursion})-(\ref{eq:Tzerorecursionhb}).
The largest such set will certainly be no bigger than $\Gamma^{bound}$,
and can be discovered by a recursive pruning process,
removing irrelevant fields.

Consider a mapping from a set $\Gamma^{(t)}$ to a set $\Gamma^{(t+1)}$ defined
\begin{equation}
\Gamma^{(t+1)} = T_{\Gamma}(\Gamma^{(t)}) = \{ \vec{h}^E | \vec{h}^E = {\hat T}^{E}(\{\vec{h}_c^E\}),\;\hbox{for some}\; \{\vec{h}_c^E\} \in \Gamma^{(t)}, C \}
\;,
\label{eq:SCS}
\end{equation}
where ${\hat T}^E$ is the mapping defined by $\lambda$ and $C$ (\ref{eq:Tzerorecursion})-(\ref{eq:Tzerorecursionhb}), and $C$ can be any connectivity within the graph ensemble.
This mapping may be iterated and assuming convergence is achieved, $\Gamma^{(t+1)}=\Gamma^{(t)}$,
the set derived will be a self-consistent one.

Such an iterative procedure is not guaranteed to converge for an arbitrary choice of initial set,
but taking $\Gamma^{(0)}=\Gamma^{bound}_{C_{max},|\lambda|}$,
in effect all fields as the initial condition,
the mapping is a contraction and must converge.
For the linear ensemble with maximum connectivity $C_{\max}$,
minimum connectivity $C_{\min}$ and coupling ratio $\lambda$,
we obtain the set $\Gamma^{4}_{C_{max},C_{min},\lambda}$.
For the symmetric set of fields we can similarly define $\Gamma^{2}_{C_{max},C_{min},\lambda}$,
which is arrived at by the same process but taking $\Gamma^{(0)}=\Gamma^{bound}_{C_{max},|\lambda|}$.

By construction $\Gamma^4$ is the largest set of fields self-consistent under the mapping $T_\Gamma$,
it is generally a much smaller set than $\Gamma^{bound}$.
The bounds of section \ref{appssec:bounds_on_fields_describing_symmetry_broken_states} predict a scaling of $O(C^4)$ for the redundant set $\Gamma^R$ on which our most precise methods rely,
and we are unable to study sufficiently large set sizes to establish a more favorable scaling.
By contrast $\Gamma^{2}$ has a simple description for the linear ensemble.
It can be derived analytically,
for ensembles in which $C_{max}\leq C_{min}+1$,
\begin{equation}
J_{j \rightarrow i} \in \left\lbrace
\begin{array}{ll}
\left[ -(C_{max} - 1 - |\lambda|), C_{max} - 1 - |\lambda| \right] &\; \hbox{if}\; |\lambda|\leq C_{min} - 1 \\
\left[ C_{min} - 1 - |\lambda|, C_{max} - 1 - |\lambda| \right] &\; \hbox{if}\; |\lambda| \geq C_{min} - 1 \;.
\end{array} \right.
\label{eq:2statebound2}
\end{equation}
The derivation of (\ref{eq:2statebound2}) is outlined as follows.
The coupling $J_{j \rightarrow i}$ is given by $[E_{j \rightarrow i}(1,-1),E_{j \rightarrow i}(1,1)]/2$.
Let $L$ and $H$ be the lower and upper bounds of the couplings respectively.
When the couplings $J_{j \rightarrow i}$ reaches the bounds,
it is reasonable to expect that the descendant couplings also reach the bounds.
Suppose among the $C-1$ descendants,
$m$ of them take the upper bound and $n$ of them the lower bound.
The cavity energy is reduced to
\begin{equation}
E_{j \rightarrow i}(1,S_i) = \lambda S_i + (S_i + \lambda)(x + y) + \frac{1}{2}(x+y)^2 - H x - L y\;,
\end{equation}
where $x=\sum_{k|J_{k\rightarrow j}=H}S_k$ and $y=\sum_{k|J_{k\rightarrow j}=L} S_k$.
When $\lambda \leq C-1$,
empirical observations show that $H \geq 0 \geq L$.
Hence the minimum energy is given by
\begin{equation}
\fl E_{j \rightarrow i}(1,S_i) \!=\! \left\lbrace
\begin{array}{ll}
\lambda S_i + (H - L) m - \frac{(L - \lambda - S_i)^2}{2} &\; \hbox{if}\; m - n \leq L - \lambda - S_i \\
\lambda S_i + (H - L) m - \frac{(H - \lambda - S_i)^2}{2} &\; \hbox{if}\; m - n \geq H - \lambda - S_i \\
\lambda S_i \!-\! H m \!+\! L n \!+\! (S_i \!+\! \lambda)(m \!-\! n) \!+\! \frac{(m\! -\! n)^2}{2} \!&\; \hbox{otherwise}\;.
\end{array} \right.
\end{equation}
Respectively,
the above three cases are obtained at the $x=m$ boundary,
the $y=n$ boundary and the corner $(x,y)=(m,-n)$.
These energy expressions result in the couplings summarized by
\begin{equation}
J_{j \rightarrow i}= \left\lbrace
\begin{array}{ll}
 -L &\; \hbox{if}\; m - n \leq L-\lambda \\ %
-\lambda + m - n &\; \hbox{if}\; L-\lambda \leq m-n \leq H-\lambda \\ %
-H &\; \hbox{if}\; H - \lambda \leq m-n %
\end{array} \right.
\end{equation}
Since $m-n$ is bounded by $\pm (C+1)$ we see that $H=-L=C-1-|\lambda|$.
Generalizing the result to multiple connectivities,
the first case of (\ref{eq:2statebound2}) is obtained.

When $|\lambda|\geq C_{max}-1$,
$L=H=0$ for $C=C_{max}$.
This means that the next neighbor interactions (represented by $C_{max}-1$) are completely neutralized by the nearest neighbor interactions (represented by $|\lambda|$).
The energy minimum starts to be dominated by nearest neighbor interactions,
with descendant spins aligning anti-parallel to $S_j$ for positive $\lambda$,
and parallel for negative $\lambda$.
Hence we consider the solution with $x+y=1-C$ for positive $\lambda$ and $C-1$ for negative $\lambda$ and obtain the second case of (\ref{eq:2statebound2}).

The set size
\begin{equation}
|\Gamma_{C,\lambda}|= 2\max\left\lbrace C_{max}-C_{min},C_{max}-1-|\lambda|\right\rbrace+1
\;,
\label{eq:2statemodelsetsize}
\end{equation}
increases linearly with $C$ for fixed small $|\lambda|$,
and decreases linearly with $|\lambda|$ to $2(C_{max}-C_{min})+1$.
For large $|\lambda|$ the largest self-consistent sets are trivial (single valued) for regular connectivity.

\subsection{Generation and stability of smaller set representations}
The problem with the preceeding method for generating a self-consistent basis for the 1RSB method is twofold.
Firstly we must begin by manipulating a potentially large set,
the set $\Gamma^{bound}$,
considering all possible states of $C-1$ descendant fields.
The method becomes unfeasible if $C_{max}$ is too large,
greater than about five,
since the number of field recombinations to be processed grows as a large power of $C$ in (\ref{eq:firstboundSetSize}).
For fields restricted to the two-state model description the scaling is more reasonable and creates no problem numerically up to $C_{max} \sim 10$ in a brute force evaluation.
However,
it may be that the largest self-consistent set,
which is the one we establish, may contain many fields irrelevant both to the cavity method solution and its local stability properties.
There may exist some smaller yet sufficient sets that can be the basis of analysis.

A sampling approach might be developed to address these problems.
We identify a procedure that leads to a hierarchy of sets of increasing size,
allowing greater control in the computational complexity.
The method proposed relies on the recursive expansion of two sets.
The first is $\Gamma_S$ which is to be developed as the support for the solution,
the second is a disjoint set of support perturbations $\Gamma_P$,
which will be contracted to remove fields irrelevant to any local stability consideration.
We define a mapping from these two sets to a third set $\Gamma_N$ as follows,
\begin{equation}
\fl \Gamma_N = T_\Gamma (\Gamma_S,\Gamma_P) = \Gamma_S \cup \{\vec{h}^E | \vec{h}^E = {\hat T}(\{h^E_k\}) \;, \vec{h}^E_{C-1} \in  \Gamma_P, \vec{h}^E_1 \ldots \vec{h}^E_{C-2} \in \Gamma_S\}
\label{eq:SCSpert}
\;.
\end{equation}
In the mapping only one of the descendants takes a field from the set of perturbations $\Gamma_P$.
The set of fields in the perturbation set are considered bugs in an analogous manner to the type II stability analysis standard in 1RSB~\cite{Montanari:NLT},
and do not interact.
If under such a mapping $\Gamma_S=\Gamma_N$ we say the set $\Gamma_S$ is linearly stable towards support bugs in the set $\Gamma_P$.

The procedure is therefore to select a simple set,
using (\ref{eq:SCS}) to derive a self-consistent set $\Gamma_{S'}$,
then using (\ref{eq:SCSpert}) to discover the subset of perturbations that can survive the linearized recursion, $\Gamma_{P'}$.
In principle we can now solve the 1RSB equations on the set $\Gamma_{S'}$ and test the stability in the expanded space $\Gamma_N=\Gamma_{S'}\cup\Gamma_{P'}$.

Assuming we find the solution is locally unstable,
we can seek a larger support that might be sufficient to support the stable solution.
Thus we can propose to repeat the process choosing $\Gamma_S=\Gamma_{N}$.
We summarize our method:
\begin{itemize}
\item[1.] Begin with an estimated set $\Gamma_S$.
\item[2.] Set $\Gamma^{(t=0)}=\Gamma_S$; derive a self consistent set recursively, applying (\ref{eq:SCS}) until convergence. Call the derived set $\Gamma_{S'}$. If it does not converge repeat step 1 with a different set.
\item[3.] Take a set $\Gamma_P$ of support bugs.
\item[4.] Set $\Gamma^{(t=0)}=\Gamma_P$; derive new sets recursively $\Gamma^{(t+1)}= T_{\Gamma,2}(\Gamma_{S'},\Gamma^{(t)}) \setminus \Gamma_{S'}$ (\ref{eq:SCSpert}), until convergence. Call the derived set $\Gamma_{P'}$. If it does not converge repeat step 3 with a different set.
\item[5.] Return the pair $\{\Gamma_{S'},\Gamma_{P'}\}$, solve the energetic 1RSB method on $\Gamma_{S'}$ and test 1RSB instability across the support $\Gamma_{S'}\cup\Gamma_{P'}$.
\item[6.] If $\Gamma_{P'}$ is the empty set, the support $\Gamma_{S'}$ is linearly stable against support bugs, terminate. Otherwise set $\Gamma_S = \Gamma_{S'} \cup \Gamma_{P'}$, and begin from step 1, to generate a new pair.
\end{itemize}
Two types of stability analysis may be considered for the 1RSB solution,
called type I and type II~\cite{Montanari:NLT},
and our method is able to identify the additional support necessary to rule out these linear instabilities that might go beyond the basis describing the solution.
The art in the method comes from selecting sets that converge,
avoiding cyclical behaviour that arises from symmetry breaking or some other effects.
In fact,
we find that the recursive generation procedure always produce converging sets provided one makes reasonable initial choices for the sets.

The choice for $\Gamma_S$ is more involved,
the simplest choice might be a single field meeting symmetry requirements and bounds.
For ensembles with all nodes of connectivity at least $|\lambda|+1$ it is found that $\Gamma_S=\{(0,0,0)\}$ is self-consistent,
and meets the requirements of stability against any perturbative set.
This observation on the support alone implies that there must exist a locally stable trivial solution to the energetic RSB method in these ensembles.

To generate non-trivial and asymmetric sets,
those capable of describing a spin glass solution,
we can consider the choice $\Gamma_S=\Gamma^{bound,2}$ as a basis.
This can describe the symmetric solution,
and with a reasonable choice of perturbations can be expanded to include the most relevant symmetry breaking fields.
An alternative choice is $\Gamma^{frozen} = \{(f^+_{j \rightarrow i},f^-_{j \rightarrow i},b_{j \rightarrow i}) | (f^+_{j \rightarrow i},f^-_{j \rightarrow i},b_{j \rightarrow i}) =(X_1,X_2,X_3), X_i = \pm X,0\}$ with $X\rightarrow\infty$,
which is referred to as a frozen set since the effect of the infinite fields is to freeze the spin variables.
This set initially violates the bounds, but converges upon iteration to a subset within the bounds.

A choice for $\Gamma_P$,
sufficient to consider any instability, is one that includes all possible self consistent fields,
i.e. $\Gamma^4$ if it is known,
or else $\Gamma^{bound}$,
or all fields up to a symmetry constraint,
i.e. $\Gamma^{2}$ or $\Gamma^{bound,2}$.
These are relatively large sets of perturbations,
but since we need to consider only one of $C-1$ descendants taking values from this large set,
the combinatorial complexity is only linear in this number.

Under the recursive set expansion procedure with either the frozen or asymmetric initial condition the sets converge at every stage,
and following several iterations the process halts,
in many cases with $\Gamma_{S'}=\Gamma^{4}$,
or the trivial set $\Gamma_{S'}=\{(0,0,0)\}$.
Interestingly from the frozen set,
we find some other symmetry broken self-consistent sets,
and also self-consistent sets smaller than $\Gamma^{4}$ for which the support is locally stable.

Some results for the sizes of sets derived by our method are shown in the table,
we denote as $\Gamma_{F}$ the self-consistent set derived from $\Gamma^{frozen}$ using (\ref{eq:SCS}),
and $X^+$ the set $\Gamma_{S'}$ found by recursively applying our expansion method from an initial pair $\Gamma_S=X$ and $\Gamma_P=\Gamma^{bound}$.
After several iterations,
with respect to the same perturbation set,
a set of maximum size is found.
A comparison of the support generated by various methods is given in table \ref{table:setsize}.
\begin{table}
\begin{center}
\begin{tabular}{|c|c|c|c|c|c|c|c|}
\hline
$\lambda$ & $C_{min}$ & $C_{max}$ & $\Gamma^{bound}$ & $\Gamma^{4}$ & $\Gamma^{2+}$ & $\Gamma^{F}$ & $\Gamma^{frozen+}$ \\
\hline
1 & 2 & 3 & 375 & 89 & 89 & 89 & 89 \\%(225)
1 & 3 & 3 & 375 & 89 & 89 & 79 & 89 \\%(225)
1 & 3 & 4 & 1421 & 611 & 611 & 451 & 451 \\
1 & 4 & 4 & 1421 & 593 & 593 & 87 & 451 \\
2 & 2 & 3 & 525 & 49 & 49 & 49 & 49 \\%(225)
2 & 3 & 3 & 525 & 43 & 1 & 14 & 43 \\%(225)
2 & 3 & 4 & 1911 & 314 & 314 & 314 & 314 \\%(1127)
2 & 4 & 4 & 1911 & 297 & 297 & 297 & 297 \\%(1127)
\hline
\end{tabular}
\end{center}
\caption{\label{table:setsize} The support for the energetic fields.}
\end{table}
Where the expanded sets are of equal size to $\Gamma^{4}$ they are the same set,
where the set is of size one it is the trivial set $\{(0,0,0)\}$.
We find that $\Gamma^{2+}$ always converges to one of these cases for $\lambda\in[-5,5]$ and ${\bar C}<5$,
the trivial set only occurs for regular graphs in the special ratio (unlocked) regime $\lambda = C-1$.
It is a trivial consequence,
that the paramagnetic solution is locally stable in these cases at the level of energetic fields.
However,
since $|\Gamma^4|>1$ a more complicated solution may exist but cannot be found by a permutation procedure on the symmetric set of fields.

\subsubsection{Low energy simplifications of the field description}

In so far as the 1RSB analysis is concerned,
alternative simplifications can be proposed in order to allow manipulation of smaller sets.
Firstly we can note that a very important special case of the 1RSB analysis involves the configuration parameter $\mu \rightarrow \infty$.
The g-spin ancestor takes one of $4$ states,
and it is sufficient to consider only whether the state is excited (forbidden) or not excited,
since any excited state contributes a vanishing weight to the pure-state distribution.
The mappings of energetic fields (\ref{eq:Tzerorecursion}) can be modified and the notions of self-consistency and perturbative sets can be carried over to a set of size only $2^4-1$.
Since the size of this set does not scale with connectivity,
or $\lambda$,
it is very transferrable between ensembles.

More generally $\mu$ may be non-zero in the thermodynamically relevant solutions,
and also we wish to describe metastable states.
However,
excited states are strongly penalized in general,
and it can be expected that the phase space structure is mostly sensitive to only the local ground state structure.
For this reason we can propose the auxiliary Hamiltonian (\ref{eq:FactorHami2}).
This removes all structures in the excitations,
under the assumption that second level excitations and higher ones will have a weak effect on the energetic properties of the solution.

For the auxiliary Hamiltonian we can carry out the same procedures as for the Quadratic Hamiltonian with significantly reduced set sizes.
For this Hamiltonian a tighter set of bounds are immediately apparent,
$f^{\pm}_{j \rightarrow i} \in \{0,\pm 1\}$,
and hence any symmetric solution has a support of three,
and this can be combined with a modified bound $H^b=C_{max}+1$ (\ref{eq:4statebound}) which describes the symmetry broken solutions.
A comparison of the set generation methods for the auxiliary Hamiltonian is shown in table \ref{table:setsizeAux}.
\begin{table}[h!]
\begin{center}
\begin{tabular}{|c|c|c|c|c|c|c|c|}
\hline
$\lambda$ & $C_{min}$ & $C_{max}$ & $\Gamma^{bound}$ & $\Gamma^{4}$ &  $\Gamma^{2+}$ & $\Gamma^{F}$ & $\Gamma^{frozen+}$ \\
\hline
1 & 2 & 3 & 63 & 39 & 39 & 39 & 39 \\
1 & 3 & 3 & 63 & 39 & 39 & 39 & 39 \\
1 & 3 & 4 & 81 & 53 & 53 & 53 & 53 \\
1 & 4 & 4 & 81 & 53 & 53 & 47 & 47 \\
2 & 2 & 3 & 28 & 20 & 20 & 20 & 20 \\
2 & 3 & 3 & 28 & 12 & 1 & 12 & 12 \\%(20)
2 & 3 & 4 & 81 & 57 & 57 & 57 & 57 \\
2 & 4 & 4 & 81 & 57 & 57 & 57 & 57 \\
\hline
\end{tabular}
\end{center}
\caption{\label{table:setsizeAux} The support for the energetic fields, using the auxiliary Hamiltonian.}
\end{table}
It can be seen that for all linear ensembles the largest self-consistent set ($\Gamma^4$),
used in the results section,
is significantly smaller with the Auxiliary Hamiltonian.
A comparison of results obtained based on the two Hamiltonians is made in section \ref{ssec:energetic_1RSB_solutions}.

We considered also the method of this Appendix applied for ensembles with connectivity distribution supported on $[2,C_{max}]$,
for $C_{max}>3$,
but not too large.
This would describe a cut-Poisson ensemble with a truncated maximum connectivity,
phenomena were found to be similar to the linear ensemble.

\section{Implementation of the Cavity Method}
\label{app:population_dynamics_implementation}

\subsection{Population dynamics initial condition for RS and 1RSB}
An important issue is the initial condition.
For our finite temperature RS analysis we take fields to be $O(1)$ random numbers,
with symmetry broken.

The insight on the support gained by the methods of \ref{app:zero_temperature_field_support} can be applied equally to zero temperature RS and 1RSB field recursions.
However,
our zero-temperature RS results are developed with a population dynamics method,
initializing every energetic field components independently from $\{0,\ldots,X\}$.
The observables,
stability and thresholds were not found to be particularly sensitive to the choice of $X$, provided a sufficient fraction of fields were distinct from $(0,0,0)$,
$X$ is taken to be very large to obtain the results of this paper.
A population of $10^4$ fields is used in all RS results.

In the 1RSB population dynamics,
we consider a population of field distributions rather than a population of fields.
In our experiments we chose $10^3$ field distributions,
the support of each distribution being the set $\Gamma^4$ identified in \ref{app:zero_temperature_field_support}.
Each member of our population is a vector of dimension $|\Gamma^4|$,
each component of the vector being the probability of the corresponding field $\vec{h}_i$.
The components each distribution are initialized by sampling uniformly from the interval $[0,1]$,
finally each distribution is normalized.

\subsection{Population dynamics convergence}
\label{appssec:convergence}
Beginning from the initial population we assume an ergodic regime will finally be reached in which the population moments are stable.
For the replica symmetric solutions a minimum of $200$ and maximum of $1200$ iterations are allowed for convergence of the population.
The spin glass order parameter $q_F$ in (\ref{eq:q_F}) and dimer magnetization $m_D$ in (\ref{eq:m_D}),
with sums replaced by population sampling,
are expected to have a systematic trend during the transient phases of the population dynamics.
This trend is towards $0$ in $q_D$ if the symmetric solution is the unique solution.
We can consider the evolution of these observables over $\sim 50$ population updates,
and allow additional iterations unless they have settled on a fixed mean up to statistical fluctuations.

Similar criteria are applied to the 1RSB populations,
the expressions of $q_F$ and $m_D$ are obtained by differentiating the 1RSB free energy,
resulting in reweighted versions of (\ref{eq:q_F}) and (\ref{eq:m_D}) under the hypothesis of many pure states.

\subsection{The RS stability analysis for regular graphs}
\label{ssec:regular_graph_stability}

Section \ref{ssec:stability_and_symmetry_breaking} discusses a population dynamics approach to establishing solution stability, on a regular graph we have a local homogeneity that allows a simplification of the analysis.
In this case the solution consistent with a single pure state, the RS or paramagnetic solution, is concentrated on $\vec{h}^*$. The joint distribution including perturbations can be described by
\begin{equation}
 P(\vec{h},\delta\vec{h})=\delta(\vec{h} - [\vec{h}^* + \delta\vec{h}])P(\delta\vec{h})
\end{equation}
Suppose the distribution of perturbations is uniform over the population, that is, the shifts from $(f^+_{j \rightarrow i},f^-_{j \rightarrow i},b_{j \rightarrow i})=(0,0,0)$ are variations in the mean without any random component. We can then expand the mapping, a special case of (\ref{eq:linearperturbation}) in which the perturbations on each kind of component are identical, and determine the new perturbation under iteration
\begin{equation}
\delta h^{t+1}_x = \sum_{y=1}^3 M_{x y} \delta h_y^{t} \label{eq:hlin}
\end{equation}
where
\begin{equation}
M_{x y} = \lim_{\vec{h} \rightarrow \vec{h}^*} \frac{d {\hat T}_x (\{\vec{h}\})}{d \vec{h}_y} \label{eq:Mxy}
\end{equation}
Under this mapping the perturbations may grow or decay exponentially, as determined by the largest real part of any eigenvalue associated with the $3$ by $3$ matrix $M$.

It is sufficient to test only the stability of the mean and variance of the distribution to determine whether it is stable. Assuming that the mean of the distribution is stable we describe the remaining perturbations on ancestors and descendants by Gaussian distributions parameterized by $\Sigma$. Each component $\Sigma_{yz}$ describes the expectation value of the random quantity $\delta h_y \delta h_z$ for the corresponding descendant or ancestor. Assuming the randomness to be of a homogeneous type on descendants, we can find the moments of the distribution (\ref{eq:RS_population_recursion}), and hence the mapping from descendant to ancestor covariance
\begin{equation}
\Sigma_{wx}^{t+1} = \sum_{y=1}^3 \sum_{z=1}^3 \frac{1}{C-1} M_{x y} M_{wz} \Sigma_{yz}^t
\end{equation}
As in the linear stability case, the principal eigenvalue (in the real part) of a 9 by 9 matrix determines this non-linear stability, towards random zero mean perturbations.

\subsection{The 1RSB stability analyses overview}
The stability analysis is limited in our study to the ensembles ${\bar C}=3,\lambda=1$ with the quadratic and auxiliary Hamiltonians,
and ${\bar C}=4,\lambda=2$ with the auxiliary Hamiltonian.
These are locked problems of regular connectivity,
and we consider the stability of the solution found with respect to the support $\Gamma^{4}$ described in \ref{app:zero_temperature_field_support}.
Since for a regular graph ensemble there is no variation in the locally tree like neighborhoods,
the 1RSB solution can be described by a single distribution rather than a population of distributions.
The standard type I and type II analyses are conducted~\cite{Montanari:NLT}.
We test the type I stability by adding perturbations to the distribution that evolve according to a linearized mapping.
We monitor the first and second moments of the perturbations under iteration and find that the moments always decay under iteration,
indicating stability.

For type II instability,
also called bug proliferation,
we monitor how bugs in the distribution evolve on iteration.
The bugs are described by infinitesimal probabilities $P(\vec{h}_a,\vec{h}_b)$ that $\vec{h}_a$ is "mistaken" for $\vec{h}_b$.
Hence,
on iteration the evolution of bugs is determined by a matrix $|\Gamma^4|^2$ by $|\Gamma^4|^2$,
relating the probabilities $P(\vec{h}_a,\vec{h}_b)$ before and after an iteration at a linearized level.
The largest eigenvalue of this matrix determines the local stability.
Rather than calculating this directly,
we establish instability by randomly initializing a vector of the bug probabilities and measuring the sum of the variances of parturbations at successive iterations.
Divergence of this sum implies at least one eigenvalue exceeds one,
and the solution is unstable.
We find that the sum diverges for every configurational parameter $\mu$,
and for both the quadratic and auxiliary Hamiltonian.
Therefore the ground state solution, and metastable solutions,
are locally unstable.
Thus our solutions are type I stable,
but type II unstable,
for $C=3,\lambda=1$ and $C=4,\lambda=2$.

\section{High Temperature Expansion in the $\lambda \gg 1$ Limit}
\label{Appendix:expin1lambda}

This Appendix considers the case in which the next nearest neighbor interactions are much stronger than the nearest neighbor ones, thereby deriving the asymptote (\ref{eq:nnTc2}) in the high temperature limit. Substituting the cavity probability distribution in (\ref{eq:Pjtoi_siteindependent}) into the recursion relation (\ref{eq:PSiSjrecursion}), we obtain the recursions identical to Eqns. (\ref{eq:TzerorecursionJ}) to (\ref{eq:Tzerorecursionhb}), with the energy $D(S_j,S_i,\{\vec{h}_{k\rightarrow j}\})$ substituted by the free energy $F(S_j,S_i,\{\vec{h}_{k\rightarrow j}\})$ where
\begin{equation}
\fl \begin{array}{lll}
F(S_j,S_i,\{\vec{h}_{k\rightarrow j}\}) &=& -\frac{1}{\beta}\log \mathrm{Tr}_{\{S_k\}}\exp\left\lbrace -\beta \left[ \frac{1}{2}\left(S_i + \sum_{k=1}^{C-1} S_k \right)^2 \right. \right. \\ &+& \left.\left. \sum_{k=1}^{C-1}\left(\left(\lambda - J_{k \rightarrow j}\right)S_j - h^b_{k\rightarrow j} \right) S_k \right]\right\rbrace\;. \label{appeq:FSSh}
\end{array}
\end{equation}
We note in passing that in the zero temperature limit, (\ref{appeq:FSSh}) reduces to (\ref{eq:D}). Here we utilize Eq (\ref{appeq:FSSh}) in the opposite limit of high temperature, in which the quadratic term in the Boltzmann factor is approximated as
\begin{equation}
\exp\left\lbrace -\beta \left[\frac{1}{2}\left(S_i + \sum_{k=1}^{C-1} S_k \right)^2 \right]\right\rbrace \approx 1 - \frac{\beta}{2}\left(S_i + \sum_{k=1}^{C-1} S_k \right)^2 + \ldots\; \label{appeq:bQuad}
\end{equation}
This enables us to consider the effective Hamiltonian as one with purely nearest neighbor interactions, and the next nearest neighbors terms in (\ref{appeq:bQuad}) as a thermodynamic average in the presence of such an effective Hamiltonian, that is,
%\begin{equation}
%\begin{array}{lll}
\begin{eqnarray}
\fl F(S_j,S_i,\{\vec{h}_{k\rightarrow j}\}) &\approx& -\frac{1}{\beta}\log \mathrm{Tr}_{S_k}\exp\left\lbrace -\beta  \sum_{k=1}^{C-1}\left(\left(\lambda - J_{k \rightarrow j}\right)S_j - h^b_{k\rightarrow j} \right) S_k \right\rbrace \nonumber\\
&+& \frac{1}{2}\left\langle\left(S_i + \sum_{k=1}^{C-1} S_k \right)^2 \right\rangle\;, \label{appeq:FSSh2}
\end{eqnarray}
%\end{array}
%\end{equation}
where the thermodynamic average in the last term is taken in the nearest neighbor model. Hence, apart from constant terms, we obtain
%\begin{equation}
%\begin{array}{lll}
\begin{eqnarray}
F(S_j,S_i,\{\vec{h}_{k\rightarrow j}\}) &\approx& -\frac{1}{\beta}\sum_{k=1}^{C-1} \log \cosh\left( \beta \left(\lambda - J_{k \rightarrow j}\right)S_j - \beta h^b_{k\rightarrow j}\right) \nonumber\\ &-& S_i \sum_{k=1}^{C-1}\tanh\left( \beta \left(\lambda - J_{k \rightarrow j}\right)S_j - \beta h^b_{k\rightarrow j}\right) \nonumber\\&+& \frac{1}{2}\sum_{k<l = 1}^{C-1} \left[ \tanh\left( \beta \left(\lambda - J_{k \rightarrow j}\right)S_j - \beta h^b_{k\rightarrow j}\right)\right. \nonumber \\ &\times& \left. \tanh\left( \beta \left(\lambda - J_{l \rightarrow j}\right)S_j - \beta h^b_{l\rightarrow j}\right)\right]\;. \label{appeq:FSSh3}
\end{eqnarray}
%\end{array}
%\end{equation}
To consider the stability of the paramagnetic phase, we consider cases that $h^f_{j\rightarrow i}$,$h^b_{j\rightarrow i}$ are small. Following Eqs. (\ref{eq:TzerorecursionJ}) to (\ref{eq:Tzerorecursionhb}), we have
\begin{equation}
J_{j\rightarrow i} = -\lambda + \sum_{k=1}^{C-1} \tanh(\beta(\lambda-J_{k\rightarrow j}))\;,\label{appeq:Jjtoi}
\end{equation}
\begin{equation}
h^f_{j\rightarrow i} = -\beta \sum_{k=1}^{C-1} h^b_{k\rightarrow j}\sech^2(\beta(\lambda-J_{k\rightarrow j}))\;,
\end{equation}
%\begin{equation}
\begin{eqnarray} h^b_{j\rightarrow i} &=& \sum_{k=1}^{C-1}h^f_{k\rightarrow j} - \sum_{k=1}^{C-1}h^b_{k\rightarrow j} \tanh(\beta(\lambda-J_{k\rightarrow j})) \nonumber \\&+& \beta \sum_{k=1}^{C-1} h^b_{k\rightarrow j} \sech^2(\beta(\lambda-J_{k\rightarrow j})) \sum_{l\neq k}\tanh(\beta(\lambda-J_{k\rightarrow j})\;.
\end{eqnarray}
%\end{equation}
Eq (\ref{appeq:Jjtoi}) yields the high temperature paramagnetic state given by $(J_{j\rightarrow i},h^f_{j\rightarrow i},h^b_{j\rightarrow i})$ = $(J_0,0,0)$ where $J_0=-\lambda+(C-1)\tanh(2\beta \lambda)$. We are now ready to study the ferromagnetic and spin glass instabilities as described in \ref{ssec:regular_graph_stability}. The $3 \times 3$ matrix (\ref{eq:Mxy}) can be expanded to linear order in $\beta$, to give the matrix $M=M^0 + \beta M^1$. Abbreviating $t=\tanh(2\beta \lambda)$ the two non-zero components of the matrix $M^0$ are
\begin{equation}
M^0_{32} = (C-1)\qquad \hbox{and}\qquad M^0_{33} = -(C-1)t\;,
\end{equation}
implying that the non-zero eigenvalue at the leading order is $e_0 = -(C-1)t$. This eigenvalue is perturbed according to the elements of $M^1$, whose non-zero elements are
\begin{equation}
\fl M^1_{11} = M^1_{23} = -(C-1)(1-t^2)\qquad \hbox{and}\qquad M^1_{33} = (C-1)(2 C - 3) t (1-t^2)\;.
\end{equation}
To establish the perturbation in the eigenvalue, we consider the matrix $M' = M -e_0 I$, $I$ being the identity matrix. The relevant eigenvalue of $M'$ is of the order $\beta$. Hence it suffices to consider the characteristic polynomial of $M'$  to the first order in $\beta$, given by
\begin{equation}
C(x) = (-e_0)^2\left(\beta M^1_{33} - x\right) - (-e_0)\beta M^1_{23}M^0_{32}\;.
\end{equation}
The eigenvalue is
\begin{equation}
\beta e_1 = \frac{\beta}{t}(C-1)(1-t^2)[1 + (2 C -3) t^2]\;.
\end{equation}
The ferromagnetic instability is determined by equating the eigenvalue $e_0 + \beta e_1$ to $1$. Expanding $\lambda$ about its zeroth order value, and inverting its dependence on temperature, we find
\begin{equation}
T_c = - \frac{2}{\atanh((C-1)^{-1})}\left(\lambda + \frac{C^2 -2}{2(C-1)} \right)\;.
\end{equation}
We find that the Curie temperature is always reduced by the presence of next nearest neighbor antiferromagnetic interactions, and by $O(C^2)$.

To calculate the spin glass instability, we determine the largest eigenvalue of the $9 \times 9$ matrix $M_{(ij),(kl)} = M_{ik}M_{jl}/(C-1)$, as was outlined in \ref{ssec:regular_graph_stability}. The leading order of the eigenvalue is determined to be $e_0 = M^0_{(33),(33)}=(C-1)t^2$. To $O(\beta)$, the eigenvalue $e_0 + \beta e_1$ is determined by the equation
\begin{equation}
\fl (-e_0)^8(C-1)^{-1}(2\beta M^0_{33} M^1_{33} - e_1) - (-e_0)^7(C-1)^{-1}2\beta M^0_{32}(M^0_{33})^2 M^1_{23} = 0\;,
\end{equation}
whose solution is
\begin{equation}
e_1 = 2(C-1)(1-t^2)[1+(2C-3)t^2]\;.
\end{equation}
By equation the eigenvalue to $1$, the critical curve (\ref{eq:nnTc2}) is thereby obtained.

\cut{
\section{Expansion in small $\beta$ with $\beta\lambda$ finite}
\label{Appendix:expin1lambda}
The Appendix considers the solutions for which next nearest neighbor interactions are much smaller than nearest neighbors, thereby we derive the asymptote (\ref{eq:nnTc2}) and find several analytical results in support of the conclusions in section \ref{sec:paramagneticsolution}. We can derive the leading order perturbation to the paramagnetic solution within the 2-state model. We can then consider stability, by a linear expansion of the methods described in section \ref{ssec:regular_graph_stability} for regular graphs.

\subsection{The paramagnetic solution}
Since we can determine the solution for the full model from the solution to the two state model we work in that simplified framework.
It is first useful to note that in the absence of next nearest neighbor terms we have the following fixed point of the two state model $P_{j\rightarrow i}({\tilde S})$
\begin{equation}
P^*_{j\rightarrow i}({\tilde S}) = \frac{\exp(- \beta \lambda {\tilde S})}{2\cosh(\beta\lambda)} \label{eq:Bethefixedpoint}
\end{equation}
This is a solution in every cavity graph, independent of local connectivity distribution. We will in this section therefore consider a small parameter expansion about this solution $\beta J_{j\rightarrow i} = -\Lambda + \beta A_{j\rightarrow i}$ (\ref{eq:2stateparameterization}), with $\beta$ small but $\Lambda = \beta \lambda$ finite.
Using (\ref{eq:4to2state}) we identify the corresponding triplet of fields $(\beta J, \beta h^f, \beta h^b)$ $=$ $(-\Lambda + A ,0,0)$ with the paramagnetic solution of the full model.

Replacing every pair of spins $S_j S_k$ by the 2-state dimer variable ${\tilde S}_{ij}$, along with the transformation of field distributions (\ref{eq:4to2state}), one can find the recursion for the two state model on a graph to be
\begin{equation}
P_{j\rightarrow i}({\tilde S}_{ji}) \propto
\sum_{{\tilde S}_{\partial_j} \setminus {\tilde S}_{ji}} \prod_{k\in \partial_j \setminus i} \left[P_{k \rightarrow j}({\tilde S}_{jk})\right] \exp\left(-\Lambda\sum_{k \in \partial_j}{\tilde S}_{jk} +\frac{\beta}{2}\left(\sum_{k \in \partial_j} {\tilde S}_{jk} \right)^2\right)\;,
\end{equation}
separated into $O(\Lambda)$ and $O(\beta)$ parts with respect to the Hamiltonian. We can expand this expression in $\beta$, both in the Hamiltonian term and field perturbations $A_{k\rightarrow j}$. From this we determine a linear relationship amongst the $A$, which for the case of homogeneous fields can be solved as
\begin{equation}
  A_{j\rightarrow i} = \frac{1}{2\beta}\log \frac{P_{j\rightarrow i}(1)}{P_{j\rightarrow i}(-1)} + \Lambda = \frac{1}{2\beta}\log \frac{\mathbb{E}[(1 - \beta \sum_{k\in\partial_j\setminus i}{\tilde S}_{jk})+O(\beta^2)]}{\mathbb{E}[(1 - \beta \sum_{k\in\partial_j\setminus i}{\tilde S}_{jk})+O(\beta^2)]}
\end{equation}
where the expectations are with respect to the unperturbed ($\beta=0$) probability distribution. Abbreviating $t=\tanh(-2\Lambda)$ the result for the case of Homogeneous connectivity is $A_{j\rightarrow i} = A = -(C-1)t$. The free energy $f$ of the two state model is, up to constant terms, written
\begin{equation}
  - \beta f = \sum_j \log\left(\prod_{k \in \partial_j}P_{k\rightarrow j}({\tilde S}_{jk})\exp \left(-\beta H_2({\tilde S})\right) \right) - \sum_{(ij)}\log(P_{i\rightarrow j}({\tilde S}_{ij})P_{j\rightarrow i}({\tilde S}_{ij}))
\end{equation}
Expanding in $A$ and $\beta$ to second order, and normalizing by $N$ we obtain the free energy density
\begin{equation}
  - \beta f = - \beta f_0(\beta) - \beta(1 + A) \mathbb{E} \left[\left(\sum_k {\tilde S}_k\right)^2/2\right] + \beta^2/2 \mathrm{Var} \left[\left(\sum_k {\tilde S}_k\right)^2/2\right] - \frac{C(1-t^2) A^2}{2}
\end{equation}
where $f_0(\beta)$ is the free energy in which the only dependence on $\beta$ is through $\Lambda$ (the model without next nearest neighbor interactions).  The expectation and variance are calculated with respect to the a factorized distribution over the spins, each probability of the form (\ref{eq:Bethefixedpoint}). The energy is obtained by partial derivative with respect to $\beta$, and entropy by the relation $\beta(f-e)$.

A positive energy change is apparent at $O(\beta)$, and an entropy change that is negative at $O(\beta^2)$. Since the energy and entropy of the full model is affine to the two state model, we can conclude that throughout the paramagnetic solution for entropy and energy follow the same trend. This result implies that negative entropy may occur at relatively higher temperature compared to the nearest neighbor model.

\subsection{Paramagnetic solution stability for regular graphs}
We can now study the linear and non-linear stability as described for the regular case in \ref{ssec:regular_graph_stability}. The $3$ by $3$ matrix (\ref{eq:Mxy}) can be expanded to linear order in $\beta$, to give a leading and subleading order parts $M=M^0 + \beta M^1$. The two non-zero components of the matrix $M^0$ are
\begin{eqnarray}
M_{33} = (C-1)t\; \qquad\hbox{and}\qquad M_{32} = (C-1) \;.
\end{eqnarray}
Implying the non-zero eigenvalue at leading order: $e_0 = (C-1)t$. This eigenvalue is perturbed according to the elements of $M^1$, we must describe
\begin{equation}
 M^1_{33} = -{2(C-1) \choose 2}t \left(1- t^2\right) \qquad \hbox{and}\qquad M^1_{23} = -(C-1)\left(1- t^2\right)
\end{equation}
although other non-zero components exist, they do not contribute at leading order in the perturbation of the eigenvalue.

To establish the perturbation in the eigenvalue consider the matrix $\delta M=M-e_0 I$, $I$ being the identity matrix. We can write the characteristic equation for the new matrix in $x$, the roots of which determine the eigenvalues. Since we are only interested in the eigenvalue closest to $0$, at $O(\beta)$, we can take $x=\beta e_1$ and expand the characteristic equation only to first order
\begin{equation}
  C_{\delta M}(x) = (-e_0)^{2}(\beta M^1_{33} -\beta e_1) - (-e_0) \beta M^1_{32}M^0_{23} + O(\beta^2)
\end{equation}
The eigenvalue is
\begin{equation}
  e_1 = \left[{2 C -2 \choose 2} t + \frac{C-1}{t} \right](1-t^2)
\end{equation}
and describes the perturbation of $e_0$ at $O(\beta)$.
Now, we are interested in the solutions to the equation $e_0(\Lambda)+\beta e_1(\Lambda,\beta)=1$, which determines the linear instability of the model. Solving at leading order $\beta=0$, and then expanding about this value to leading order in $\beta$ we obtain the critical value for the nearest neighbor coupling $\lambda$ at fixed $\beta$.
\begin{equation}
  \Lambda_{c} = \beta \lambda_{c} = - \frac{\atanh(1/(C-1))}{2} + \beta \frac{C^2-2}{2(C-1)} \;
\end{equation}
This relationship is easily inverted to give $\beta_c$ as a function of $\lambda$: we find that the critical temperature is always reduced, and by $O(C^2)$.

For positive $\lambda$ we can anticipate a paramagnetic to spin glass solution indicated by the non-linear susceptibility.
As was outlined in \ref{ssec:regular_graph_stability} to calculate the non-linear instability we must determine the principle eigenvalue of the nine by nine matrix $M_{(ij),(kl)}=M_{ik}M_{jl}/(C-1)$, fortunately the matrix structure is similar to that evaluated in determining the linear instability. The leading order eigenvalue is determine to be $e_0=M^0_{(33),(33)}=(C-1)t^2$. Then we can again simply find the root of the characteristic equation keeping only the terms linear in $\beta$
\begin{eqnarray}
  C_{\delta M}(x) &=& (-e_0)^{8}(M_{(33),(33)} -\beta e_1) - (-e_0)^{7} (M_{(33),(23)}M_{(23),(33)} + M_{(33),(32)}M_{(32),(33)})\nonumber\\
  &=& (-e_0)^{7}\left[-e_0 (2 M^0_{33} \beta M^1_{33} -\beta e_1) - (2 (M^0_{33})^2 M^0_{32} \beta M^1_{23})\right]/(C-1)\nonumber
\end{eqnarray}
Solving in $e_1$ the root of this equation we have the eigenvalue
\begin{equation}
e_1 = 2 (C-1) (1-t^2)\left(1  + (2 C - 3) t^2\right)
\end{equation}
We thereby proceed as for the linear instability, expanding about the solution of $e_0(\lambda)=1$. The critical curve thereby obtained is (\ref{eq:nnTc2}).
The effect for positive $\lambda$ of the antiferromagnetic term is at leading order to reduce the critical temperature, by $O(C)$.
}
%\section*{Acknowledgements}
\section{Acknowledgements}
\ack
We thank Bill Yeung and David Saad for fruitful discussions, and Stefan Boettcher for providing source code that was adapted for this investigation.
The work is supported by Research Grants Council of Hong Kong (grant numbers 604008 and 605010).
% ----------------------------------------------------------------
%\bibliographystyle{unsrt}

%\bibliography{Bibliography}

\section*{References}
\begin{thebibliography}{10}

\bibitem{Mezard:SGT}
M.~M\'{e}zard, G.~Parisi, and M.A Virasoro.
\newblock {\em Spin Glass Theory and Beyond}.
\newblock World Scientific, Singapore, 1987.

\bibitem{Nishimori:SP}
H.~Nishimori.
\newblock {\em Statistical Physics of Spin Glasses and Information Processing}.
\newblock Oxford Science Publications, Oxford, UK, 2001.

\bibitem{Krzakala:GS}
F.~Krzakala, A.~Montanari, F.~Ricci-Tersenghi, G.~Semerjian, and
  L.~Zdeborov\'{a}ँ.
\newblock {Gibbs states and the set of solutions of random constraint
  satisfaction problems}.
\newblock {\em PNAS}, 104(25):10318--10323, 2007.

\bibitem{Montanari:NLT}
A.~Montanari and F.~Ricci-Tersenghi.
\newblock On the nature of the low-temperature phase in discontinuous
  mean-field spin glasses.
\newblock {\em Eur. Phys. J. B}, 33:339, 2003.

\bibitem{Selke:ANNNI}
W.~Selke.
\newblock The \mbox{ANNNI} model- theoretical analysis and experimental
  application.
\newblock {\em Phys. Rep.}, 170(4):213--264, 1988.

\bibitem{Domb:2D}
C.~Domb and R.B. Potts.
\newblock A two-dimensional model with first and second interactions.
\newblock {\em Proc. Roy. Soc. Lond. A.}, 210:125 -- 141, 1951.

\bibitem{Stephenson:IM}
J.~Stephenson and D.~D. Betts.
\newblock \mbox{I}sing model with antiferromagnetic next-nearest-neighbor
  coupling. ii. ground states and phase diagrams.
\newblock {\em Phys. Rev. B}, 2(7):2702--2706, 1970.

\bibitem{Reimers:OBD}
J.N. Reimers and A.J. Berlinsky.
\newblock Order by disorder in the classical \mbox{H}eisenberg kagom\'e
  antiferromagnet.
\newblock {\em Phys. Rev. B}, 48(13):9539--9554, Oct 1993.

\bibitem{Bak:MFT}
M.H. Jensen and P.~Bak.
\newblock Mean-field theory of the three-dimensional anisotropic \mbox{I}sing
  model as a four-dimensional mapping.
\newblock {\em Phys. Rev. B}, 27(11):6853--6868, Jun 1983.

\bibitem{Vannimenus:PDIM}
J.~Vannimenus.
\newblock Phase diagram of an \mbox{I}sing model with competitive interactions
  on a \mbox{H}usimi tree and its disordered counterpart.
\newblock {\em Z. Phys. B}, 43:141 -- 148, 1981.

\bibitem{Moreira:MS}
J.~G. {Moreira} and S.~R. {Salinas}.
\newblock Modulated structures in the \mbox{I}sing model with competing
  interactions on the \mbox{C}ayley tree.
\newblock {\em Phys. Rev. B}, 47:778--786, January 1993.

\bibitem{Yokoi:SA}
C.~S.~O. Yokoi, M.~J.~de Oliveira, and S.~R. Salinas.
\newblock Strange attractor in the \mbox{I}sing model with competing
  interactions on the \mbox{C}ayley tree.
\newblock {\em Phys. Rev. Lett.}, 54(3):163--166, Jan 1985.

\bibitem{Inawashiro:IM}
S.~Inawashiro, C.J. Thompson, and G.~Honda.
\newblock \mbox{I}sing model with competing interactions on a \mbox{C}ayley
  tree.
\newblock {\em J. Stat. Phys.}, 33(2):419 -- 436, 1983.

\bibitem{Katsura:BL}
S.~Katsura and M.~Takizawa.
\newblock \mbox{B}ethe lattice and the \mbox{B}ethe approximation.
\newblock {\em Prog. Theor. Phys.}, 51(1):82--98, 1974.

\bibitem{Ganikhodjaev:ES}
N.N. Ganikhodjaev, C.H. Pah, and M.R.B. Wahiddin.
\newblock Exact solution of an \mbox{I}sing model with competing interactions
  on a \mbox{C}ayley tree.
\newblock {\em J. Phys. A}, 36(15):4283--4289, 2003.

\bibitem{daSilva:IMBL}
C.R. da~Silva and S.~Coutinho.
\newblock \mbox{I}sing model on the \mbox{B}ethe lattice with competing
  interactions up to the third-nearest-neighbor generation.
\newblock {\em Phys. Rev. B}, 34(11):7975--7985, Dec 1986.

\bibitem{Ganikhodjaev:PD}
N.~Ganikhodjaev, F.~Mukhamedov, and C.H Pah.
\newblock Phase diagram of the three states potts model with next nearest
  neighbour interactions on the \mbox{B}ethe lattice.
\newblock {\em Phys. Lett. A}, 373(1):33 -- 38, 2008.

\bibitem{Melin:SGB}
R.~M\'elin and S.~Peysson.
\newblock Spin glass behavior upon diluting frustrated magnets and spin
  liquids: a \mbox{B}ethe-\mbox{P}eierls treatment.
\newblock {\em Eur. Phys. J. B}, 14(1):169--176, 2000.

\bibitem{Monroe:PD}
J.L. Monroe.
\newblock Phase diagrams of \mbox{I}sing models on \mbox{H}usimi trees ii. pair
  wand multisite interaction systems.
\newblock {\em J. Stat. Phys.}, 67(5):1185--1200, 1992.

\bibitem{Chandra:SL}
P~Chandra and B~Doucot.
\newblock Spin liquids on the \mbox{H}usimi cactus.
\newblock {\em Journal of Physics A: Mathematical and General}, 27(5):1541,
  1994.

\bibitem{Ostilli:PDIM}
M.~Ostilli, F.~Mukhamedov, and J.F.F. Mendes.
\newblock Phase diagram of an \mbox{I}sing model with competitive interactions
  on a \mbox{H}usimi tree and its disordered counterpart.
\newblock {\em Physica A: Statistical Mechanics and its Applications},
  387(12):2777 -- 2792, 2008.

\bibitem{Moreira:IMTNN}
J.G. Moreira and S.R. Salinas.
\newblock \mbox{I}sing model with third-neighbour interactions on the
  \mbox{C}ayley tree.
\newblock {\em J. Phys. A}, 20(6):1621, 1987.

\bibitem{Wong:MU}
K.Y.M. Wong and D.~Saad.
\newblock Minimizing unsatisfaction in colourful neighbourhoods.
\newblock {\em J. Phys. A}, 41(32):324023 (25pp), 2008.

\bibitem{Bounkong:CRG}
S.~Bounkong, J.~van Mourik, and D.~Saad.
\newblock Coloring random graphs and maximizing local diversity.
\newblock {\em Phys. Rev. E}, 74(5):057101, 2006.

\bibitem{Pelizzola:SMS}
A.~Pelizzola, M.~Pretti, and J.~van Mourik.
\newblock Palette-colouring: a belief-propagation approach.
\newblock arXiv:1104.4024, 2011.

\bibitem{Kearns:GMGT}
M.~Kearns, M.~Littman, and S.~Singh.
\newblock Graphical models for game theory.
\newblock In {\em Proceedings of the Seventeenth Conference Annual Conference
  on Uncertainty in Artificial Intelligence (UAI-01)}, pages 253--260, San
  Francisco, CA, 2001. Morgan Kaufmann.

\bibitem{Ramezanpour:SPGG}
{Ramezanpour, A.}, {Realpe-Gomez, J.}, and {Zecchina, R.}
\newblock Statistical physics approach to graphical games: local and global
  interactions.
\newblock {\em Eur. Phys. J. B}, 81(3):327--339, 2011.

\bibitem{Mezard:CMZT}
M.~M\'{e}zard and G.~Parisi.
\newblock The cavity method at zero temperature.
\newblock {\em J. Stat. Phys.}, 111(1-2):1--34, 2003.

\bibitem{yedidia}
J.~S. Yedidia, W.~T. Freeman, and Y.~Weiss.
\newblock Constructing free energy approximations and generalised belief
  propagation algorithms.
\newblock Technical Report TR2002-35, Mitsubishi Electric Research
  Laboratories, 2002.

\bibitem{Kschischang:FG}
F.R. Kschischang, B.J. Frey, and Hans-Andrea Loeliger.
\newblock Factor graphs and the sum-product algorithm.
\newblock {\em IEEE Trans. on Info. Theory}, 47(2):498--518, 2001.

\bibitem{Zdeborova:CSPIS}
L.~Zdeborov\'{a} and M.~M\'{e}zard.
\newblock Constraint satisfaction problems with isolated solutions are hard.
\newblock {\em J. Stat. Mech.}, 2008(12):P12004, 2008.

\bibitem{Janson:RG}
A.~Janson, T.~Luczak, and A.~Rucinski.
\newblock {\em Random Graphs}.
\newblock John Wiley \& sons, New York, NY, USA, 2000.

\bibitem{Boettcher:NWO}
S.~Boettcher and A.~Percus.
\newblock Nature's way of optimizing.
\newblock {\em Artif. Intell.}, 119(1-2):275--286, 2000.

\bibitem{Boettcher:NR}
S.~Boettcher.
\newblock Numerical results for ground states of mean-field spin glasses at low
  connectivities.
\newblock {\em Phys. Rev. B}, 67:060403, 2003.

\bibitem{Baxter:ESM}
R.J. Baxter.
\newblock {\em Exactly Solved Models in Statistical Mechanics}.
\newblock Academic Press, New York, NY, USA, 1982.

\bibitem{Mezard:BLSG}
M.~M\'{e}zard and G.~Parisi.
\newblock The \mbox{B}ethe lattice spin glass revisited.
\newblock {\em Eur. Phys. J. B}, 20(2):217--233, 2001.

\bibitem{Montanari:IOS}
A.~Montanari, G.~Parisi, and F.~Ricci-Tersenghi.
\newblock Instability of one-step replica-symmetry-broken phase in
  satisfiability problems.
\newblock {\em Journal of Physics A: Mathematical and General}, 37(6):2073,
  2004.

\bibitem{Bayati:ECM}
M.~Bayati, C.~Borgs, J.~Chayes, and R.~Zecchina.
\newblock On the exactness of the cavity method for weighted b-matchings on
  arbitrary graphs and its relation to linear programs.
\newblock {\em Journal of Statistical Mechanics: Theory and Experiment},
  2008(06):L06001, 2008.

\bibitem{Zdeborova:NMRG}
L.~Zdeborov\'{a} and M.~M\'{e}zard.
\newblock The number of matchings in random graphs.
\newblock {\em Journal of Statistical Mechanics: Theory and Experiment},
  2006(05):P05003, 2006.

\bibitem{Kanter:MFT}
I.~Kanter and H.~Sompolinsky.
\newblock Mean-field theory of spin-glasses with finite coordination-number.
\newblock {\em Phys. Rev. Lett.}, 58(2):164--167, 1987.

\end{thebibliography}
\end{document}